\documentclass[english,showpacs,twocolumn,superscriptaddress,showpacs,showkeys]{revtex4-1}
\usepackage{graphicx}
\usepackage[export]{adjustbox}
\usepackage{dcolumn}
\usepackage{bm}
\usepackage[usenames,dvipsnames]{color}
\usepackage{amsmath}
\usepackage{amssymb}
\usepackage{subcaption}
\usepackage[utf8]{inputenc}
\usepackage{url}
\usepackage{hyperref}
\usepackage[table]{xcolor}
\usepackage{appendix}

\graphicspath{{./Files/}}

\begin{document}


\title{Long time fate of 2-dimensional incompressible High Reynolds number Navier-Stokes
	turbulence: a quantitative comparision between theory and simulation }

\author{Shishir Biswas}
\email{shishirbeafriend@gmail.com}
\affiliation{Institute for Plasma Research, Bhat, Gandhinagar, Gujarat  382428, India}
\affiliation{Homi Bhabha National Institute, Training School Complex, Anushaktinagar, Mumbai 400094, India}

\author {Rajaraman Ganesh}
\email{ganesh@ipr.res.in}
\affiliation{Institute for Plasma Research, Bhat, Gandhinagar, Gujarat 382428, India}
\affiliation{Homi Bhabha National Institute, Training School Complex, Anushaktinagar, Mumbai 400094, India}
\date{\today}
\begin{abstract} 
Predicting the long time or late time states of 2-dimensional incompressible, high Reynolds number, slowly decaying turbulence has been one of the long standing problems. Using ``point vortices'' as ``inviscid'' building blocks,  but which do not respect incompressibility, statistical mechanical models conserving only total energy and zero total circulation result in the well known sinh-Poisson relation between vorticity and stream function.  On the other hand, statistical mechanics of ``inviscid patch'' vortices, which respects incompressibility by conserving regions of zero and nonzero vorticity, predicts a generalized relaxed state, which has never been systematically compared with direct numerical simulations (DNS).  In this study, starting from highly packed regions of nonzero initial vorticity, we demonstrate using high resolution, high Reynolds number DNS, that the late time states agree with predictions patch vortex models. As total circulation is reduced or diluted, we show that late time states of our DNS systematically and unambiguously leads to sinh-Poisson relationship between vorticity and stream function. We believe that our quantitative findings solve one of the long standing problems in 2-dimensional turbulence.
\end{abstract}
\maketitle

\section{Introduction}
Decaying incompressible 2-dimensional Navier-Stokes turbulence has been well explored via experiments, theory as well as by numerical simulation \cite{kraichnan:1980,brachet:1988,montgomery_d:1991}. Turbulence may be  defined as the nonlinear stage of a dynamical system with infinite degree of freedom, which, for example, could be a consequence of linear instabilities. Thus turbulence is typically unpredictable except certain features, such as its late time spatial energy spectra, velocity correlations \textcolor{black} {and so on} \cite{frisch_turbulence:1995}. In fully developed turbulence, several eddies and vortices of different sizes and strengths form, merge, transporting energy from a forcing scale to viscous scales via inertial scales. To understand late-time turbulence properties, certain macroscopic quantities such as enstrophy (square of vorticity in the volume/area of fluid flow), energy (Kinetic and/or Magnetic) and Circulation (vorticity contained in a volume/area of the fluid flow) are often invoked. For example, following Fjortoft's theorem for the 2-dimensional hydrodynamic turbulence, enstrophy is known to  cascade from smaller to larger $k$ (Direct Cascading) and the kinetic energy is known to  cascade to smaller $k$ values (Inverse cascade)\cite{ganesh_thesis:1998}, where $k$ is a typical wave number. Hence nonlinear energy transfer from one mode to another mode is mediated through the dynamics and merging of vortices, thus essentially regulating the time evolution of the fluid or plasma. \textcolor{black} {For 2-dimensional steady turbulence, Kraichnan showed that such a system admits simultaneously,  two inertial ranges i.e $E(k)\propto k ^{-3}$ and $E(k) \propto k^{-\frac{5}{3}}$ \cite{ganesh_thesis:1998}, where $k$ is a typical wave number. As can be expected, spectral indices for slowly decaying Navier-Stokes turbulence have been reported to become steeper with time \cite{gomez:1996}.} These are well established ideas supported by vast literature \cite{Pointin_point:1976, Thess_sheet:1994,Ganesh_patch:2002, Rupak_patch:2019}.

 For 2-dimensional Navier-Stokes turbulence, it is well known  that  the  ratio  of  enstrophy  to  kinetic   energy  is  a non  increasing  function  of  time, suggesting that the  enstrophy can decay significantly ``faster'' while the energy decays by a relatively negligible amount for high enough Reynolds number. This ``selective decay" \cite{Hasegawa:1985, selective_decay:1980, selective_decay:2005}  process  and its generalizations have been introduced as a possible explanation for relaxation and to predict the late time state of  ``slowly'' decaying 2-dimensional turbulence, in fluids, plasmas and in magnetofluids.
 
 Yet another school of thought to predict the late time fate of 2-dimensional slowly decaying Navier-Stokes turbulence is that of entropy extremization subject to conservation of kinetic energy and circulation in high Reynolds number Navier-Stokes turbulence \cite{montgomery_d:1991,sinh:1992,montgomery_prl:1991}. In particular, point vortices have been used as ``building blocks'' for statistical mechanical models, leading to the famous ``sinh-Poisson'' equation \cite{sinh:1992} connecting the late time vorticity and stream function. This relationship has been extensively tested using direct simulation of Navier-Stokes equations \cite{sinh:1992, gomez:1996}.  However, while the continuum fluid is regarded \textcolor{black}{as} incompressible, \textcolor{black}{whereas} point vortex  model  can accomodate any amount of kinetic energy as point vortices can be brought arbitrarily close to each other.
 
 To properly account for ``incompressibility'' in a statistical mechanical model of 2-dimensional Navier-Stokes turbulence, Kuz'min \cite{Kuzmin:1982}, Miller \cite{Miller_PRL:1990} and later Roberts \& Sommeria \cite{robert_sommeria:1991}, rather independently proposed that ``patch vortices''' or vortices with finite size, if used as ``building blocks'',  should be able to account for incompressibility effects. For example, for small values of total initial circulations of a given type (i.e, $+$ or $-$) viz $C_{\pm} =\int \omega_{\pm} dx dy$, where $\omega_{\pm}(x,y,t)$ is the local vorticity function of $+$ and $-$ type quantifying the $+$ and $-$ vortices at a given time $t$, the regions of zero circulation dominate and thus statistical mechanical predictions of point vortex model should suffice. However, for large values of initial total circulations, both regions of zero and nonzero circulation would play important role - for example, two vortex patches cannot occupy the same space at the same time (a classical exclusion principle) due to incompressibility, thus, resulting in the statistical mechanical generalization of ``sinh-Poisson'' equation for a system with total circulation $C = C_{+} + C_{-} = 0$ (in the rest of this discussion, we dub this finite size or patch vortex statistical mechanical theories as Kuz'min–Miller–Robert–Sommeria theory or KMRS theory). Except for a single attempt in the past \cite{Montgomery_POF:2003}, this generalized relationship  between vorticity and stream function for a 2-dimensional Euler-like turbulence has not been tested systematically using direct numerical simulation at very large Reynolds number and grid sizes. 
 
 It is important to \textcolor{black} {remind oneself} that the predictions regarding late time states resulting from both point vortex model \textcolor{black}{(PV)} as well as KMRS theory of patch vortices are strictly applicable in the limit of \textcolor{black}{Reynolds number} $R_n \rightarrow \infty$ for \textcolor{black}{an} incompressible fluid or for \textcolor{black}{an} incompressible 2-dimensional Euler turbulence. Thus both incompressibility effect (or classical exclusion principle) and high $R_n$ limit - are crucial conditions to be respected. Consequently, the direct numerical simulations (DNS) of 2-dimensional Navier-Stokes turbulence also requires to satisfy the same two conditions - high $R_n$ and incompressbility (or exclusion of occupied vortex regions). While the former condition is achieved by choosing high grid sizes, the later is achieved by  a suitable set of initial conditions which makes it possible to systematically \textcolor{black} {control the fraction} of initial occupied vortex regions. In the past, one such attempt has been reported \cite{Montgomery_POF:2003} to test the generality of sinh-Poisson model, however, the DNS was performed using a grid resolution of $512^2$ at $R_n \sim 10^4$, using \textcolor{black}{a} checker-board vortex configuration as initial condition. Clearly, in this work \cite{Montgomery_POF:2003}, the parameters used were inadequate to achieve the two crucial conditions of that of high $R_n$ (which \textcolor{black}{requires} high grid resolution) and controlled initial occupied vortex regions (and consequently truly incompressible vortex dynamics) to achieve an incompressible 2-dimensional Euler-like limit of 2-dimensional NS turbulence  and to put to test the concomitant predictions of statistical mechanics theories of vortices.  In the present work, we alleviate the above said drawbacks.
 
 In the present work, using high resolution direct simulation of 2-dimensional Navier-Stokes turbulence \textcolor{black}{at the grid size of $2048^2$} and at high Reynolds number $R_n=228576$, we systematically investigate the relaxed state of 2-dimensional Navier-Stokes decaying turbulence for various initial total circulation values of $C_{+}$ and $C_{-}$ such that total circulation $C= C_{+} + C_{-} =0$.  We consider finite size alternate sign vortex strips of different circulation ( $C_+$ \& $C_-$ ) as our initial condition. As there exists strong velocity shear between the vortex layer, such initial conditions are inherently Kelvin-Helmholtz unstable  and hence leads to turbulence. Our results are compared with the statistical mechanical model of patch vortices \cite{Kuzmin:1982,Miller_PRL:1990,robert_sommeria:1991} (KMRS theory) from small to large initial circulation values and a systematic deviation from sinh-Poisson model is demonstrated for increasing values of $C_+$ \& $C_-$ or in other words, for increasing area of occupancy of nonzero circulation. \textcolor{black}{ As is well known, maximum number of scales $k_{max}$ resolvable at a grid size say $N_{grid}$ is $k_{max} \sim N_{grid}/3$ \cite{frisch_turbulence:1995}. The corresponding maximum Reynolds number is $\simeq N_{grid}^2$ \cite{grid:2009}. In ref. \cite{Montgomery_POF:2003}, the grid sizes used are $512^2$ and working value of Reynolds number in ref. \cite{Montgomery_POF:2003} was  $R_n\sim10^4$.
 	In our work, $N_{grid} \sim 2048$ along one dimension, which is typically 4 times that used in ref. \cite{Montgomery_POF:2003}, thus allowing us a meaningful access to larger Reynolds number considered in the present work.}
 
The organization of the paper is as follows. In Sec. II we present the equations. Our numerical solver, simulation details and bench-marking of the solver are described in Sec. III. In Section IV we describe relevant physical background of the problem. The initial conditions, parameter details are shown in Sec. V.  Section VI is dedicated to the simulation results that we obtain from our solver. Lastly the summary and conclusions are listed in Sec. VII.

\section{Governing Equations}
\textcolor{black}{To} address the problem mentioned above we start with the Navier-Stokes equation. \textcolor{black}{Let} $L_0$ be the characteristic length-scale, $t_0$ be a characteristic timescale and $u_0$ be the initial velocity then, one can write Navier-Stokes equation in dimensionless form as follows,

\begin{eqnarray}
&& \label{norm momentum} \frac{\partial \vec{u}}{\partial t} + (\vec{u} \cdot \vec{\nabla})\vec{u} = \frac{1}{R_n}\nabla^2\vec{u}-\frac{1}{M_s^{2}}\frac{\vec{\nabla}P}{\rho}
\end{eqnarray}
 where $\rho$ and $\vec{u}$ are the \textcolor{black}{dimensionless} density and velocity of the fluid element respectively. $M_s$ is the dimensionless sonic Mach number defined as $\frac{u_0}{C_s}$, where $C_s$ is the sound speed, which is  regarded as independent of position and time. We also define dimensionless number, $R_n = \frac{u_0L_0}{\nu}$ as fluid Reynolds number, where $\nu$ is shear viscosity. In the rest of the discussion, all quantities are to be considered here as normalized, unless stated otherwise.
 
 Considering \textcolor{black}{the} incompressible limit of the flow (i.e, $M_s \rightarrow \infty$) and taking curl on the both side of the above equation, one obtains the evolution equation of scalar  vorticity field in normalized form as,
\begin{eqnarray}
&& \label{vorticity} \frac{\partial \omega}{\partial t}  = \left[\psi,\omega \right] + \frac{1}{R_n}\nabla^2\omega
\end{eqnarray}
where $\psi(x,y,t)$ is the stream function, which relates the two dimensional velocity field by, $u_x$ = $\partial_y\psi$, $u_y$ = -$\partial_x\psi$ and $\omega(x,y,t) = {\vec \omega}\cdot {\hat z} = \partial_xu_y - \partial_yu_x$ is the scalar vorticity field, which satisfies,
\begin{eqnarray}
&& \label{poisson_eqn} \omega  = - \nabla^2\psi
\end{eqnarray}
The Poisson bracket of Eq. \ref{vorticity} is defined as,
\begin{eqnarray}
&& \label{poisson_bkt} \left[\psi,\omega \right] = \partial_x\psi\partial_y\omega - \partial_y\psi\partial_x\omega
\end{eqnarray}
\label{equations}
For solving the above Eqs. [\ref{vorticity},\ref{poisson_eqn},\ref{poisson_bkt}] at very high resolution,  a scalable numerical solver is developed and benchmarked, the details of which are given in the coming sections.
\section{Simulation Details : GPU Based Numerical Solver and Benchmarking}
 Recently, we have  upgraded an existing incompressible 2-dimensional hydrodynamic solver developed in house at Institute for Plasma Research \cite{rupak_thesis:2019}, to GPU architecture for better performance. The newly upgraded GPU based incompressible hydrodynamic solver [GHD2D] is now capable of handling $2048^2$ matrix or more, routinely. The solver uses pseudo-spectral technique, one of the most accurate computational fluid dynamics (CFD) techniques available today. In this pseudo-spectral technique, one calculates the spatial derivative to evaluate non-linear term in governing equation, followed by a standard $\frac{2}{3}$ dealiazing rule \cite{dealiasing:1971}. Eqs. \ref{vorticity},\ref{poisson_eqn},\ref{poisson_bkt} are evaluated in two dimensions, with a square doubly periodic box in cartesian coordinate. We use CUDA based FFT library [cuFFT library]\cite{cufft} to perform \textcolor{black}{F}ourier transforms and Adams-Bashforth time solver for time \textcolor{black}{integration}. For visualization an in-house developed Python based code is used which is based on an open source Python module named ``mpl toolkits.mplot3d"\cite{mpl}.
 
\begin{figure*}
	\centering
	\begin{subfigure}{0.32\textwidth}
		\centering
		\includegraphics[scale=0.38]{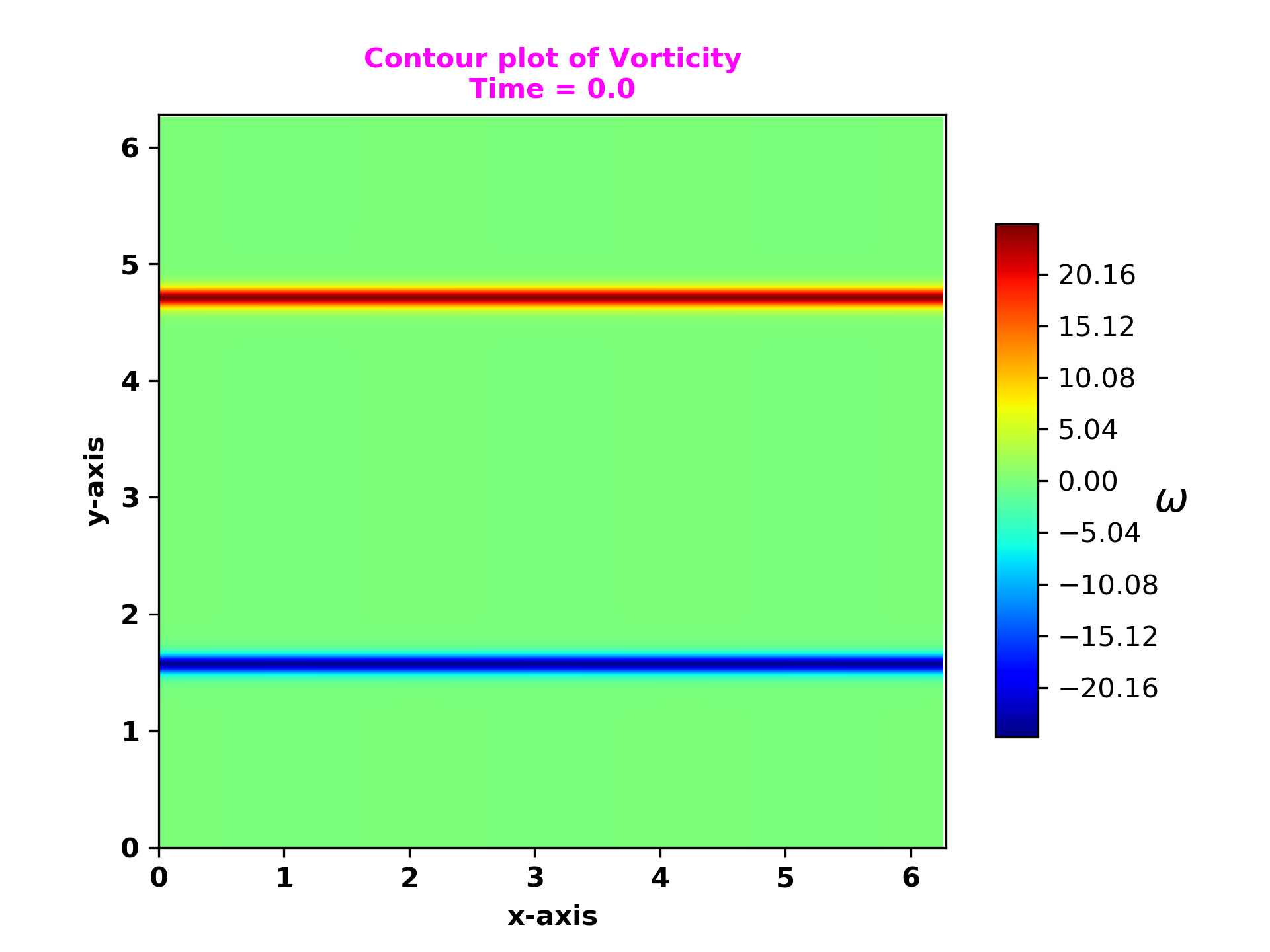}
		\caption*{(a)}
	\end{subfigure}
	\begin{subfigure}{0.32\textwidth}
		\centering
		\includegraphics[scale=0.38]{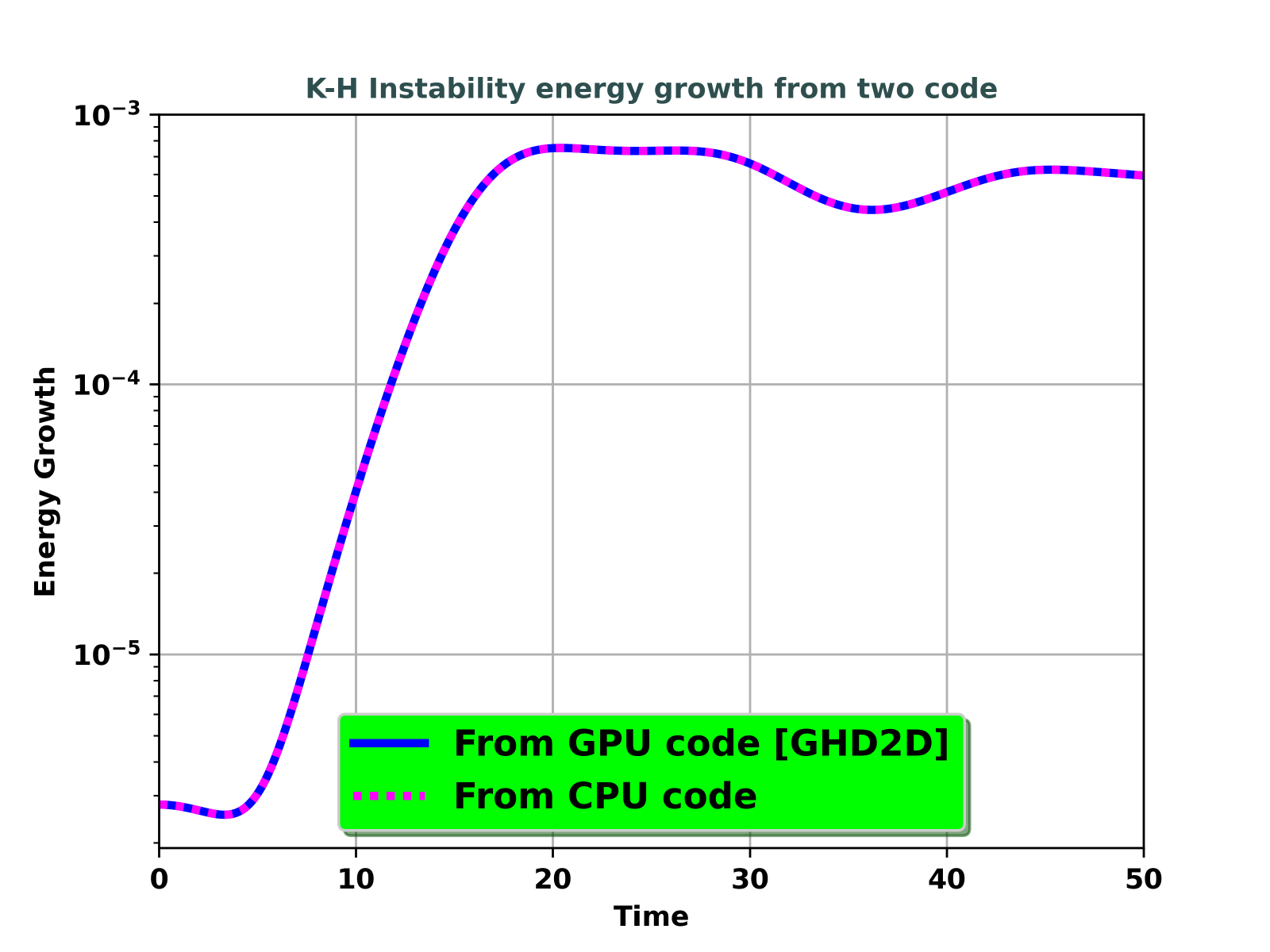}
		\caption*{(b)}
	\end{subfigure}
	\begin{subfigure}{0.32\textwidth}
		\centering
		\includegraphics[scale=0.38]{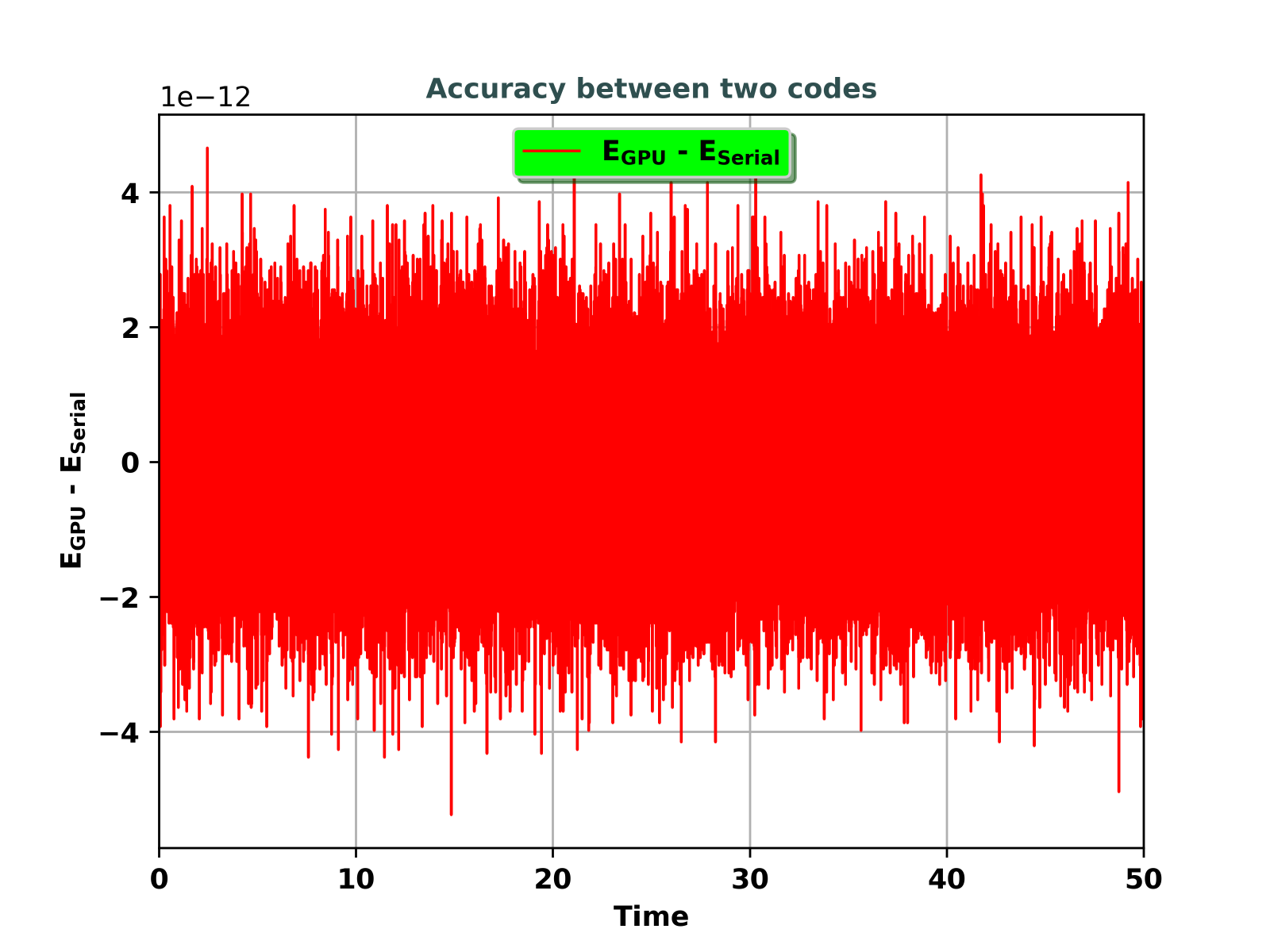}
		\caption*{(c)}
	\end{subfigure}
	\caption{(a) Initial condition : two oppositely directed jets (broken-jet) each of width $\frac{3\pi}{128}$. (b) Time evolution of kinetic energy in the direction perpendicular to the flow direction is evaluated with time.(c) Numerical accuracy between the CPU and GPU solver at $256^2$ grid resolution.}
	\label{yenergy2code}
\end{figure*}

\begin{figure}
\includegraphics[scale=0.55]{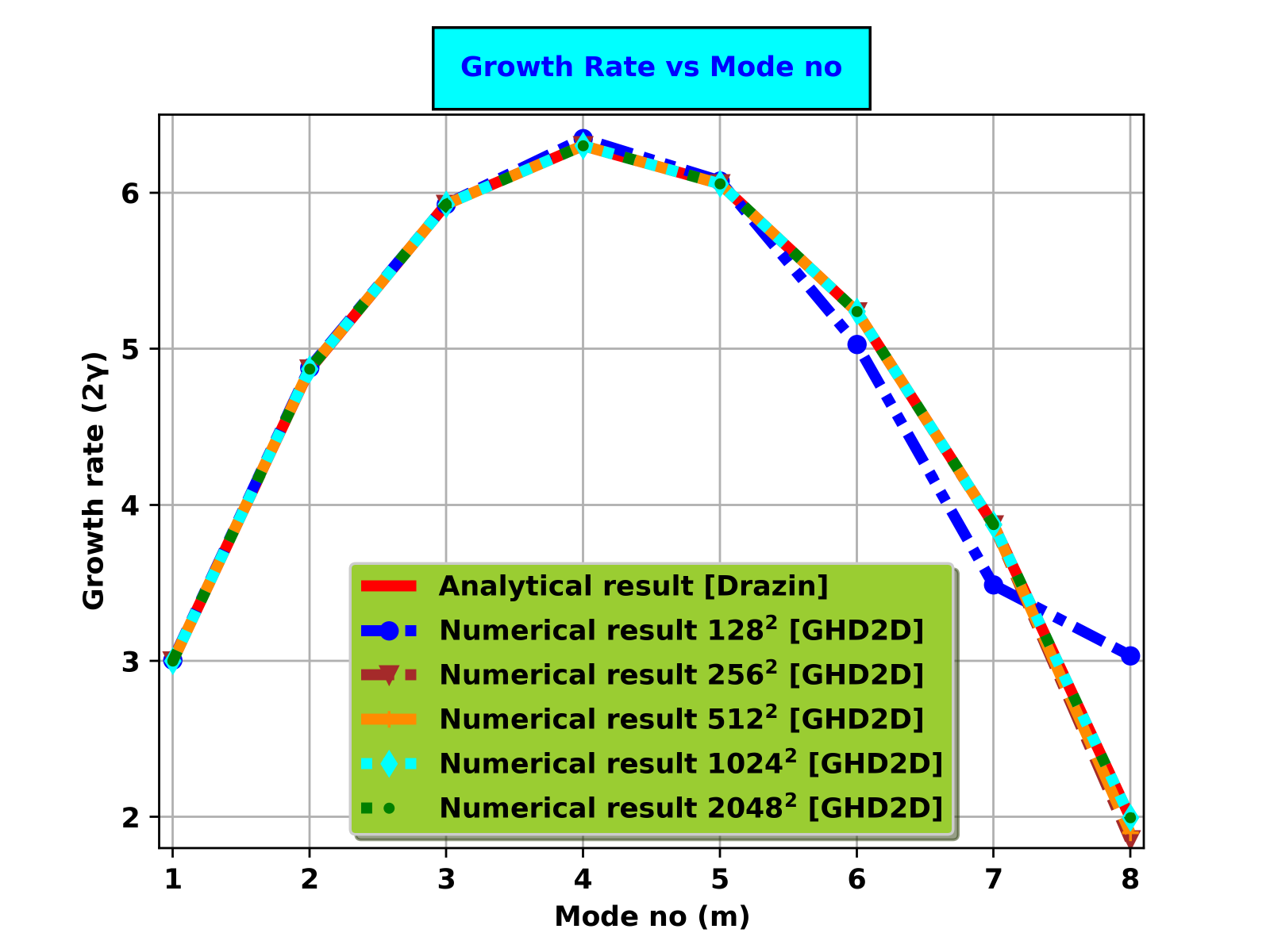}
\caption{Growth rate ($2\gamma$) of K-H instability is plotted with mode number of excitation. The solid red line is evaluated from the analytical expression obtained by Drazin \cite{drazin:1961}. The rest of the lines  represents the growth rate from GHD2D solver at different grid resolution.}
\label{growth}
\end{figure}

To cross-check the \textcolor{black}{accuracy of our solver}, we use two oppositely directed jets (i.e, broken jets) [Fig. \ref{yenergy2code}(a)] in a doubly periodic domain at \textcolor{black}{the} incompressible limit and  estimate  the growth rate of Kelvin-Helmholtz instability, using the sequential version and upgraded GPU version of the solver. It is shown that, the data accuracy between the two solvers match up-to machine precision [[Fig. \ref{yenergy2code}(b)], Fig. \ref{yenergy2code}(c)].
For further  benchmarking, we reproduce \textcolor{black}{numerically} the growth rate of Kelvin-Helmholtz instability, analytically calculated earlier by Drazin\cite{drazin:1961} for broken-jet equilibrium. We obtain identical growth rates from our newly upgraded solver as Drazin had predicted [Fig. \ref{growth}].

\section{Extremization models : Late time state prediction of 2-dimensional high Reynolds number NS turbulence }
The late time state for a 2-D Navier-Stokes turbulence is predicted via two extremization models so far. One is enstrophy extremization where enstrophy is considered to decay faster than energy, and the other is entropy extremization keeping kinetic energy as a conserved quantity.
\subsection{Enstrophy extremization:}
For freely decaying high Reynolds number two-dimensional (2-dimensional) turbulence, the most direct consequence is rapid decrease of enstrophy  ($\Omega$ = $\frac{1}{2}\int \omega^2 dxdy$), relative to energy ($E = \frac{1}{2}\int \psi\omega dxdy$). Since,
\begin{eqnarray}
\dot{E} = -2\frac{\Omega}{R_n}\\
\dot{\Omega} = -2\frac{P}{R_n}
\end{eqnarray}
where P ($P(k) = \sum k^6\left\lvert \psi_k \right\rvert^2$) is palinstrophy, it is easily verified from the above set of equations that,
\begin{equation}\label{enstrophy_to_energy}
\frac{d}{dt}\left(\frac{\Omega}{E}\right)\leq 0.
\end{equation}
Hence for all 2-dimensional decaying incompressible flows, whether it is turbulent or not, the ratio of enstrophy to energy is expected to show monotonic decay in time. This observation is the basis of the selective decay model\cite{selective_decay:1980, Hasegawa:1985, selective_decay:2005}. As the enstrophy decays faster than energy, a variational principle may be envisaged by minimizing enstrophy keeping energy constant \cite{gomez:1996}. For example a free energy, $F = \Omega - \lambda E$ may be defined with $\lambda$ as a Lagrange multiplier. Variation of $F$ with respect to $\omega$, leads to a linear relationship between extremum states of $\omega$ and $\psi$ as, 
\begin{equation}
\bar \omega = - \nabla^2 \bar \psi = \lambda \bar \psi
\end{equation}
such that extremum state vorticity ($\bar{\omega}$) is proportional to the stream function ($\bar{\psi}$) where the value of $\lambda$ is to be determined from energy E.
\subsection{Entropy extremization:}
In the past, entropy extremization principle has been invoked to predict the late time relaxed state of a 2-dimensional Navier-Stokes turbulent system. This extremization principle leads two different explicit theories.
\subsubsection{Point Vortex Theory:}
Using \textcolor{black}{a} point vortex \textcolor{black}{(PV)} model, Lundgren and Pointin \cite{Pointin_point:1976} , Montgomery et al. \cite{Montgomery_Joyce:1974} showed by extremizing entropy that one can obtain an relationship between extremized vorticity and stream function as,
\begin{equation}\label{sinh for point vortex}
\bar{\omega} = -\nabla^2 \bar{\psi} = \alpha sinh(-\beta \bar{\psi})
\end{equation}
which is the well-known sinh-Poisson equation. The coefficient $\alpha$ and $\beta$ are obtained from conservation of total circulation and total energy respectively. Later, Montgomery et al.\cite{sinh:1992} and others \cite{gomez:1996} showed by DNS that the vorticity and the stream function is described by the Sinh relationship as predicted by point vortex theory. The above-said works, considered random vortices as initial condition, with low initial total circulation for each sign of vorticity, so that the total circulation is zero. Moreover Montgomery et al.\cite{sinh:1992} considered grid resolution \textcolor{black}{of $512^2$} and Reynolds number $R_n$ $\sim10^4$ for their numerical experiment, where as, Dmitruk et al.\cite{gomez:1996} considered $96^2$ grid resolution and Reynolds number $R_n$ $\sim10^3$. As discussed earlier, to answer the question about departure of the late-time states of 2-dimensional decaying NS turbulence from sinh-Poisson model, it is required to investigate the final relaxation state for different initial vortex configurations analogues to point vortices at very high grid resolution and at maximum possible Reynolds number ($R_n$).
 
\subsubsection{Finite size vortex or patch vortex theory:}
From \textcolor{black}{a} statistical mechanics point of view, by considering the finite size vortices which take\textcolor{black}{s} into account the incompressibility effect (KMRS theory), it has been shown \cite{Kuzmin:1982,Miller_PRL:1990,robert_sommeria:1991} that vorticity and stream function  obeys the general relation for a system with zero total circulation, or in the other words, the most probable state is expressed as,
\begin{equation}\label{most_probable_finite_size}
 \bar{\omega} = \frac{Aexp(-B\bar{\psi}) - Cexp(B\bar{\psi})}{1+\left[Aexp(-B\bar{\psi}) + Cexp(B\bar{\psi})\right]}
\end{equation}
where A,B,C are the real constants. This is the most probable state based on a finite size vortex model. The corresponding Poisson equation is written as,

\begin{equation}\label{most_probable_finite_Poisson}
-\nabla^2 \bar{\psi} =  \frac{Aexp(-B\bar{\psi}) - Cexp(B\bar{\psi})}{1+\left[Aexp(-B\bar{\psi}) + Cexp(B\bar{\psi})\right]}
\end{equation}
To go to point vortex model from this finite size vortex model, we choose initial conditions such that the the circulation of each type of vortices ($C_+$ \& $C_-$) is monotonically reduced.  Or in other words, starting from KMRS theory, decreasing area of occupancy of nonzero \textcolor{black}{initial} circulation leads to the results of point vortex theory, wherein, we can neglect the exclusion principle\textcolor{black}{,} which would \textcolor{black}{then} result in reduced magnitudes of  the coefficient $A$ and $C$ with respect to ``1'' in the denominator of  Eq. \ref{most_probable_finite_size}. Consequently, the most probable state Eq. \ref{most_probable_finite_size} becomes,
\begin{equation}\label{most_probable_point}
 \bar{\omega} = Aexp(-B\bar{\psi}) - Cexp(B\bar{\psi})
\end{equation}

Eq. \ref{most_probable_point} is nothing but the Sinh relation obeyed between $\bar{\omega}$ and $\bar{\bar{\psi}}$, for comparable values of $A$ and $C$, i.e. $A = C = \frac{A'}{2}$. The corresponding Poisson equation reads as,

\begin{equation}
-\nabla^2 \bar{\psi} =  \bar{\omega} = A'sinh(-B \bar{\psi})
\end{equation}
which is the well known sinh-Poisson equation. Hence, at low circulation, KMRS model [Eq. \ref{most_probable_finite_Poisson}] reduce\textcolor{black}{s} to point vortex model [Eq. \ref{sinh for point vortex}].

In the following Sections, we discuss the direct numerical simulations (DNS) of 2-dimensional decaying, incompressible NS equation for various initial individual circulation values (or occupancy) such that the total initial circulation is always zero.  
\label{physics}
\section{Initial Condition}\label{initial condn}
\begin{figure*}
	\centering
	\begin{subfigure}{0.24\textwidth}
		\centering
		\includegraphics[scale=0.36]{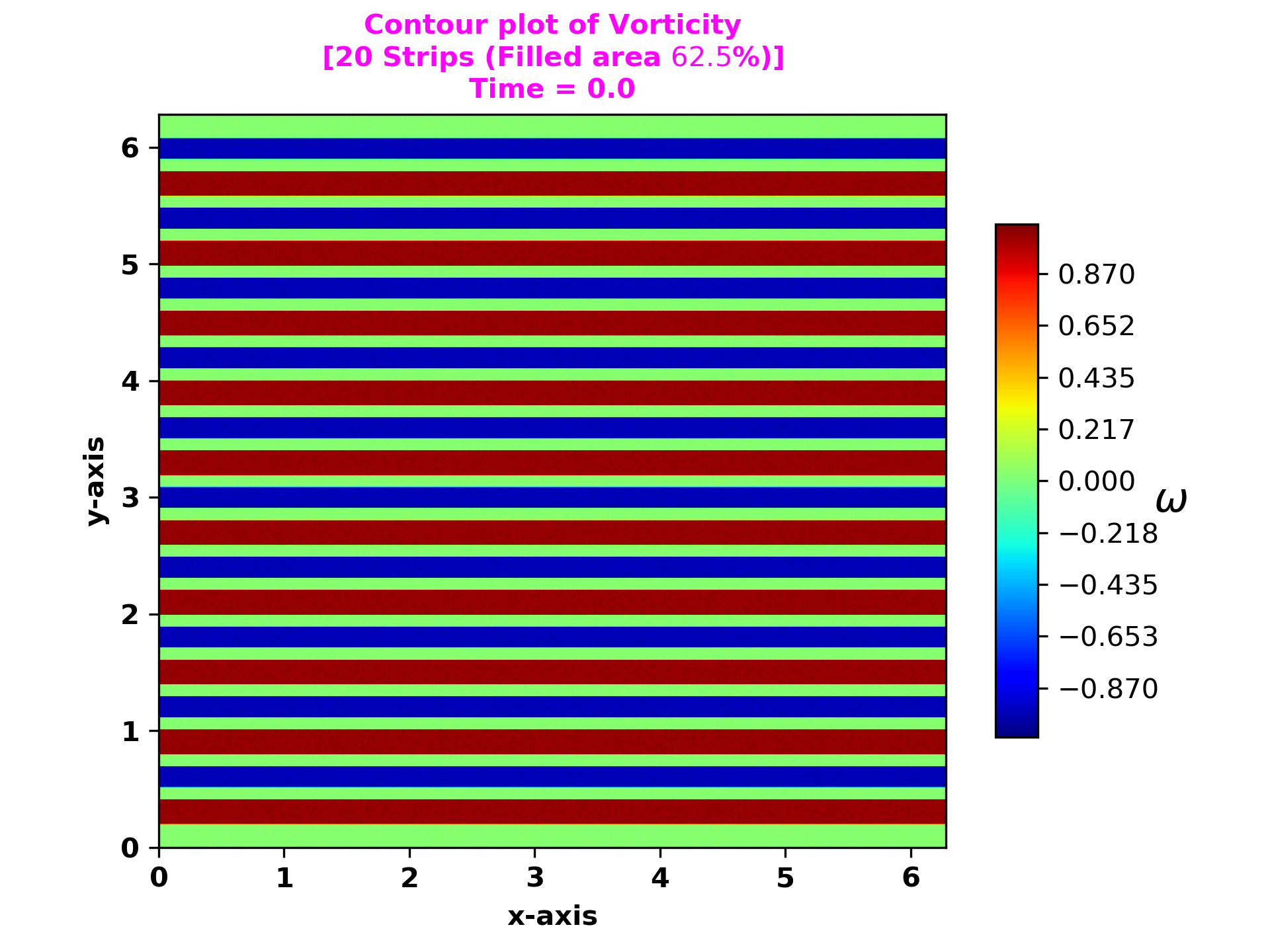}
		\caption{}
	\end{subfigure}
	\begin{subfigure}{0.24\textwidth}
		\centering
		\includegraphics[scale=0.36]{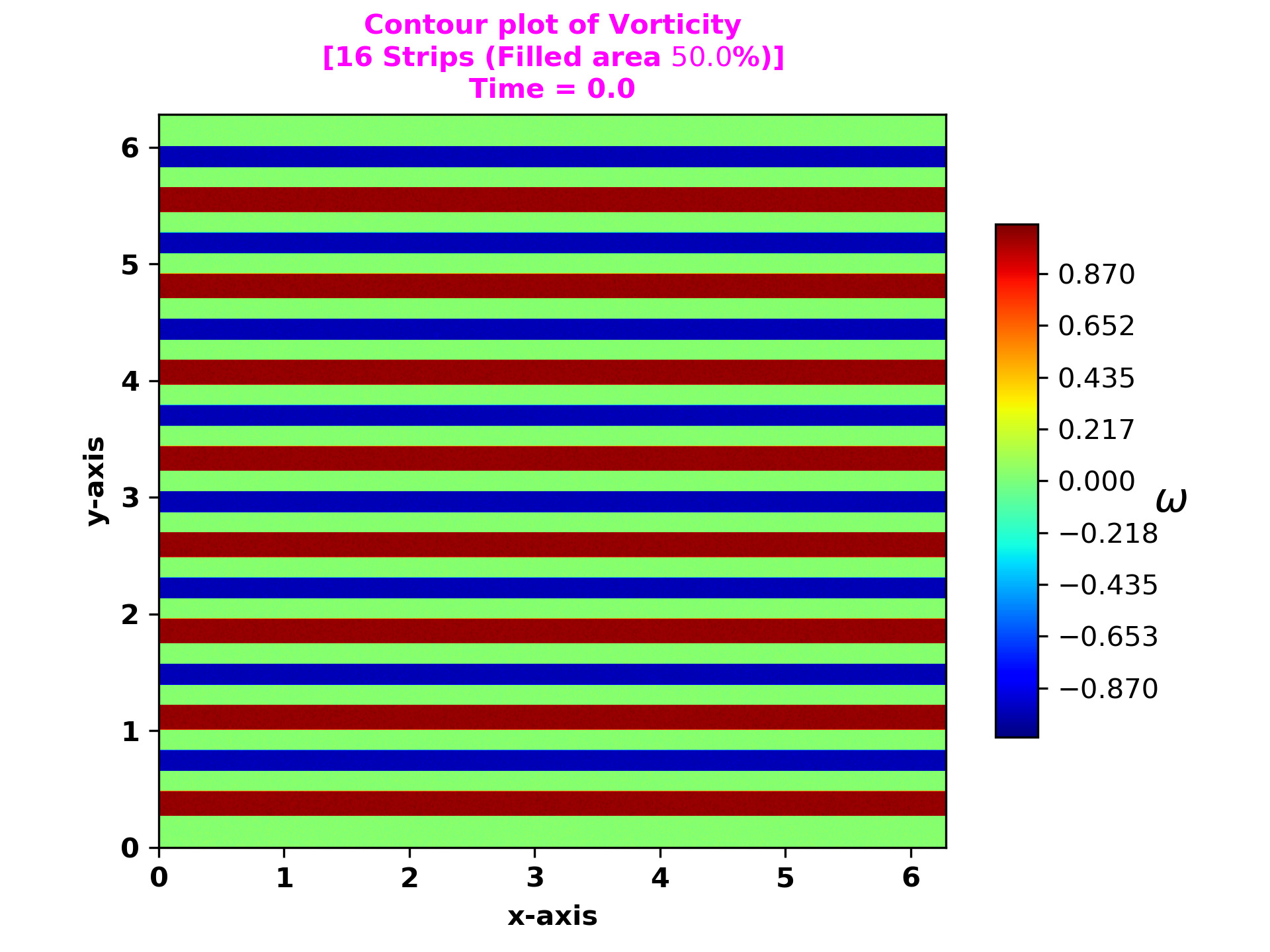}
		\caption{}
	\end{subfigure}
	\begin{subfigure}{0.24\textwidth}
		\centering
		\includegraphics[scale=0.36]{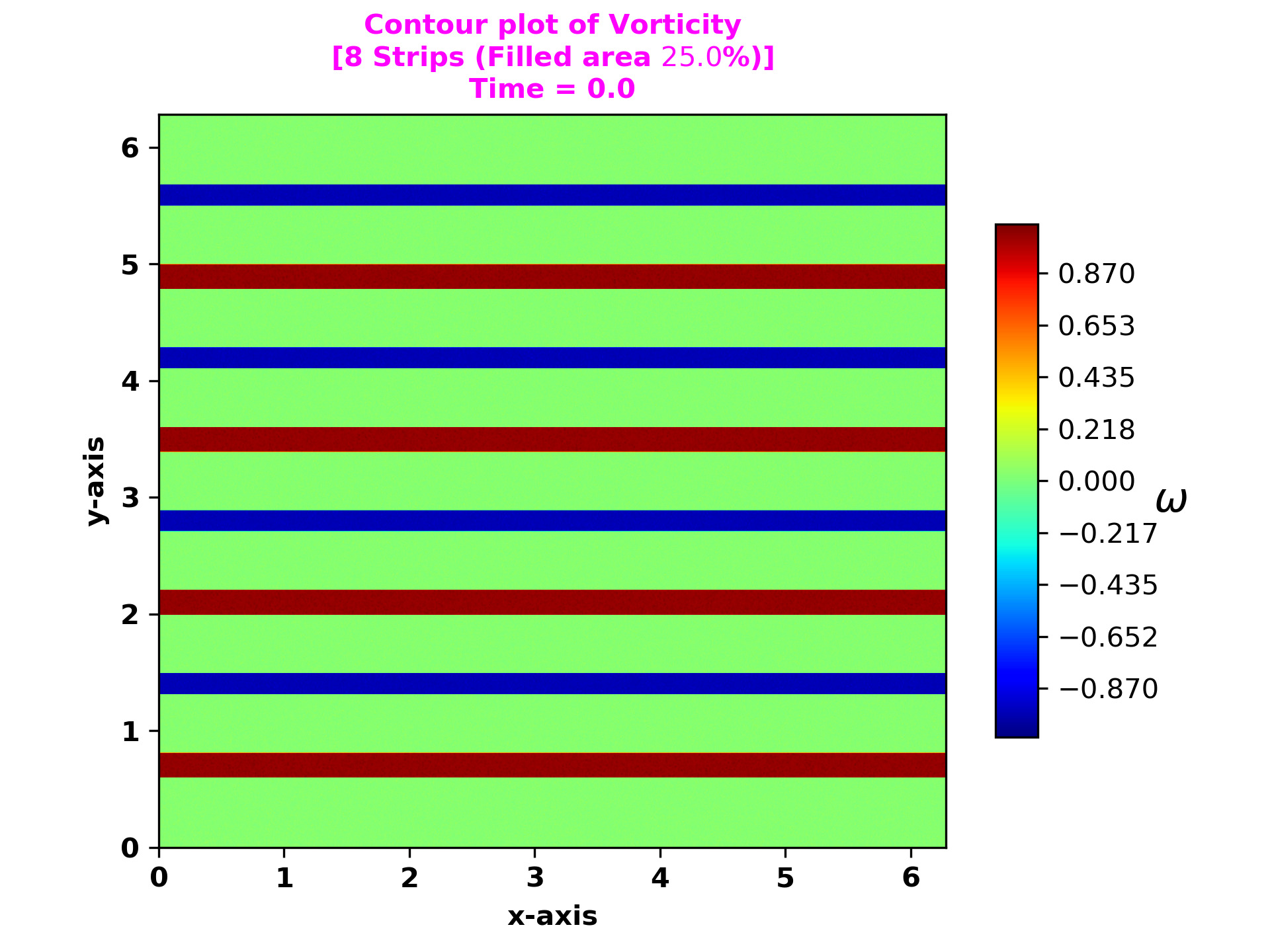}
		\caption{}
	\end{subfigure}
	\begin{subfigure}{0.24\textwidth}
		\centering
		\includegraphics[scale=0.36]{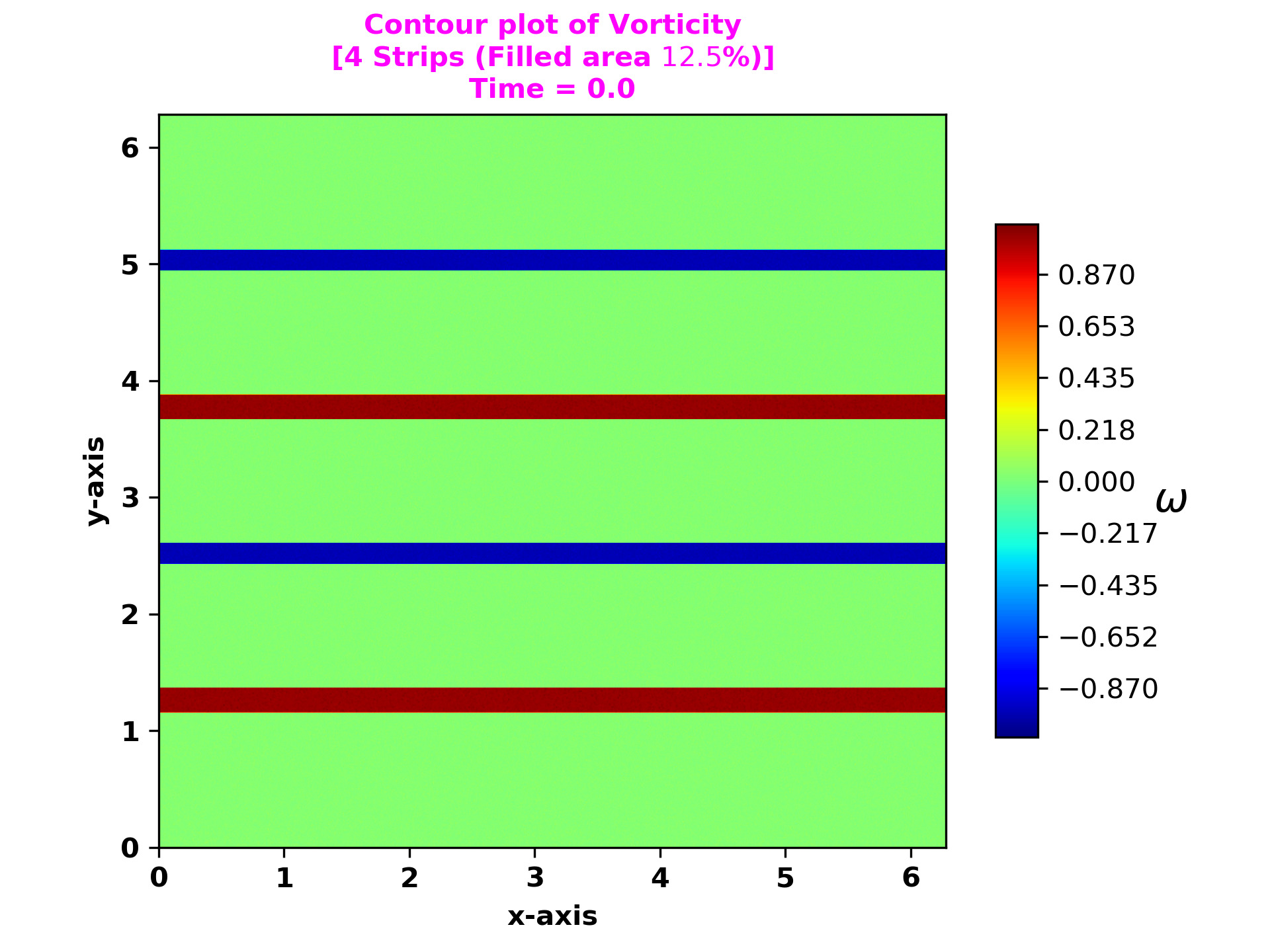}
		\caption{}
	\end{subfigure}
\caption{Initial vorticity distribution for, (a) 20 number of vortex strips each of width $\delta = \frac{\pi}{16}$ and circulation due to one kind of vortices are $\frac{5}{4}\pi^2$,  (b) 16 vortex strips each of width $\delta = \frac{\pi}{16}$ and circulation due to one kind of vortices are $\pi^2$, (c) 8 vortex strips each of width $\delta = \frac{\pi}{16}$ and circulation due to one kind of vortices are $\frac{1}{2}\pi^2$, (d) 4 vortex strips each of width $\delta = \frac{\pi}{16}$ and circulation due to one kind of vortices are $\frac{1}{4}\pi^2$. The vorticity value for blue, brown and green regions are $-1$, $+1$ and $0$ respectively.}
\label{initial}
\end{figure*}

To quantify the above discussed analytical predictions via numerical simulation,  we choose a set of initial equilibrium vortex configuration at very high resolution ($2048^2$) at reasonably large Reynolds number $R_n$ ($R_n =228576$) which are inherently unstable. As discussed in the Introduction, it is important to \textcolor{black}{keep in mind that} the predictions regarding late time states resulting from both point vortex model as well as KMRS theory of patch vortices are strictly applicable in the limit of $R_n \rightarrow \infty$ for \textcolor{black}{an} incompressible fluid or for \textcolor{black}{an} incompressible 2-dimensional Euler turbulence. Thus both incompressibility effect (or classical exclusion principle) and high $R_n$ limit - are crucial conditions to be respected. Consequently, the direct numerical simulations (DNS) of 2-dimensional Navier-Stokes turbulence also requires to satisfy the same two conditions - high $R_n$ and incompressbility (or exclusion of occupied vortex regions). While the former condition is achieved by choosing high grid sizes, the later is achieved by  a suitable set of initial conditions which makes systematically controls the occupied vortex regions. In the past one such attempt has been reported \cite{Montgomery_POF:2003} to test the generality of sinh-Poisson model, however, the DNS was performed using a grid resolution of $512^2$ at $R_n \sim 10^4$, using \textcolor{black}{a} checker-board vortex configuration as initial condition. Clearly, in this work \cite{Montgomery_POF:2003}, the parameters used were inadequate to achieve the two crucial conditions of that of high $R_n$ (which demains high grid resolution) and controlled initial occupied vortex regions (and consequently truly incompressible vortex dynamics) to achieve an imcompressible 2-dimensional Euler-like limit of NS turbulence  and to put to test the concomitant predictions of statistical mechanics theories of vortices.  In the present work, we alleviate the above said drawbacks of the DNS by choosing high grid resolution of $2048^2$, high $R_n = 228576$ and the following initial conditions:  

\begin{itemize}
	
\item\textbf{Case A:} In a $2\pi \times 2\pi$ domain, we use oppositely directed jets (i.e, broken jets) placed one after another alternately. Between two alternate jets there is a non circulating region, thus resulting in \textcolor{black}{out-of-plane} vorticity $\omega$. The width of each vortex strip thus formed is $\delta = \frac{\pi}{16}$. First we consider total 20 number of strips, of width $\delta$. Consequently, out of the total area of $2\pi \times 2\pi$, an area of  $20 \times \delta \times 2\pi$ is filled with vorticity of $\omega_{\pm}$, so the initial total vortex packing fraction is $\frac{20\times \delta \times 2\pi}{2\pi \times 2\pi}$, which is $62.5\%$ and remaining $37.5\%$ is zero vortex region [Fig. \ref{initial} (a)]. The total circulation is defined as $\int \omega dx dy$ and the same due to positive strips ($C_+$) are, $C_+ = n_+ \times \delta \times L \times \omega_+$, where $n_+$ is the number of positive strips, $\delta$ is the width of each positive strips, $L_x = L_y = L = 2\pi$ is the system length and $\omega_+$ is the strength of each positive vortex strip. Hence one obtains $C_+ = \frac{20}{16}\pi^2$, similarly the circulation due to minus strips are ($C_-$) = $-\frac{20}{16}\pi^2$. As equal number of positive  and negative strips are considered, so total initial circulation $C = C_{+} + C_{-} = \int \omega dx dy$ = 0.

\item\textbf{Case B:} Instead of 20 alternate sign vortex strips, here we consider only 16 vortex strips of the same width $\delta = \frac{\pi}{16}$. Among 16 strips 8 of them are positive vorticity and rest 8 are negative vorticity [Fig. \ref{initial} (b)]. In Case A above, the vorticity packing fraction was $62.5\%$, which is now reduced to $50.0\%$  and the remaining $50.0\%$ of the domain contains zero vorticity. Circulation due to positive vortex strips ($C_+$) are, $C_+ = \frac{16}{16}\pi^2$, and for minus vortex strips ($C_-$) = $-\frac{16}{16}\pi^2$. Here we point out that the total initial circulation is $C = C_{+} + C_{-} = \int \omega dx dy$ = 0, where as circulation for positive and negative strips are reduced.

\item\textbf{Case C:} For reducing the packing fraction further we consider 8 alternate sign vortex strips, in which 4 of them are positive and rest 4 of them are negative, keeping the width of each strips same as earlier [Fig. \ref{initial} (c)]. Consequently, the vorticity packed domain is reduced from $50.0\%$ to  $25.0\%$ such that the remaining $75.0\%$ simulation domain contains zero vorticity. Circulation for positive strips ($C_+$) for this case are, $C_+ = \frac{8}{16}\pi^2$ and for negative strips ($C_-$) = $-\frac{8}{16}\pi^2$. Here also, we \textcolor{black} {consider} that circulation for positive and negative strips are reduced further by keeping the vortex width same.

\item\textbf{Case D:} Finally we consider only 4 vortex strips. Among them 2 are positive and 2 are negative keeping the width of each strips same as earlier cases [Fig. \ref{initial} (d)]. For this last case, the vorticity filled domain is $12.5\%$ only and rest $87.5\%$ simulation domain is \textcolor{black}{of zero vorticity}. Like earlier cases, we calculate circulation for positive strips ($C_+$) as, $C_+ = \frac{4}{16}\pi^2$, similarly for negative strips ($C_-$) = $-\frac{4}{16}\pi^2$. Here also, we see that by keeping the total initial circulation zero ($\int \omega dx dy = 0$), circulation for plus and minus strips are reduced.

We therefore gradually reduce the vortex packing fraction from $62.5\%$ to $12.5\%$ keeping the total circulation zero ($C = C_{+} + C_{-} = \int \omega dx dy = 0$). High packing fraction configuration ($62.5\%$) with high individual circulation, implies large regions of nonzero vorticity, thus the effect of incompressibility or \textcolor{black}{classical} exclusion should be predominant. One may expect that the late time states of our 2-dimensional NS \textcolor{black}{h}igh $R_n$ simulations should match closely to that of the statistical mechanical predictions of finite size vortex model (KMRS theory) discussed above, resulting in deviation from sinh-Poisson results, while as the packing fraction is reduced, the findings from DNS would agree with both KMRS at low circulation and point vortex model tending towards sinh-Poisson equation.
\end{itemize}

\begin{itemize}
	
\item For all the cases addressed above, we perturb the system by the following perturbation scheme: 
\begin{equation}
\label{perturb} \omega^{Perturbation} = \sum_{m=1}^{8} 0.01 \times \cos(mx+\phi_m)
\end{equation}
where m is the mode of perturbation and $\phi_m$ is the phase in perturbation. We initialize $\phi_m$ either  as $0$ or with random white noise between $-\pi$ to $\pi$.
\end{itemize}

As described earlier, we evolve the vortex strips, with different packing fraction in a system of area $(2\pi)^2$, with grid resolution $(2048)^2$, with time steps $(10^{-4})$. Here we use very high grid resolution ($2048^2$), which is $4$ times larger than Montgomery et al. \cite{montgomery_d:1991}. Also as the Reynolds number varies as square of the grid resolution, hence, for all the runs here we use Reynolds number ($R_n$) = $228576$, which is $16$ times of that of the $R_n$ values used in Montgomery et al. \cite{montgomery_d:1991}.  A summary of  parameter details for the simulation is given in the Table \ref{parameter_run_stripvortex}. With these initial conditions and parameter spaces we present our simulation results.

\begin{table*}
\centering
\begin{tabular}{ |c|c|c|c|c|c|c|c| }
 \hline
\textbf{Case} & \textbf{Run} & \textbf{Strips} & \textbf{Initial Total Vorticity Packing Fraction} & \textbf{Grid Size} &  \textbf{$C_+ (= |C_{-}|)$} & \textbf{$R_n$} & \textbf{Phase ($\phi_m$)} \\
 \hline
 A & 1 & 20 Strips & $62.5\%$ & $2048^2$ & $\frac{5}{4}\pi^2$& $228576$ & 0 \\
 \hline 
A & 2 & 20 Strips & $62.5\%$ & $2048^2$ & $\frac{5}{4}\pi^2$& $228576$ & rand($-\pi$, $\pi$) \\
 \hline
B & 3 & 16 Strips & $50.0\%$ & $2048^2$ & $\pi^2$& $228576$ & 0 \\
 \hline
B & 4 & 16 Strips & $50.0\%$ & $2048^2$ & $\pi^2$& $228576$ & rand($-\pi$, $\pi$) \\
 \hline
C & 5 & 8 Strips & $25.0\%$ & $2048^2$ & $\frac{1}{2}\pi^2$& $228576$ & 0 \\
 \hline
C & 6 & 8 Strips & $25.0\%$ & $2048^2$ & $\frac{1}{2}\pi^2$& $228576$ & rand($-\pi$, $\pi$) \\
 \hline
D & 7 & 4 Strips & $12.5\%$ & $2048^2$ & $\frac{1}{4}\pi^2$& $228576$ & 0 \\
 \hline
D & 8 & 4 Strips & $12.5\%$ & $2048^2$ & $\frac{1}{4}\pi^2$& $228576$ & rand($-\pi$, $\pi$) \\
 \hline
\end{tabular}
\caption{Parameter details with which the simulation has been run for strip vortex problem. $C_+$ or $C_+$ such that initial $C_{total} = C_{+} + C_{-} = \int \omega dx dy = 0$ for all cases.}
\label{parameter_run_stripvortex}
\end{table*}
\section{Numerical results}
We evolve the above discussed initial conditions using our solver GHD2D. For all the cases, we use the perturbation scheme according to Eq. \ref{perturb} as indicated earlier. To make the perturbation realistic, we sum the perturbation over several mode numbers and also add random phase as well.  To establish the robustness of our findings, we present below, results for random phase and zero phase, using a sum of 8 modes in the perturbation.    
\subsection{20 Vortex Strips with total Packing Fraction $62.5\%$  - Runs 1, 2}\label{20 Strips}
As discussed in Section \ref{initial condn} we use 20 parallel strips of identical widths here as our initial vorticity distribution. From Fig. \ref{initial} (a) it is evident that this initial conditions are tightly packed vortex configurations keeping the total initial circulation is zero, i.e. $\int\omega dx dy$ = $0$. As there exists a shear between the vorticity layers, the strips are  Kelvin-Helmholtz unstable when perturbed. Eventually the vortex configuration evolves towards turbulence and system is dominated by turbulence associated with a rapid mixing of vortex layers. It is well known that in 2-dimensional Hydrodynamics, vortices of  same sign attract and merge while vortices of opposite sign repeal each other.\\

\begin{figure*}
	\centering
	\begin{subfigure}{0.32\textwidth}
		\centering
		\includegraphics[scale=0.39]{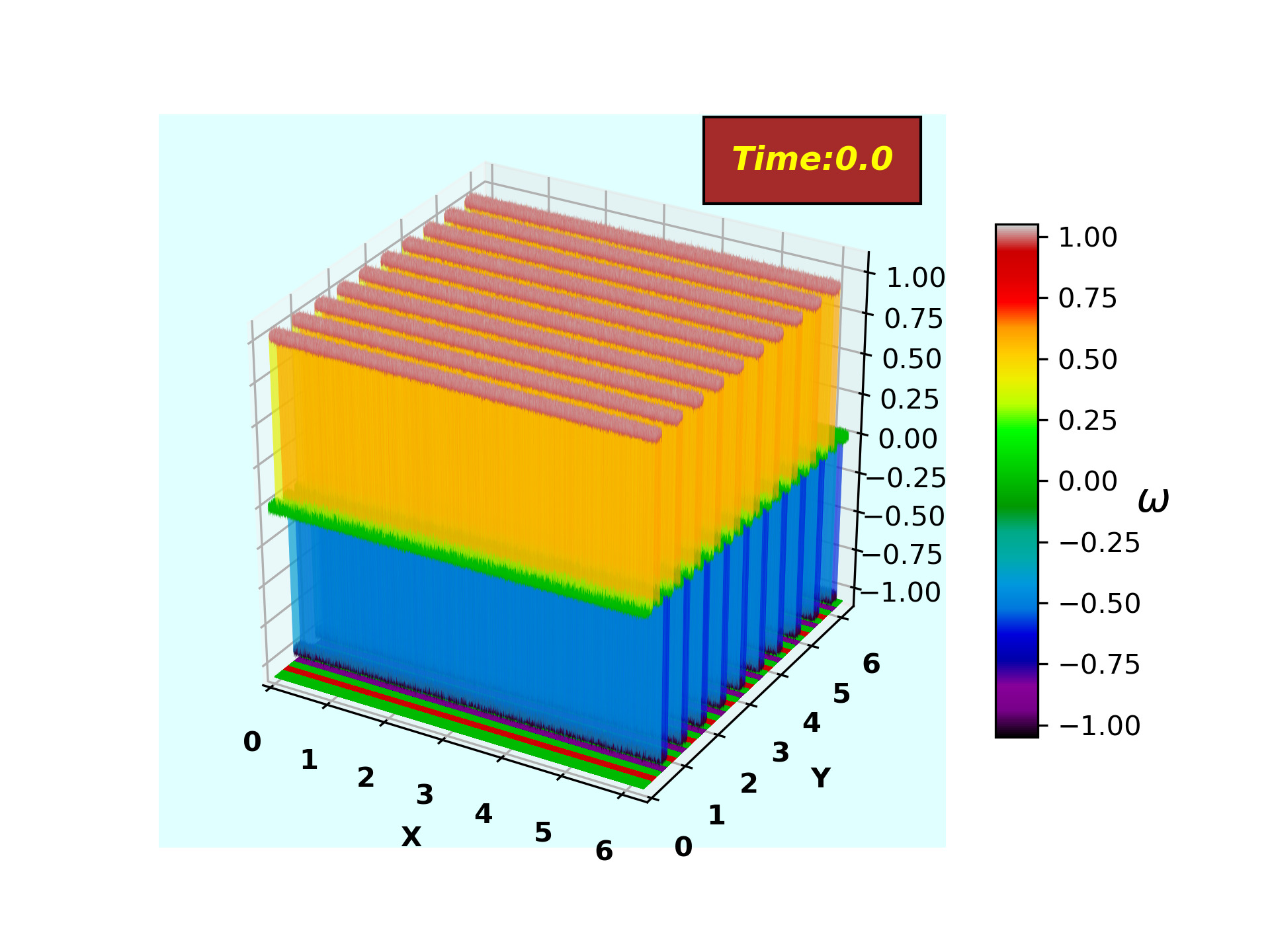}
		\caption{\textbf{Time: 0.0}}
	\end{subfigure}
	\begin{subfigure}{0.32\textwidth}
		\centering
		\includegraphics[scale=0.39]{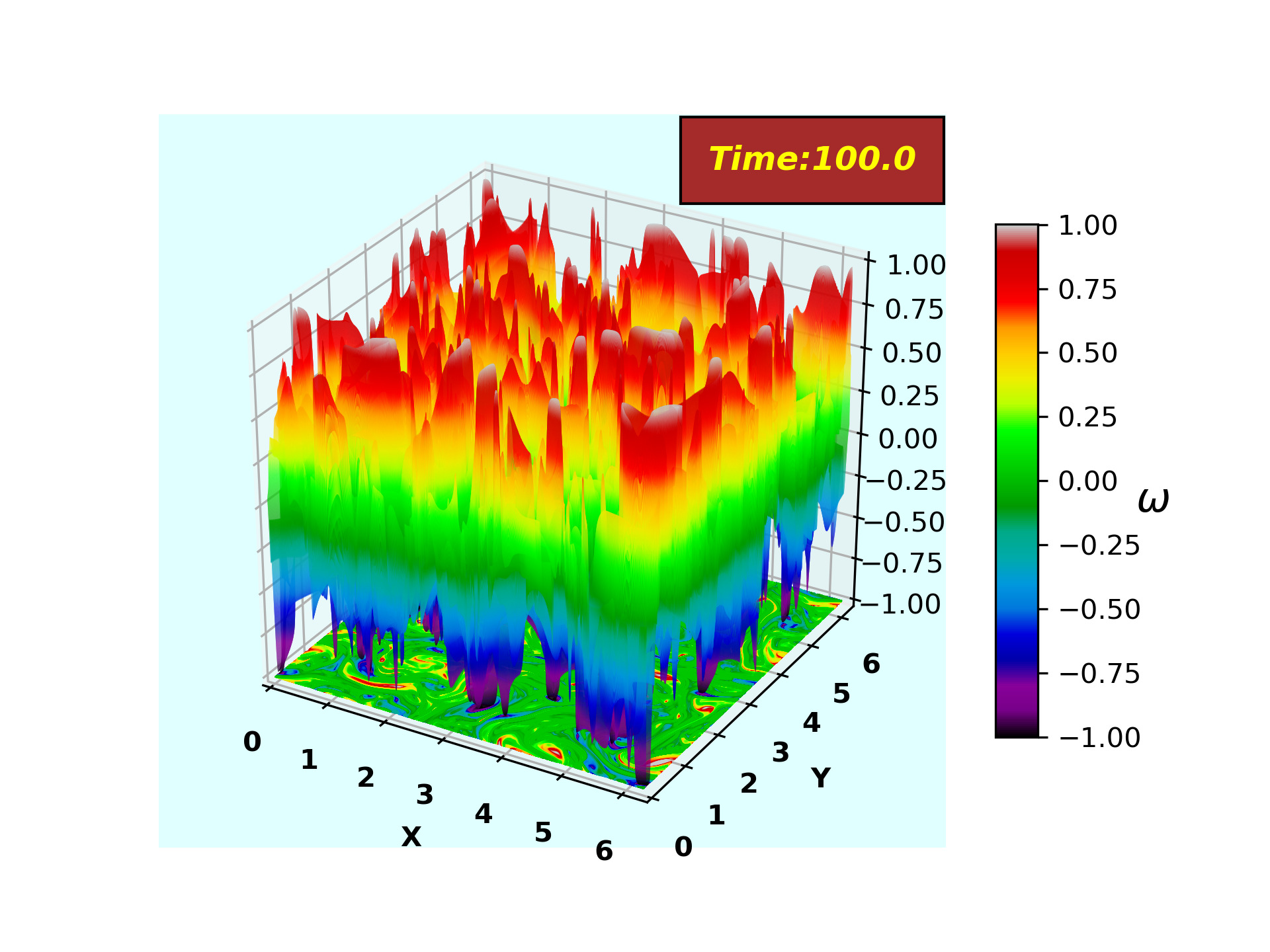}
		\caption{\textbf{Time: 100.0}}
	\end{subfigure}
	\begin{subfigure}{0.32\textwidth}
		\centering
		\includegraphics[scale=0.39]{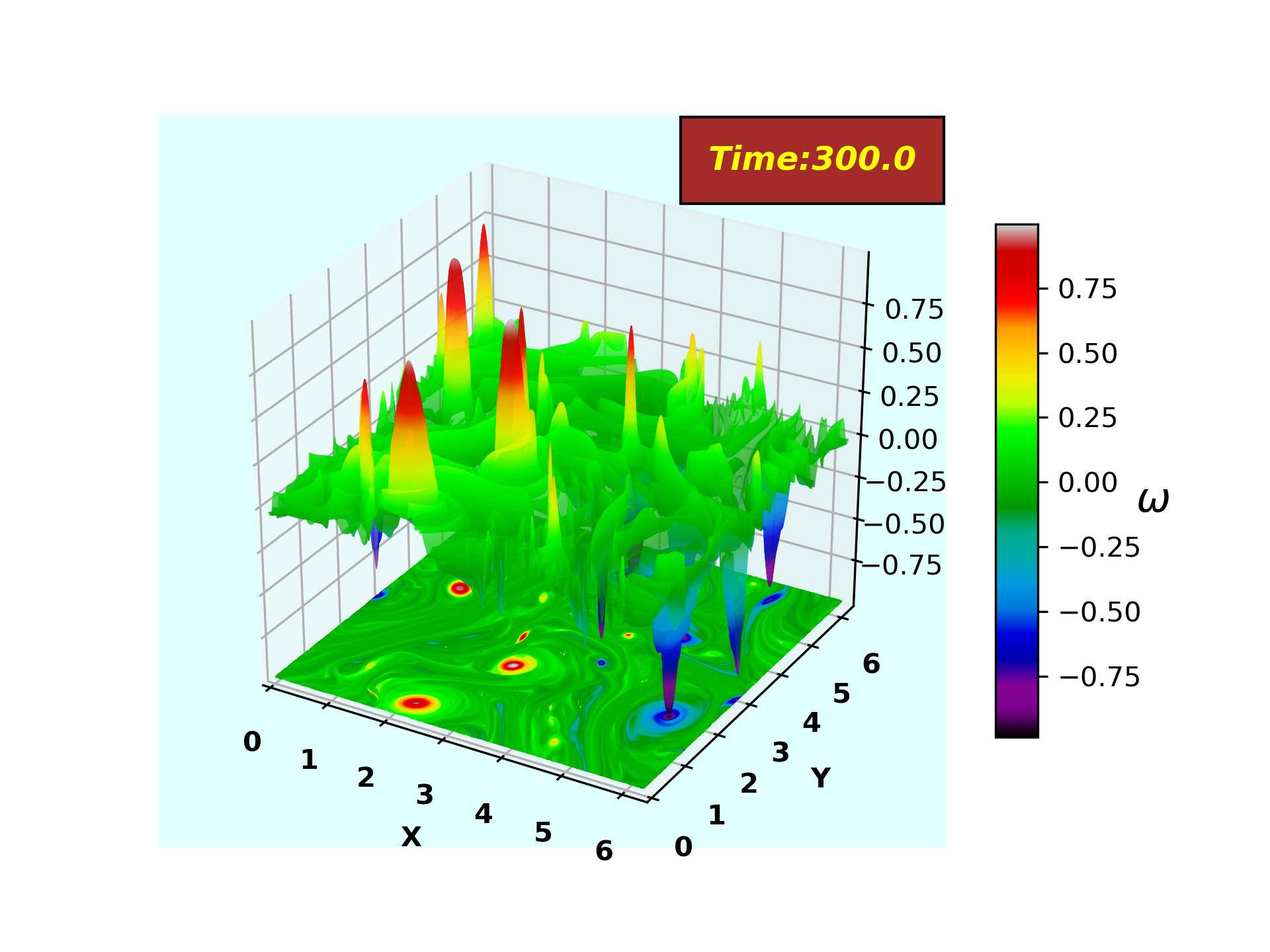}
		\caption{\textbf{Time: 300.0}}
	\end{subfigure}
	\begin{subfigure}{0.32\textwidth}
		\centering
		\includegraphics[scale=0.39]{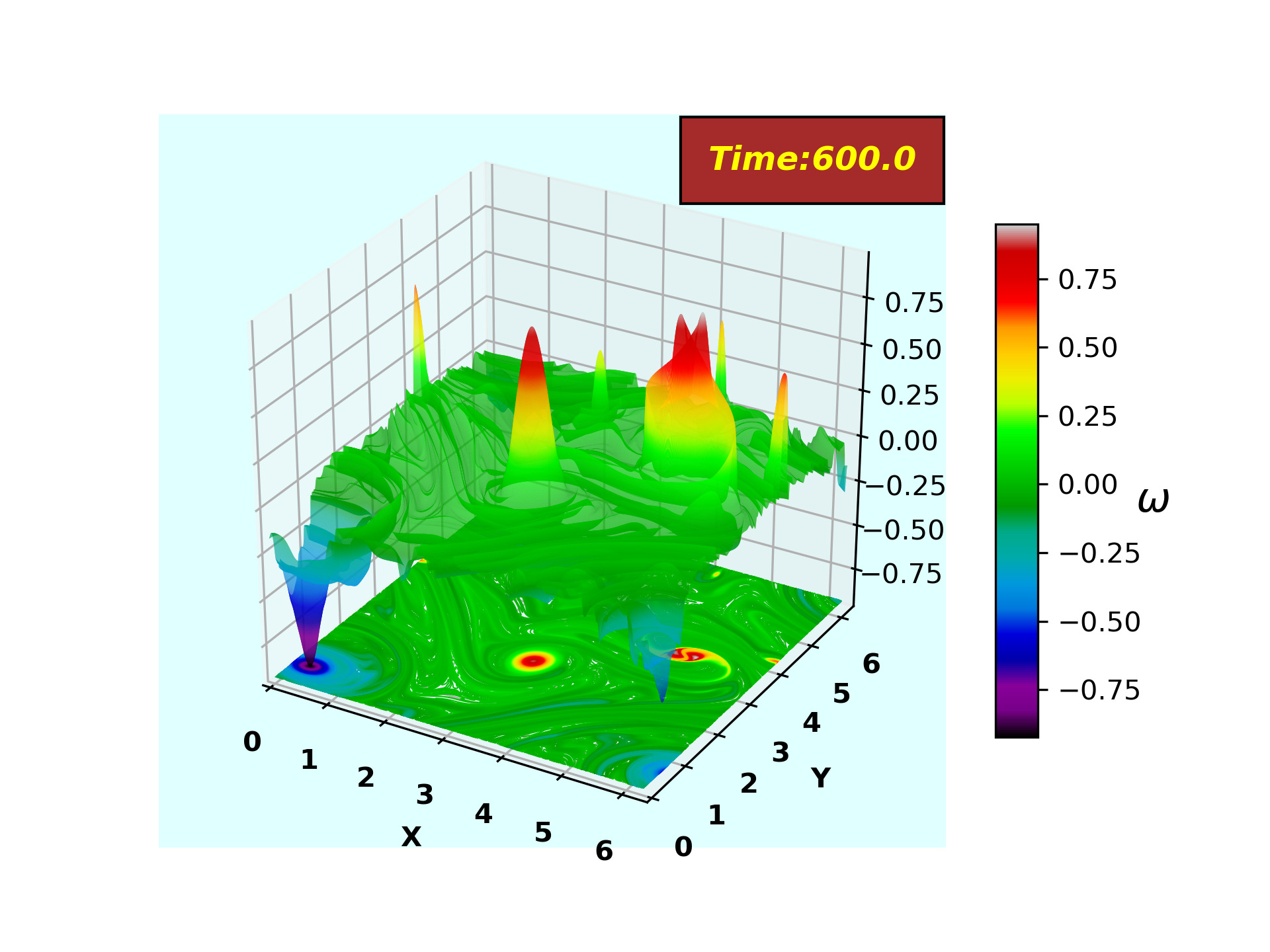}
		\caption{\textbf{Time: 600.0}}
	\end{subfigure}
	\begin{subfigure}{0.32\textwidth}
		\centering
		\includegraphics[scale=0.39]{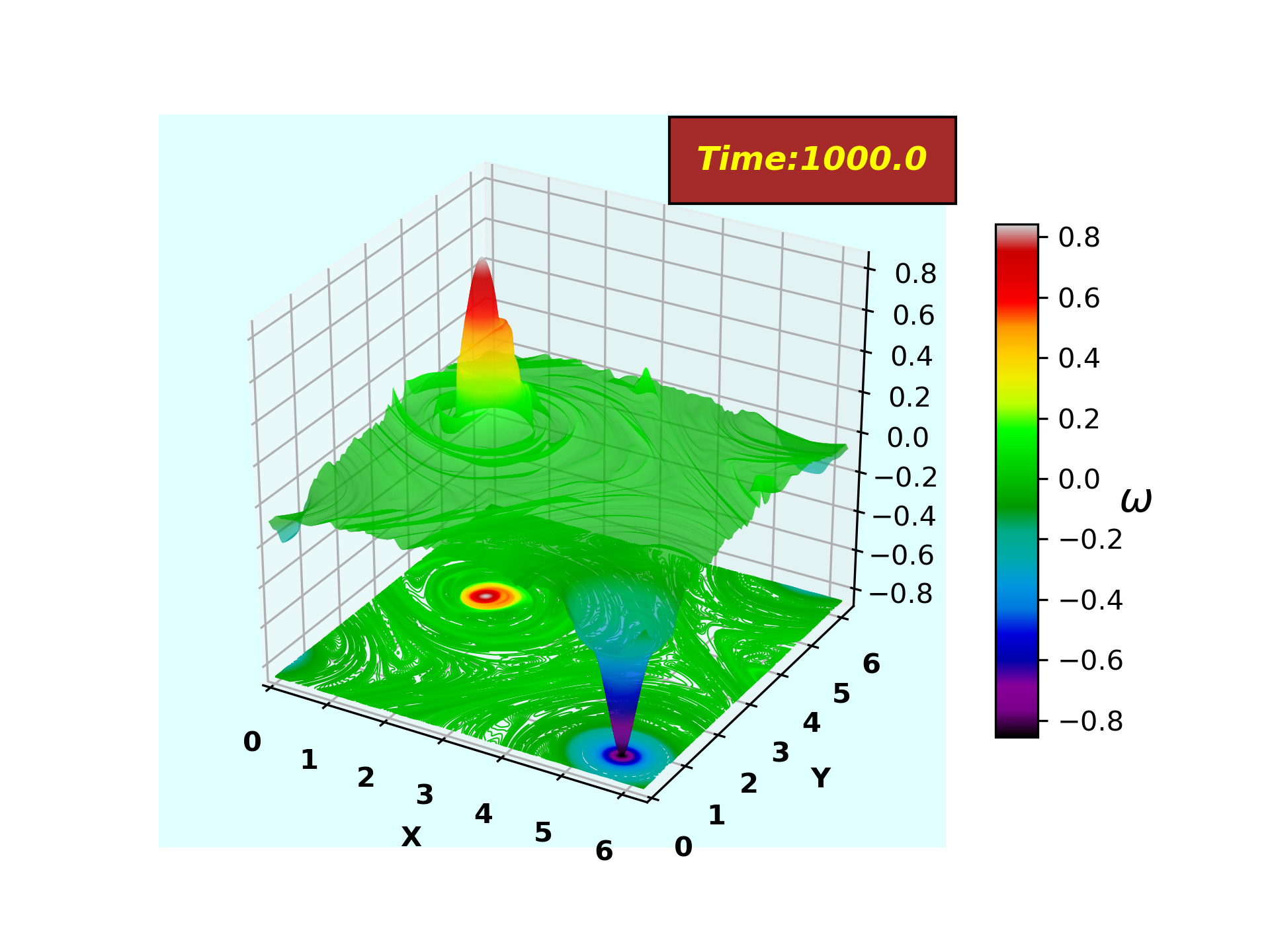}
		\caption{\textbf{Time: 1000.0}}
	\end{subfigure}
	\begin{subfigure}{0.32\textwidth}
		\centering
		\includegraphics[scale=0.39]{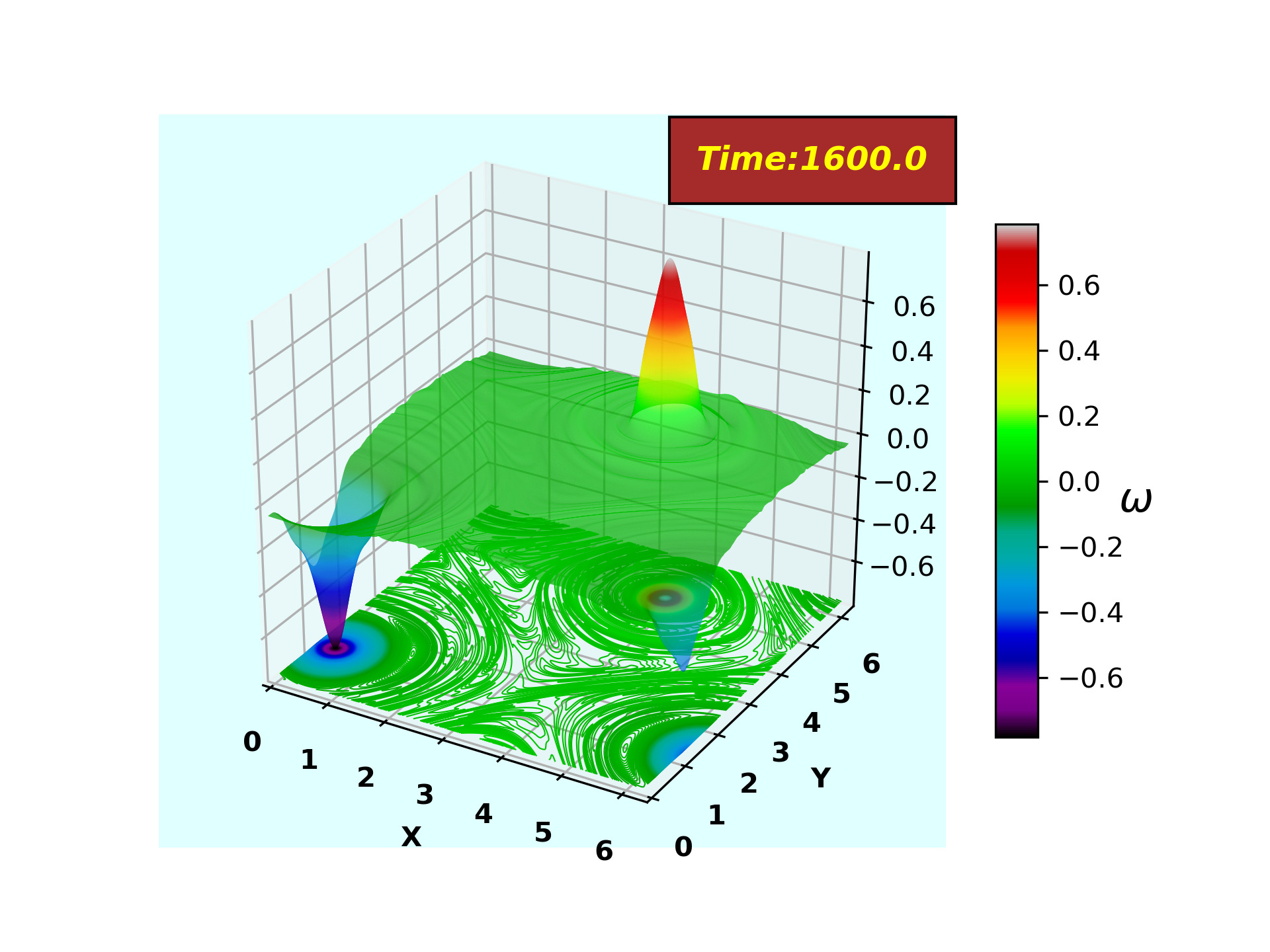}
		\caption{\textbf{Time: 1600.0}}
	\end{subfigure}
	\caption{Time evolution \textcolor{black}{[(a) Time: 0.0, (b) Time: 100.0, (c) Time: 300.0, (d) Time: 600.0, (e) Time: 1000.0, (f) Time: 1600.0]} of vorticity (3D visualization) for a tightly packed [$62.5\%$]  vortex strips. As time goes on similar polarity vortices merge and finally end up with one single vortex of each sign. Simulation details: grid resolution $2048^2$, stepping time dt = $10^{-4}$, Reynolds number = 228576. (multimedia view)}
	\label{20 Vorticity Evolution}
\end{figure*}

\begin{figure}
	\centering
	\includegraphics[scale=0.55]{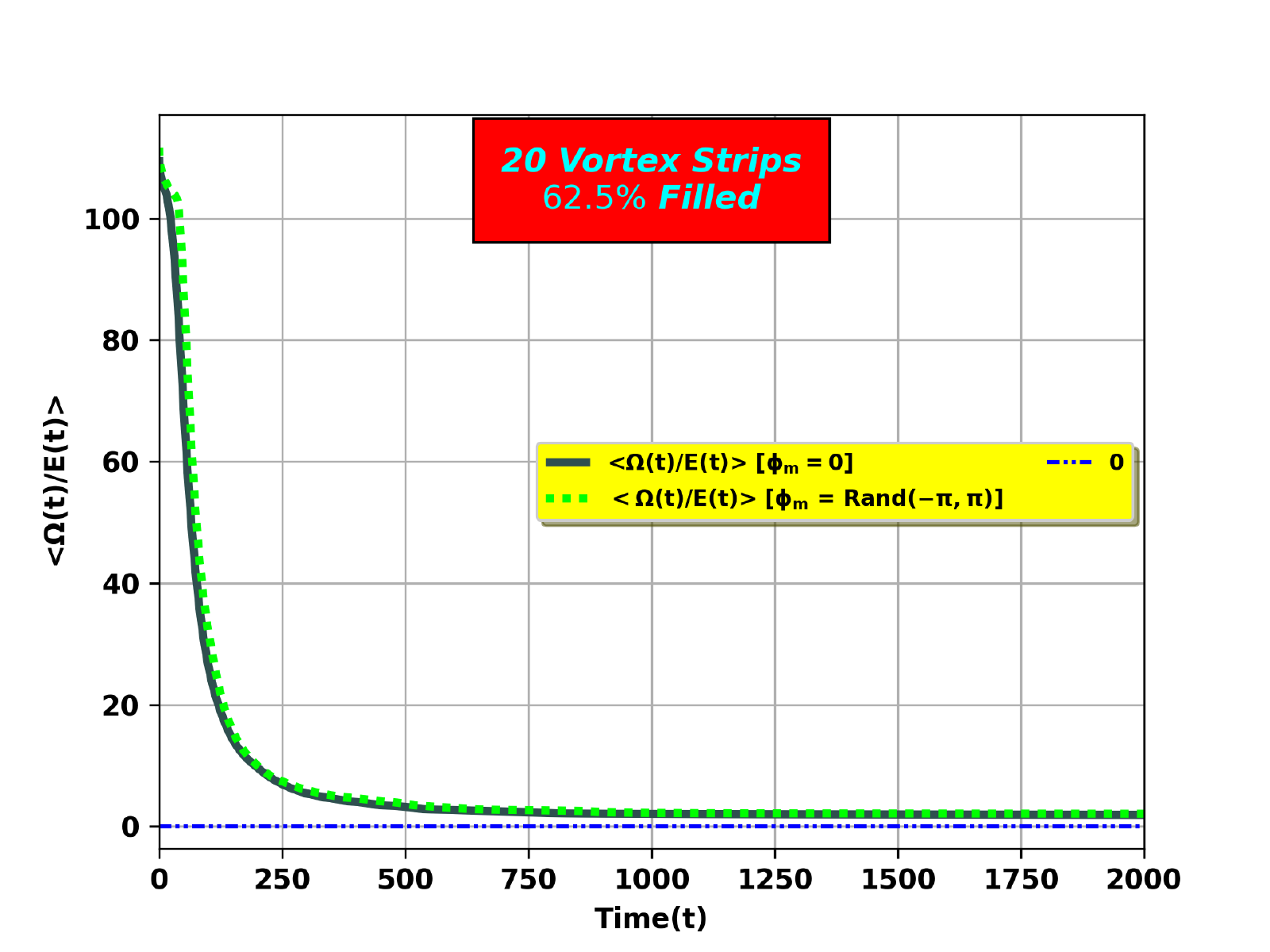}
	\caption{Enstrophy to Energy ratio as a function of time (t), for 20 vortex strips ($62.5\%$ packed) with phase and without phase in perturbation. Enstrophy decays faster than energy with time.  Simulation details: grid resolution $2048^2$, stepping time dt = $10^{-4}$, Reynolds number = 228576.}
	\label{20 Strips enstropy to energy}
\end{figure}

\begin{figure}
	\centering
	\includegraphics[scale=0.55]{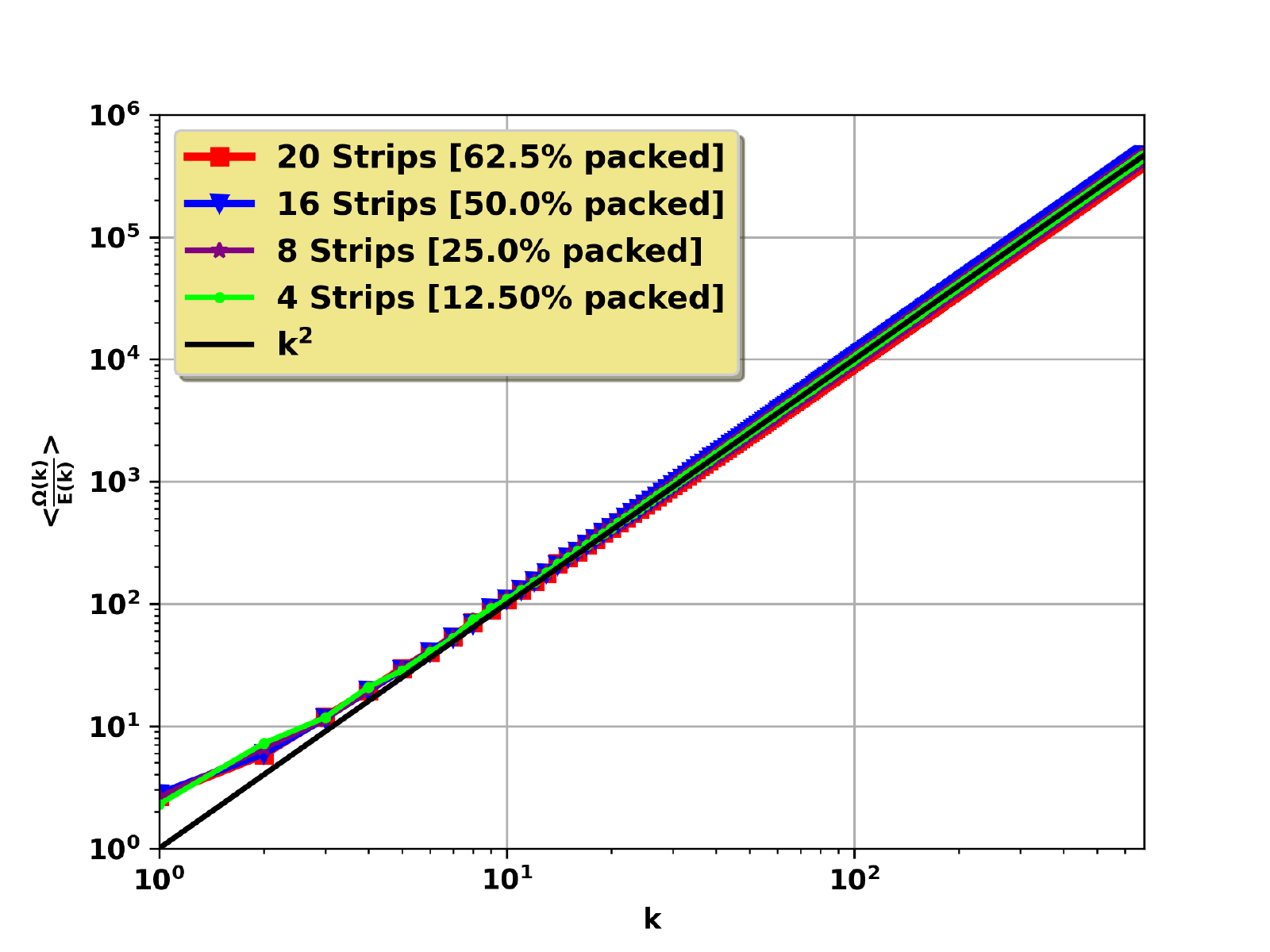}
	\caption{Log scale plot of time averaged enstrophy to energy ratio in momentum space for different initial condition. This ratio strongly indicates $k^2$ scaling.  Simulation details: grid resolution $2048^2$, stepping time dt = $10^{-4}$, Reynolds number = 228576.}
	\label{enstropy to energy}
\end{figure}

We  observe in our simulation that over the longer time, after all the possible like sign vortex capture occurs, the system ends up with one vortex of either sign in the entire 2-dimensional domain [See Fig. \ref{20 Vorticity Evolution} (multimedia view)]. However, the merging process is found to only slow down, but does not become zero. The presence of random noise in perturbation is found to not effect the final state of evolution. We perform both simulations - one with noise and one without noise and the late time dynamics is seen to be almost identical for both the cases. The late time vorticity distribution is dominated by single large vortex of either sign, which was also observed by Montgomery et al. \cite{montgomery_d:1991,sinh:1992} but, with several important differences, as will be discussed later. From the earlier discussion\textcolor{black}{,} it is known that the ratio of mean square vorticity (enstrophy) to the mean square velocity (kinetic energy) is a non increasing function of time. We calculate this ratio from our simulation for both the cases, i.e. with random noise as well as with out any noise in perturbation  and observed the monotonic decay of the $\left<\frac{\Omega (t)}{E (t)}\right>$ for both the cases [See Fig.\ref {20 Strips enstropy to energy}]. The decaying nature in Fig. \ref {20 Strips enstropy to energy} simply suggests that the enstrophy field becomes increasingly dominated by fine scales as time progresses, is consistent with Eq. \ref{enstrophy_to_energy}, however, as will be shown, the late time states do not respect enstrophy extremization.\\ 
In spectral space, energy (E), enstrophy ($\Omega$) and palinstrophy ($P$) are defined as, 
\begin{eqnarray}
E (k) = \sum k^2\left\lvert \psi_k \right\rvert^2 \\
\Omega (k) = \sum k^4\left\lvert \psi_k \right\rvert^2 \\
P (k) = \sum k^6\left\lvert \psi_k \right\rvert^2 
\end{eqnarray}
such that the ratio becomes, 
\begin{equation}
\label{enstropy_to_energy_spectral}
\left<\frac{\Omega (k)}{E (k)}\right> = \left<k^2\right>
\end{equation}
Also the monotonic decay of enstrophy to energy ratio is interpreted as the increase of average wave length suggested from Eq. \ref{enstropy_to_energy_spectral}.  In the other words, as time evolves the resultant system is dominated by the largest scale available in the system. We find from our numerical simulation that $\left<\frac{\Omega (k)}{E (k)}\right>$ scales as $\left<k^2\right>$ [See Fig. \ref{enstropy to energy}]. 

\begin{figure*}
	\centering
	\begin{subfigure}{0.45\textwidth}
		\centering
		\includegraphics[scale=0.55]{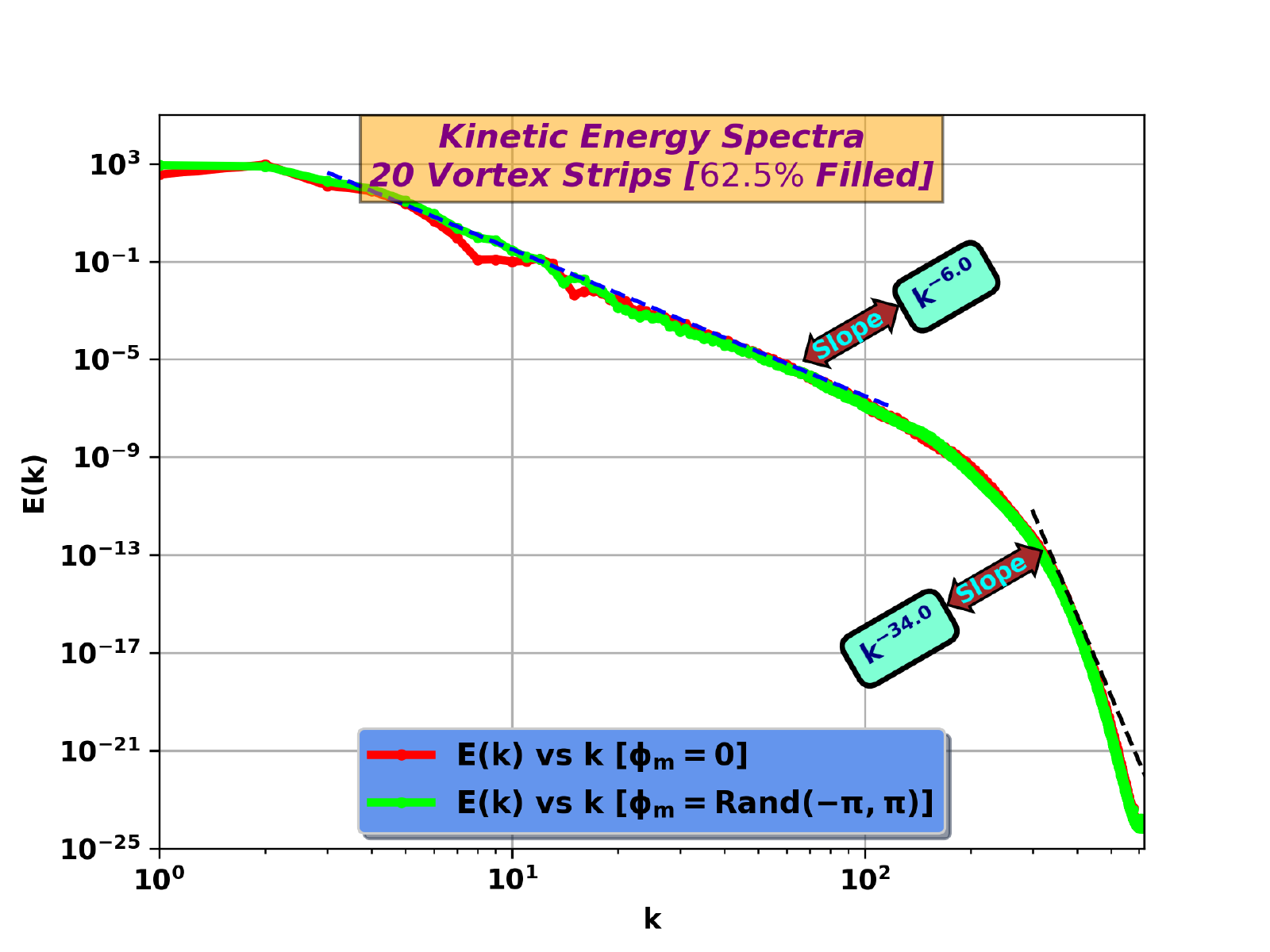}
		\caption{}
	\end{subfigure}
	\begin{subfigure}{0.45\textwidth}
		\centering
		\includegraphics[scale=0.55]{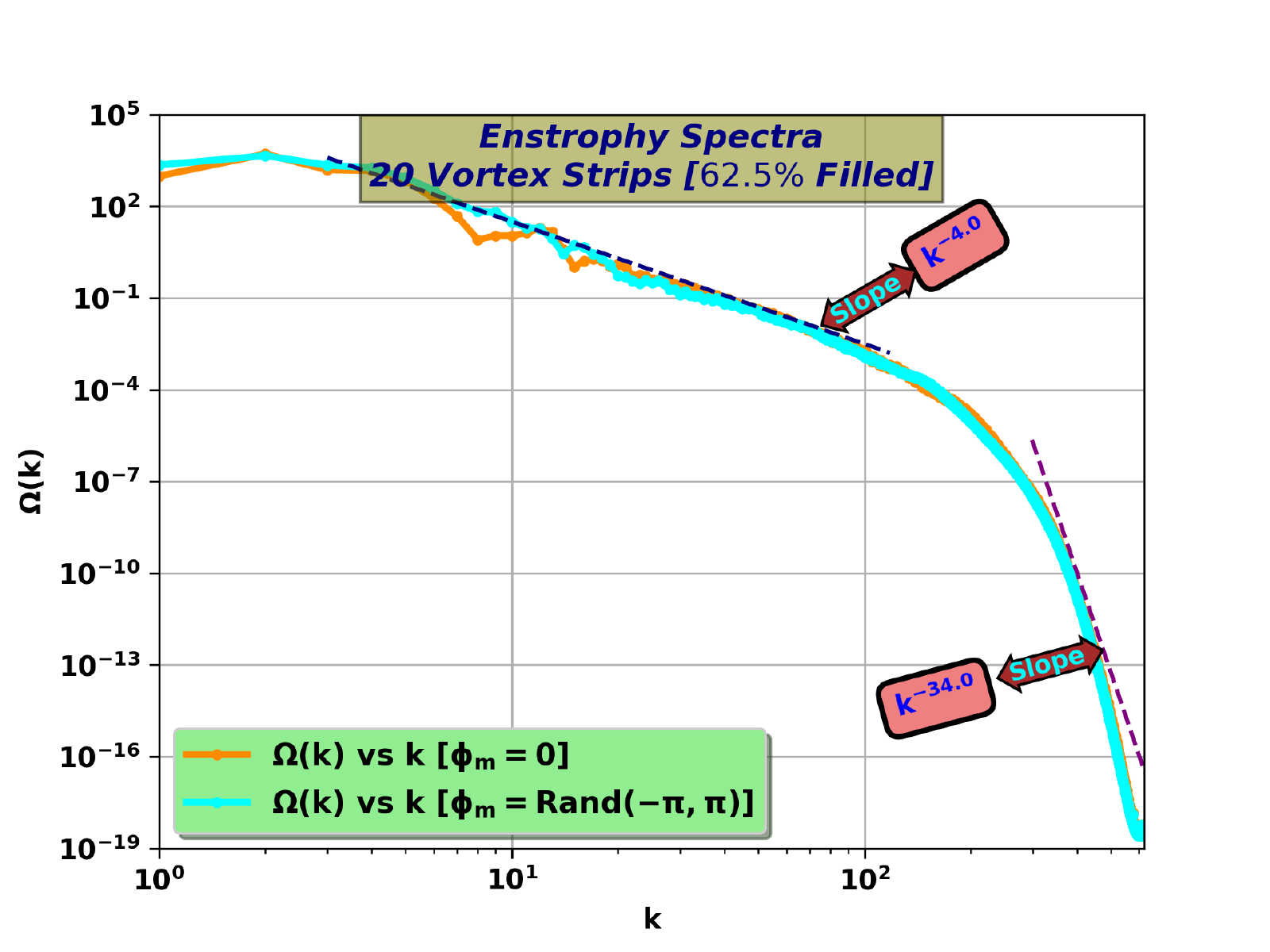}
		\caption{}
	\end{subfigure}
	\caption{(a) Time averaged (time average taken after saturation i.e from t=1000.0 to 1600.0) kinetic energy spectra [$\int_{0}^{\infty} E(k) dk$] showing inverse cascading (b) Time averaged (time average taken after saturation i.e from t=1000.0 to 1600.0) enstrophy spectra [$\int_{0}^{\infty} \Omega(k) dk$] showing direct cascading for tightly packed 20 vortex strips ($62.5\%$ packed) configuration. These k-scaling\textcolor{black}{s} for kinetic energy and enstrophy agree with the earlier work \cite{gomez:1996}. Simulation details: grid resolution $2048^2$, stepping time dt = $10^{-4}$, Reynolds number = 228576.}
	\label{20 strips Spectra}
\end{figure*}

\begin{figure*}
	\centering
	\begin{subfigure}{0.32\textwidth}
		\centering
		\includegraphics[scale=0.38]{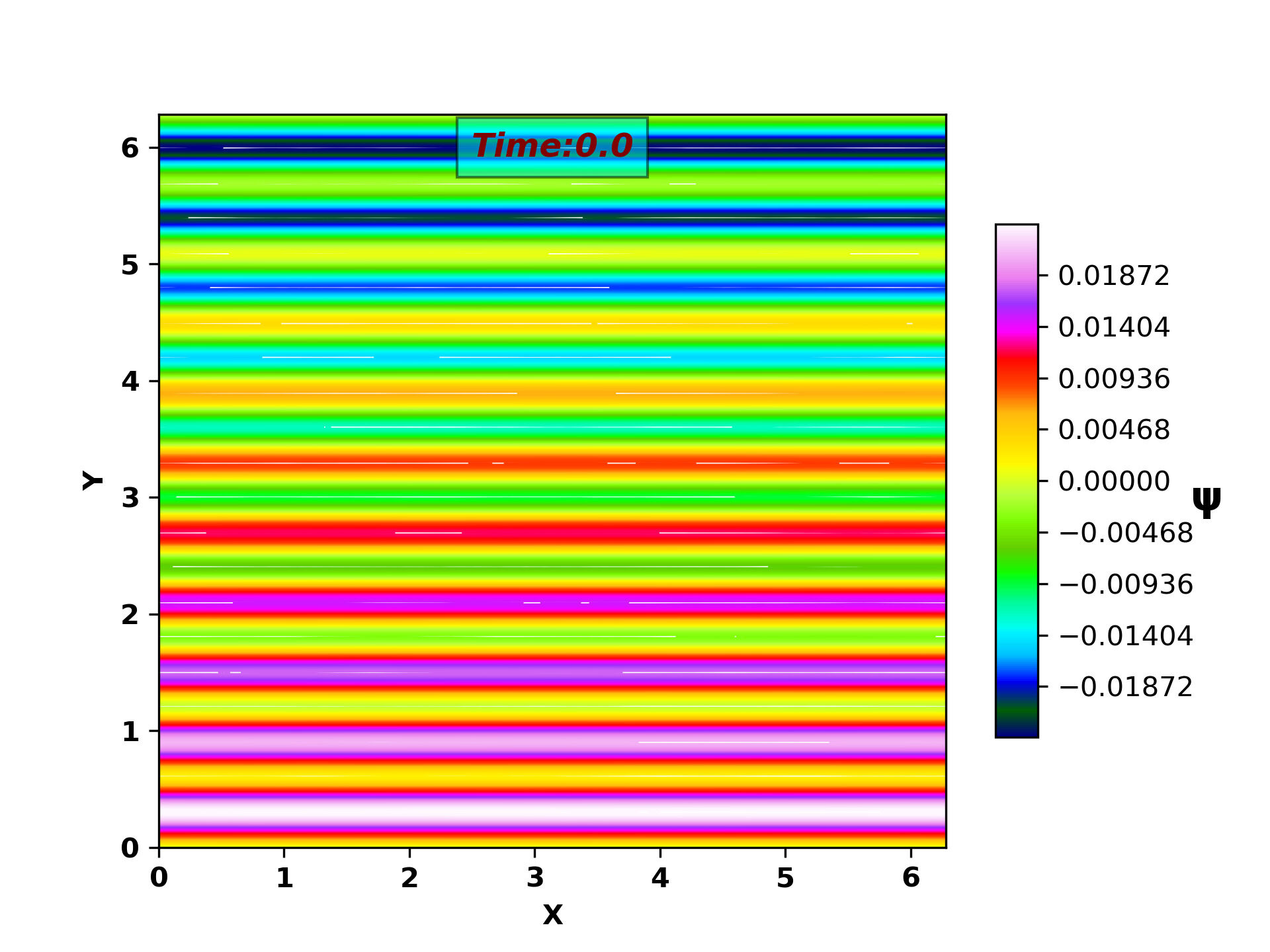}
		\caption{\textbf{Time: 0.0}}
	\end{subfigure}
	\begin{subfigure}{0.32\textwidth}
		\centering
		\includegraphics[scale=0.38]{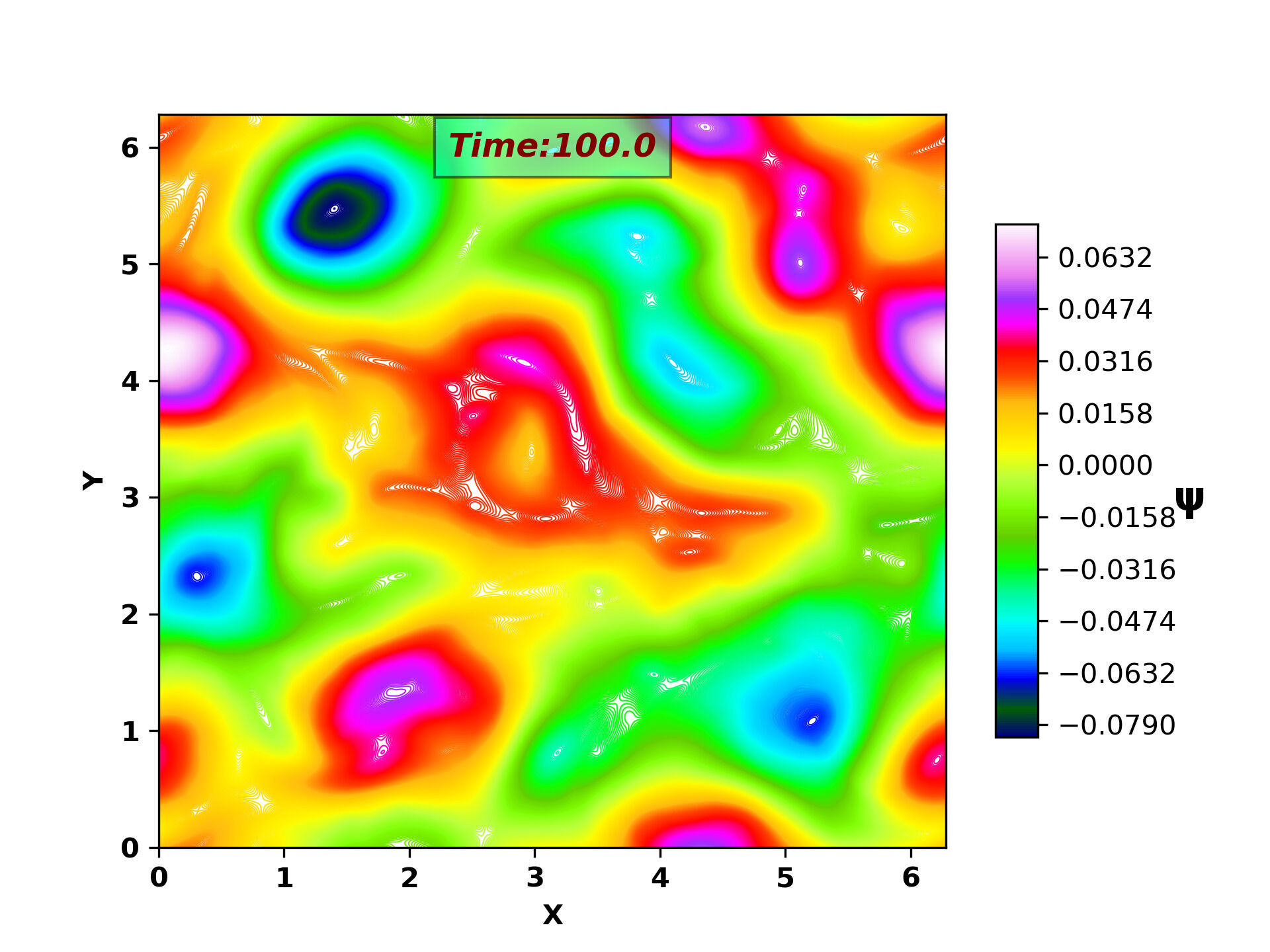}
		\caption{\textbf{Time: 100.0}}
	\end{subfigure}
	\begin{subfigure}{0.32\textwidth}
		\centering
		\includegraphics[scale=0.38]{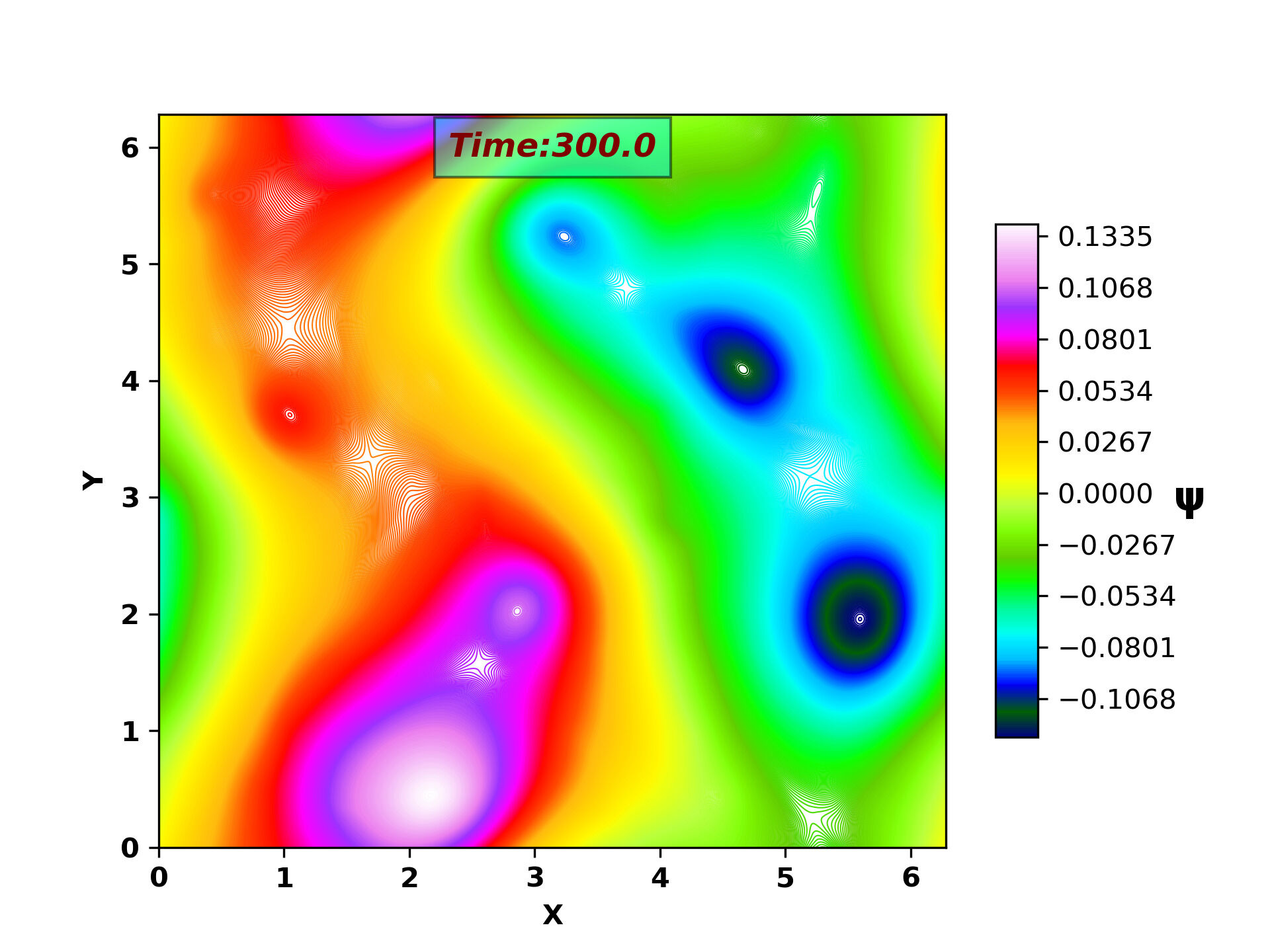}
		\caption{\textbf{Time: 300.0}}
	\end{subfigure}
	\begin{subfigure}{0.32\textwidth}
		\centering
		\includegraphics[scale=0.38]{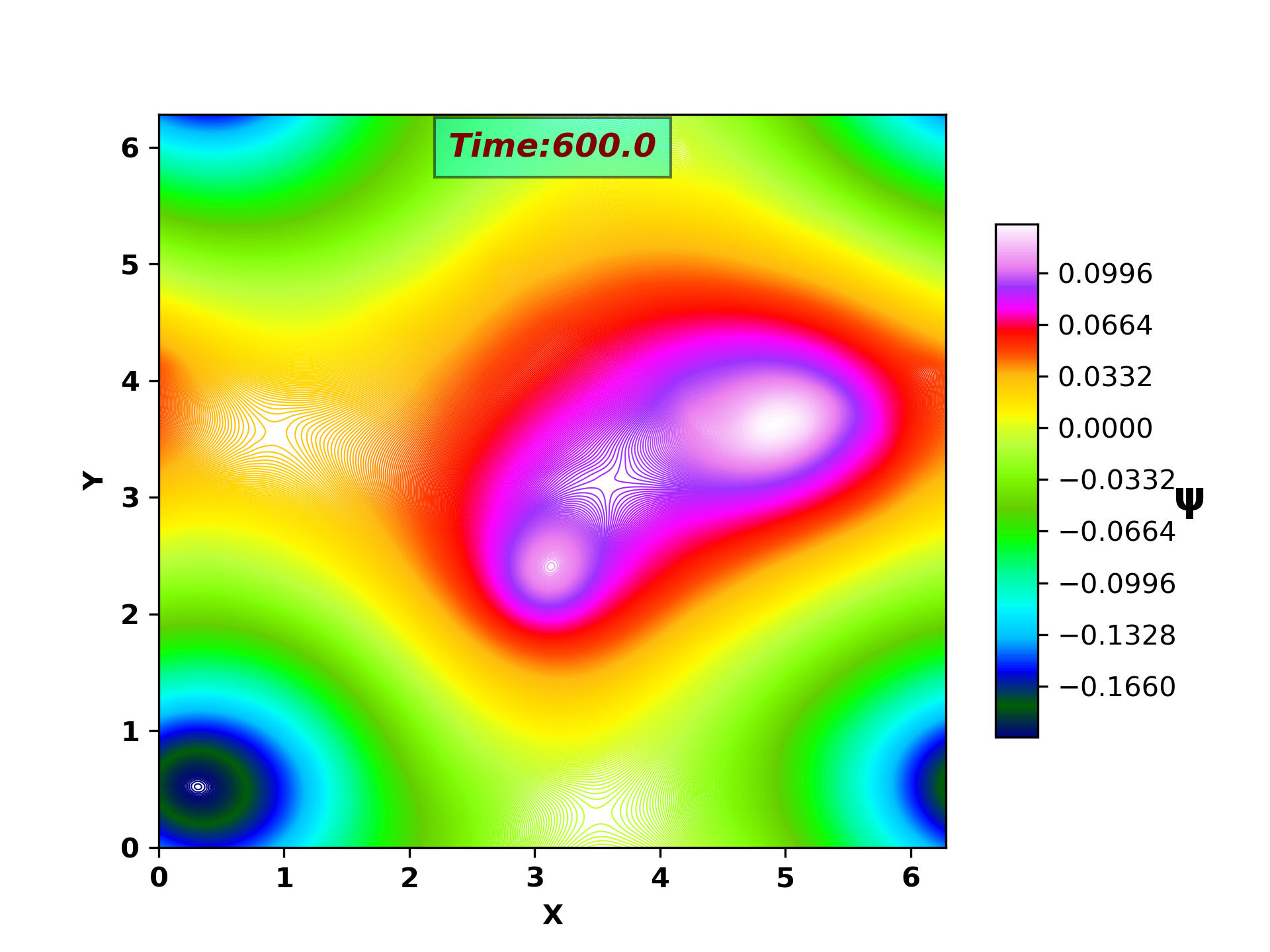}
		\caption{\textbf{Time: 600.0}}
	\end{subfigure}
	\begin{subfigure}{0.32\textwidth}
		\centering
		\includegraphics[scale=0.38]{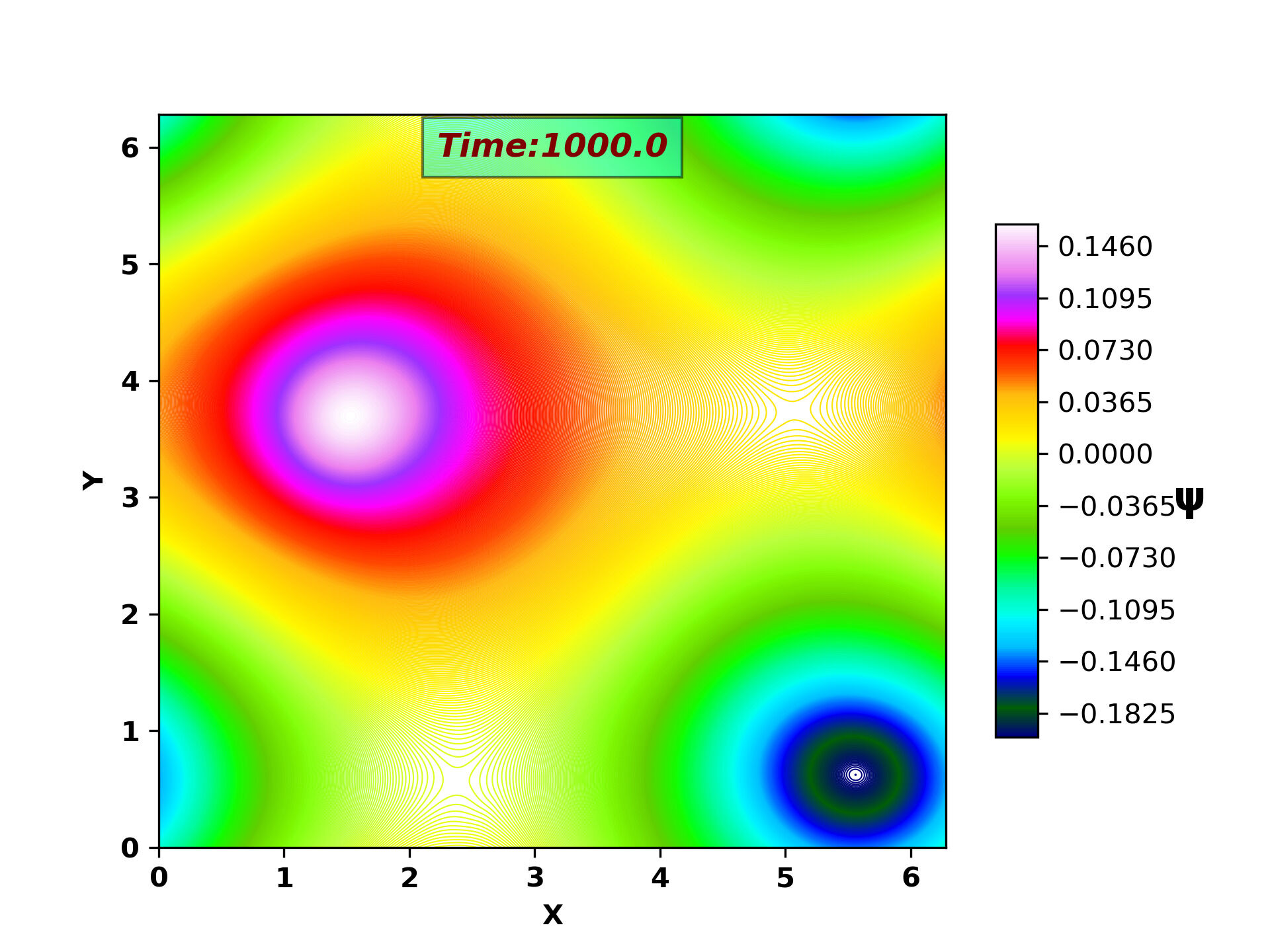}
		\caption{\textbf{Time: 1000.0}}
	\end{subfigure}
	\begin{subfigure}{0.32\textwidth}
		\centering
		\includegraphics[scale=0.38]{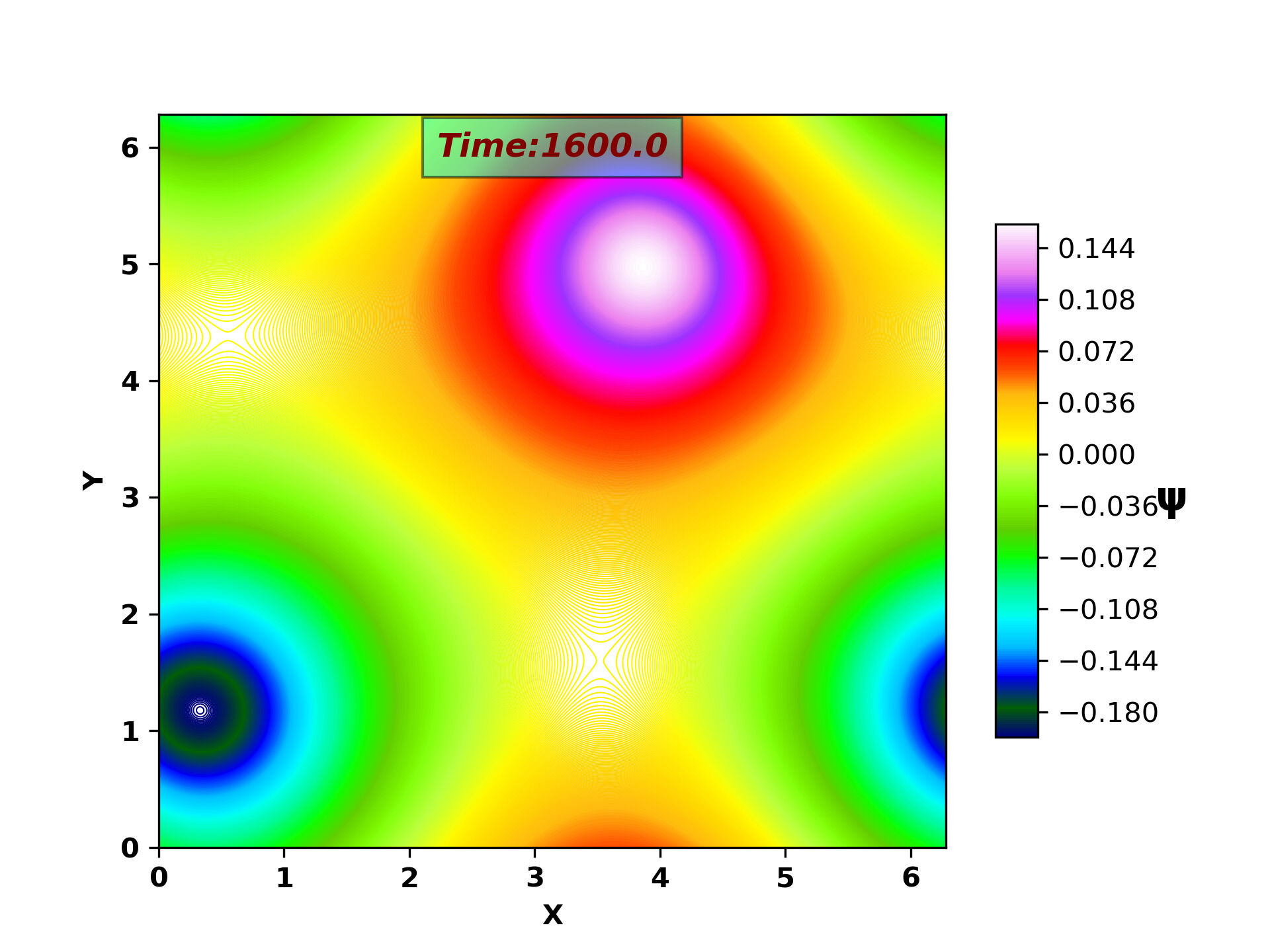}
		\caption{\textbf{Time: 1600.0}}
	\end{subfigure}
	\caption{Time evolution  \textcolor{black}{[(a) Time: 0.0, (b) Time: 100.0, (c) Time: 300.0, (d) Time: 600.0, (e) Time: 1000.0, (f) Time: 1600.0]} of stream function ($\psi$) for tightly packed [$62.5\%$]  vortex strips. After all the possible vortex mergers occur, the streamlines achieve Ewald potential like contours (with a basic cell containing two point vortices), also observed by Montgomery et al. \cite{montgomery_d:1991,montgomery_prl:1991}. Simulation details: grid resolution $2048^2$, stepping time dt = $10^{-4}$, Reynolds number = 228576. (multimedia view)}
	\label{20 psi Evolution}
\end{figure*}

\begin{figure*}
	\centering
	\begin{subfigure}{0.32\textwidth}
		\centering
		\includegraphics[scale=0.38]{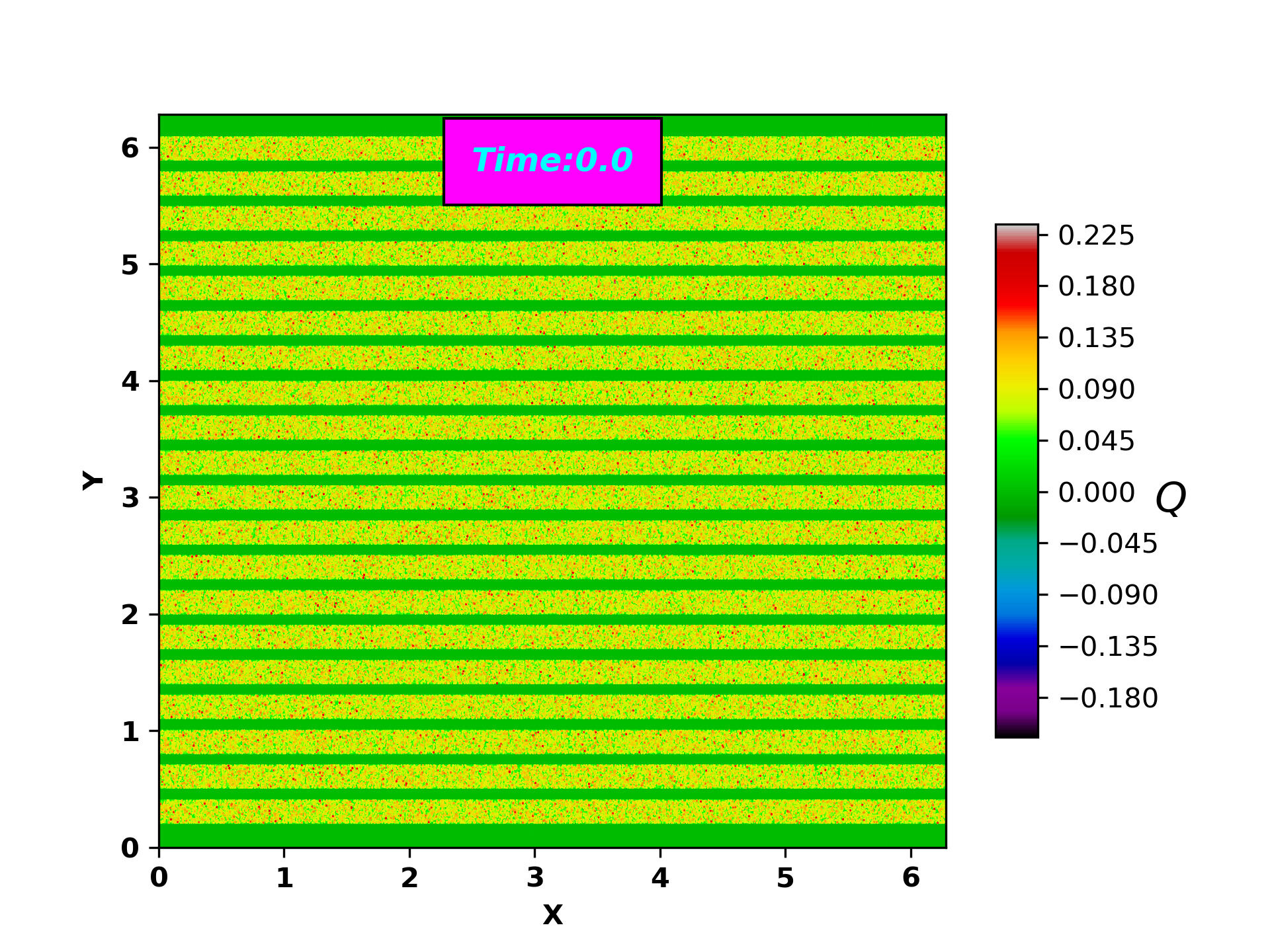}
		\caption{\textbf{Time: 0.0}}
	\end{subfigure}
	\begin{subfigure}{0.32\textwidth}
		\centering
		\includegraphics[scale=0.38]{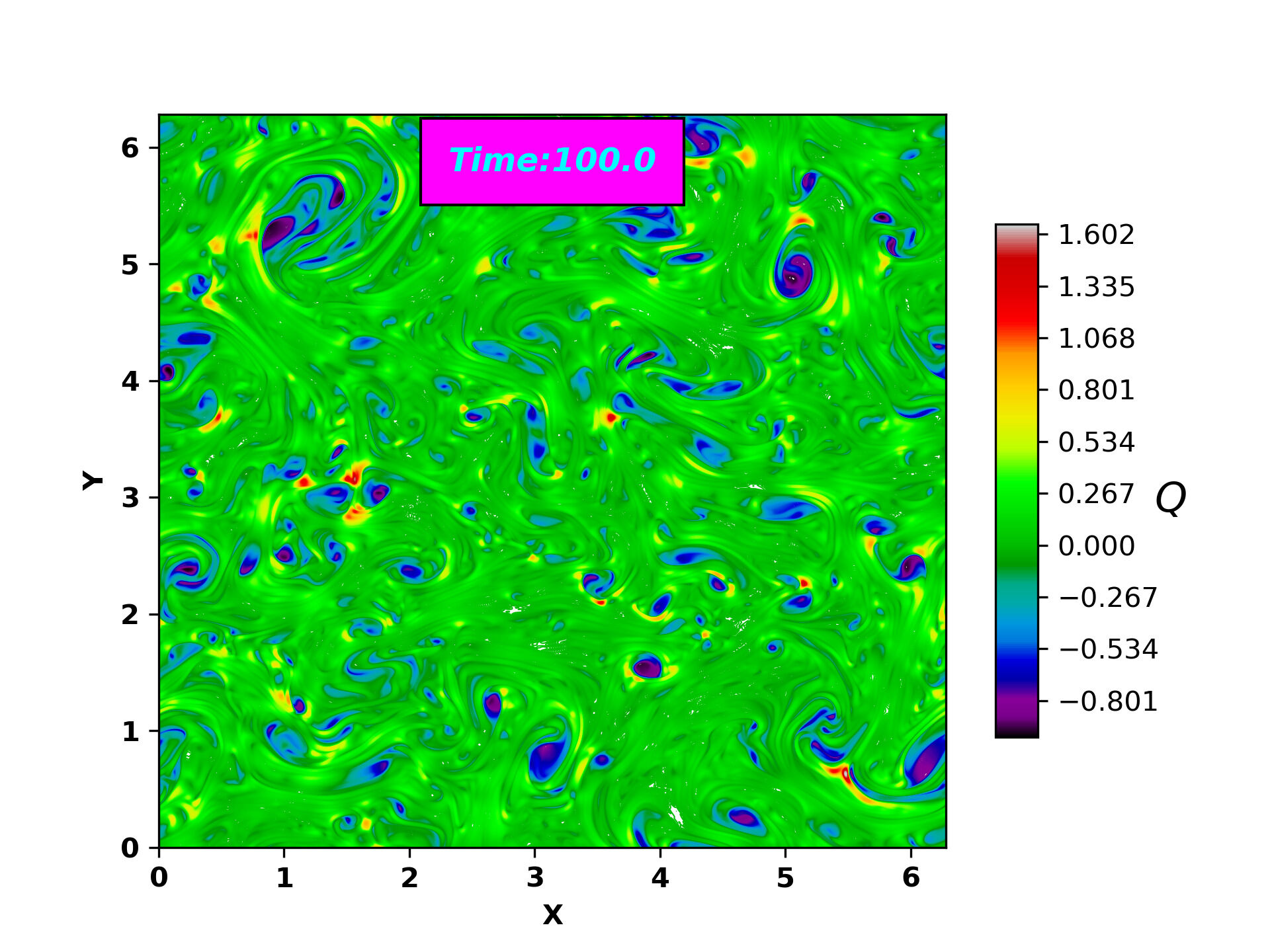}
		\caption{\textbf{Time: 100.0}}
	\end{subfigure}
	\begin{subfigure}{0.32\textwidth}
		\centering
		\includegraphics[scale=0.38]{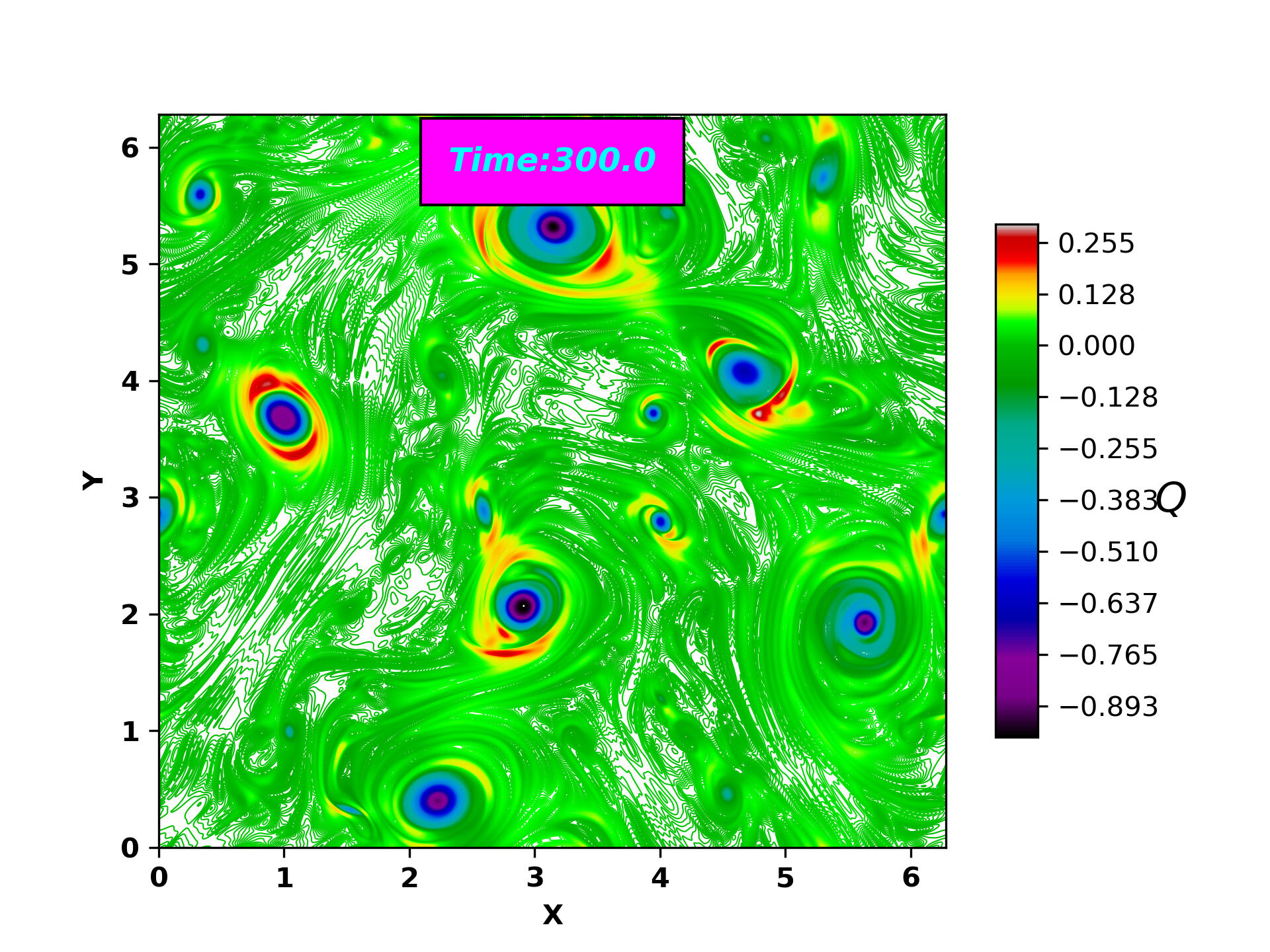}
		\caption{\textbf{Time: 300.0}}
	\end{subfigure}
	\begin{subfigure}{0.32\textwidth}
		\centering
		\includegraphics[scale=0.38]{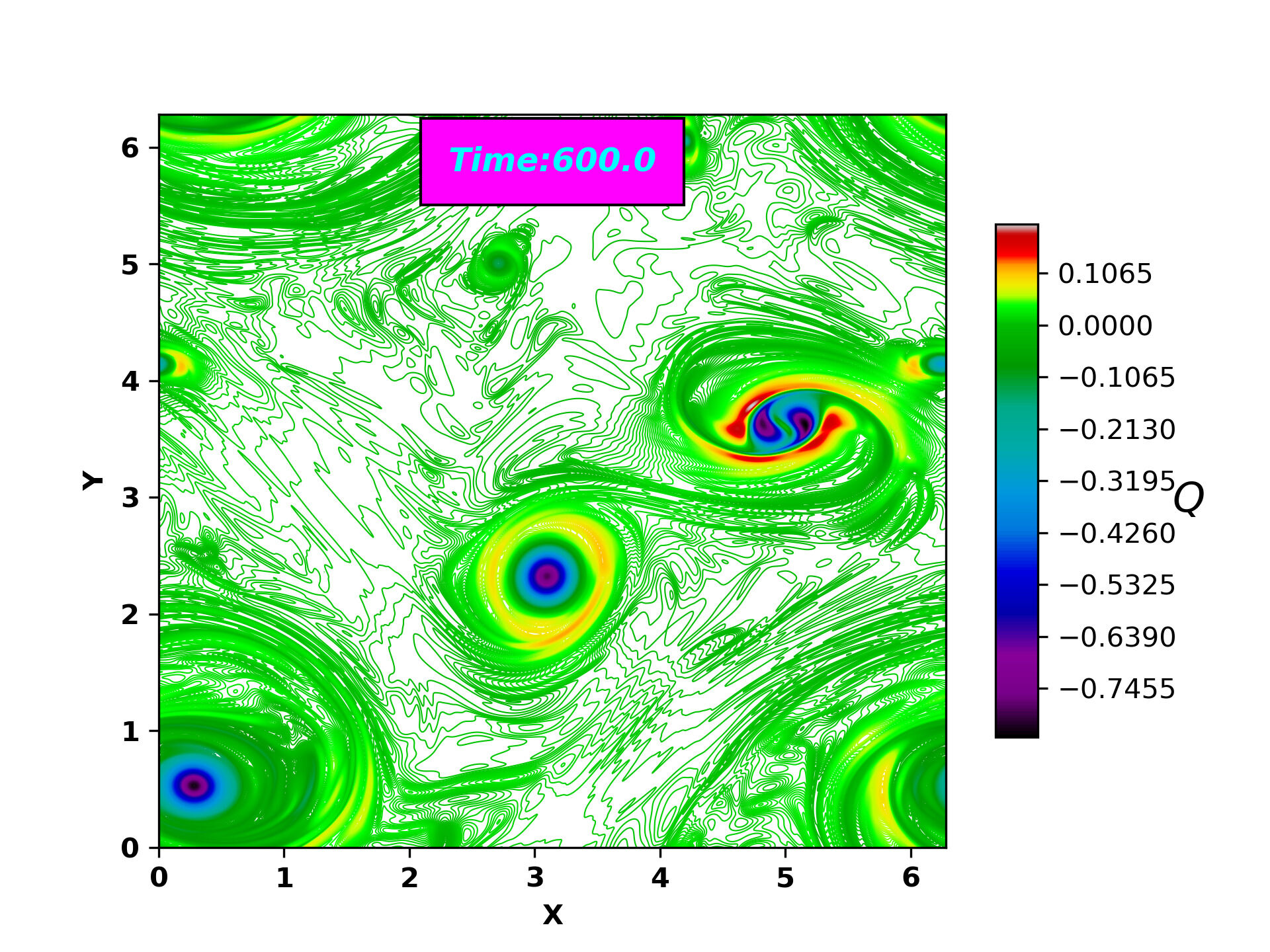}
		\caption{\textbf{Time: 600.0}}
	\end{subfigure}
	\begin{subfigure}{0.32\textwidth}
		\centering
		\includegraphics[scale=0.38]{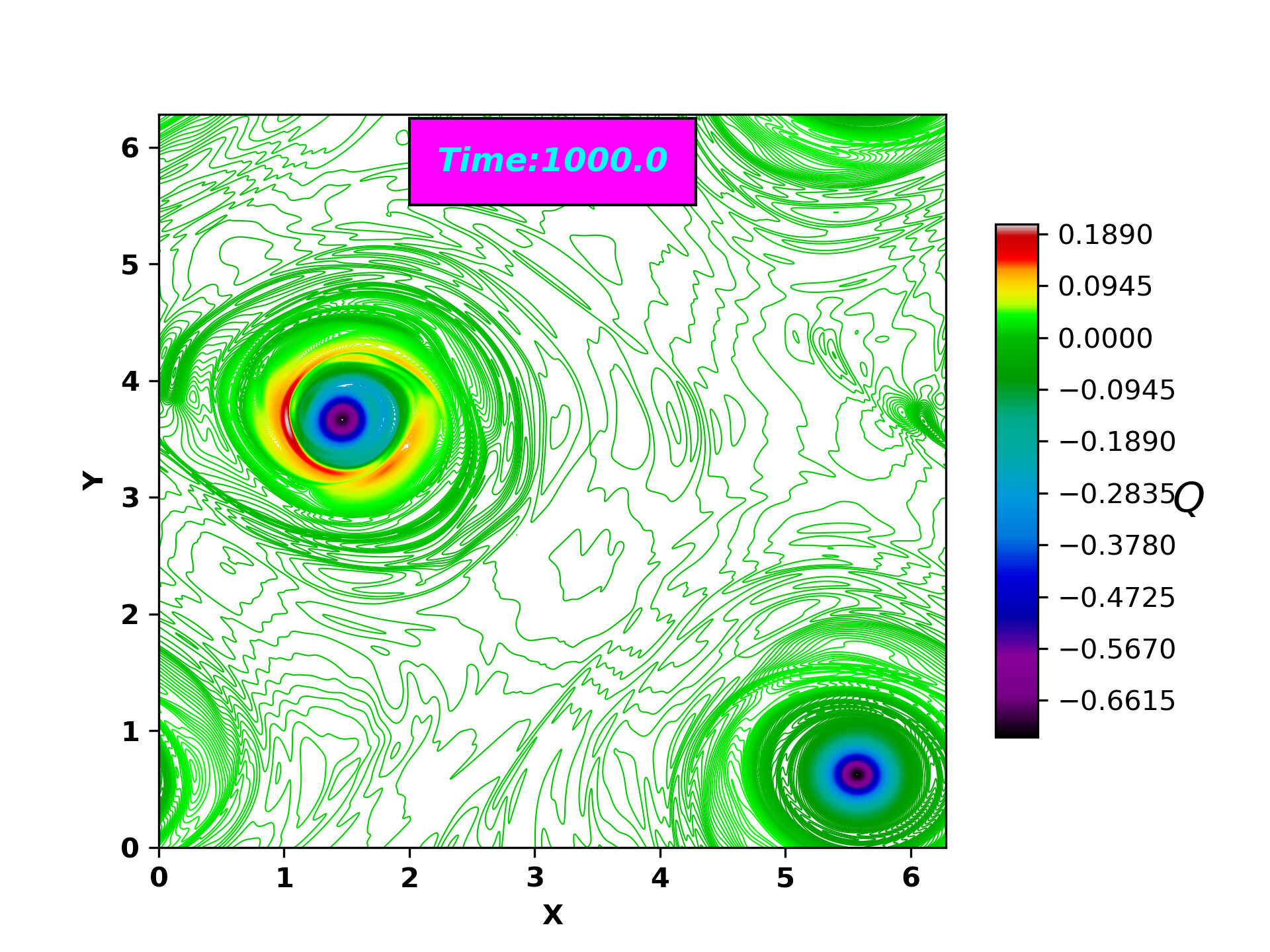}
		\caption{\textbf{Time: 1000.0}}
	\end{subfigure}
	\begin{subfigure}{0.32\textwidth}
		\centering
		\includegraphics[scale=0.38]{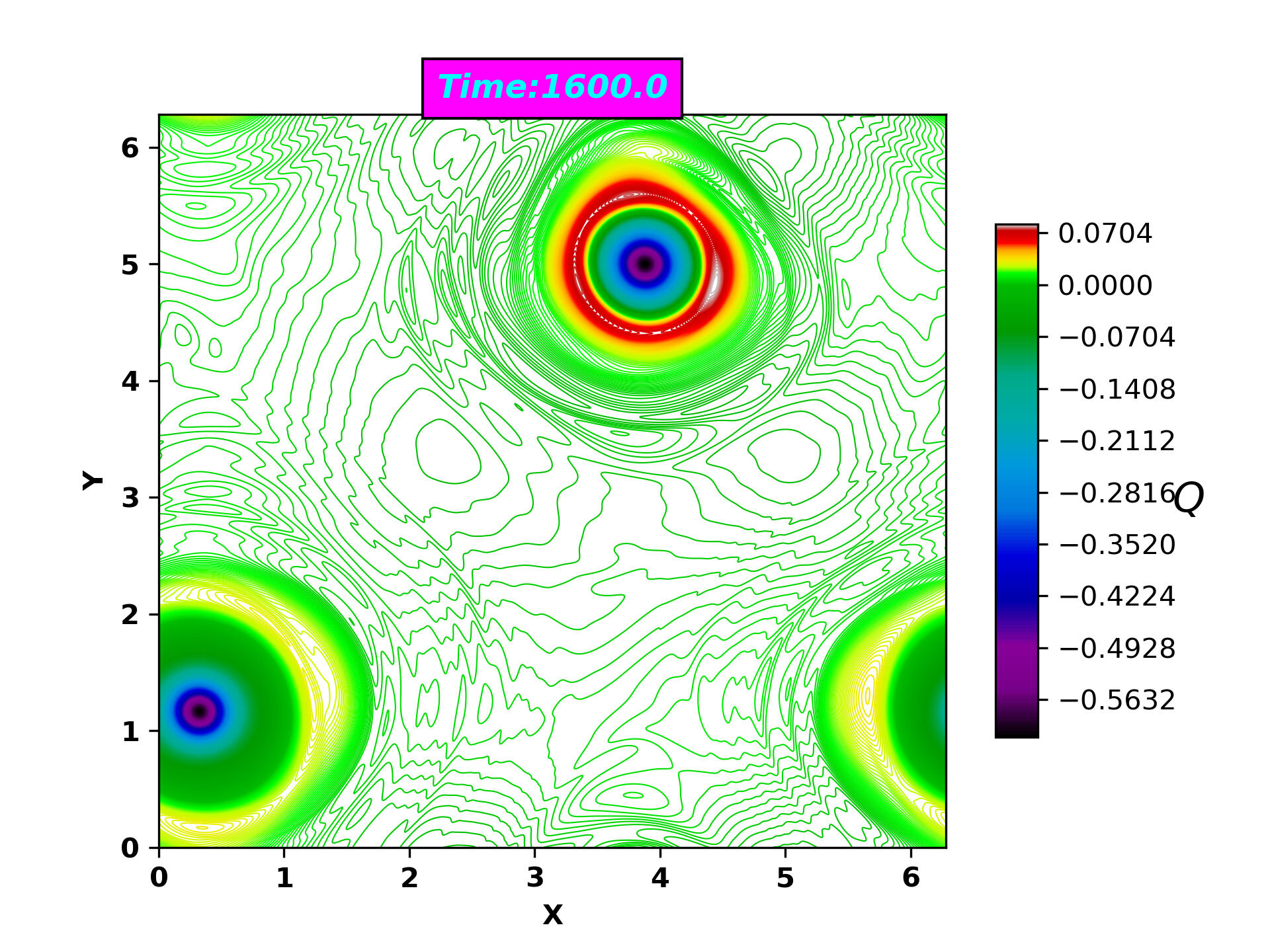}
		\caption{\textbf{Time: 1600.0}}
	\end{subfigure}
	\caption{Time evolution  \textcolor{black}{[(a) Time: 0.0, (b) Time: 100.0, (c) Time: 300.0, (d) Time: 600.0, (e) Time: 1000.0, (f) Time: 1600.0]} of Okubo-Weiss parameter ($Q(x,y,t)$) for tightly packed [$62.5\%$]  vortex strips. Two distinct regions are identified (i) Vortex cores: characterized by strong negative value of $Q(x,y,t)$. (ii) Strain cells: surrounding the vortex cores, characterized by large positive value of $Q(x,y,t)$. Simulation details: grid resolution $2048^2$, stepping time dt = $10^{-4}$, Reynolds number = 228576. (multimedia view)}
	\label{20 Q Evolution}
\end{figure*}

The characteristic wave number spectra indicates both the forward cascade to higher k for enstrophy [See Fig. \ref{20 strips Spectra} (b)] and back transfer of energy to the longest available wave length ($\frac{2\pi}{k_{min}}$) i.e inverse cascade [See Fig. \ref{20 strips Spectra} (a)]. The kinetic energy spectra shows scaling of $E(k)\propto k^{-6}$ for lower wave number and $E(k)\propto k^{-34}$ for higher wave number [See Fig. \ref{20 strips Spectra} (a)]. However the enstrophy spectra shows $\Omega(k)\propto k^{-4}$ for lower wave number and $\Omega(k)\propto k^{-34}$ for higher wave number [See Fig. \ref{20 strips Spectra} (b)]. These scaling (k-dependency) agree with earlier published works, for example, by Dmitruk et al. \cite{gomez:1996} for slowly decaying 2-dimensional NS turbulence. \\
Also it is observed that after all the possible vortex mergers occur, the vorticity achieves a particle like character, suggested by late time similarity of the streamlines with Ewald potential contours (with a basic cell containing two point vortices) [See Fig. \ref{20 psi Evolution} (multimedia view)], observed by Montgomery et al. as well\cite{montgomery_d:1991,montgomery_prl:1991}.\\
We calculate $Q(x,y,t)$ the Okubo-Weiss parameter \cite{OKUBO:1970,WEISS:1991}, which is a measure of rotation vs deformation for two dimensional turbulence. The quantity $Q(x,y,t)$ is defined as,
\begin{equation}
Q(x,y,t) = S^2 - \omega^2
\end{equation}
where $S^2 = S_1^2 + S_2^2$ and $S_1(x,y,t) = \partial_xu_x - \partial_yu_y$, $S_2(x,y,t) = \partial_xu_y + \partial_yu_x$, $\omega(x,y,t) = \partial_xu_y - \partial_yu_x$. Using this parameter $Q(x,y,t)$ as a diagnostic, two distinct domains may be identified for two dimensional turbulence namely,
\begin{itemize}
	\item ``elliptic domain" [$Q(x,y,t)<0$], where rotation dominates deformation i.e. $\omega^2>S^2$.
	\item ``hyperbolic domain" [$Q(x,y,t)>0$], where deformation dominates rotation  i.e $\omega^2<S^2$.
\end{itemize}
From our numerical simulation of $Q(x,y,t)$ \ref{20 Q Evolution} (multimedia view) we identify two regions 1. Vortex cores, characterized by strong negative value of $Q(x,y,t)$. 2. Strain cells surrounding the vortex cores, characterized by large positive value of $Q(x,y,t)$. 

\begin{figure*}
	\centering
	\begin{subfigure}{0.45\textwidth}
		\centering
		\includegraphics[scale=0.55]{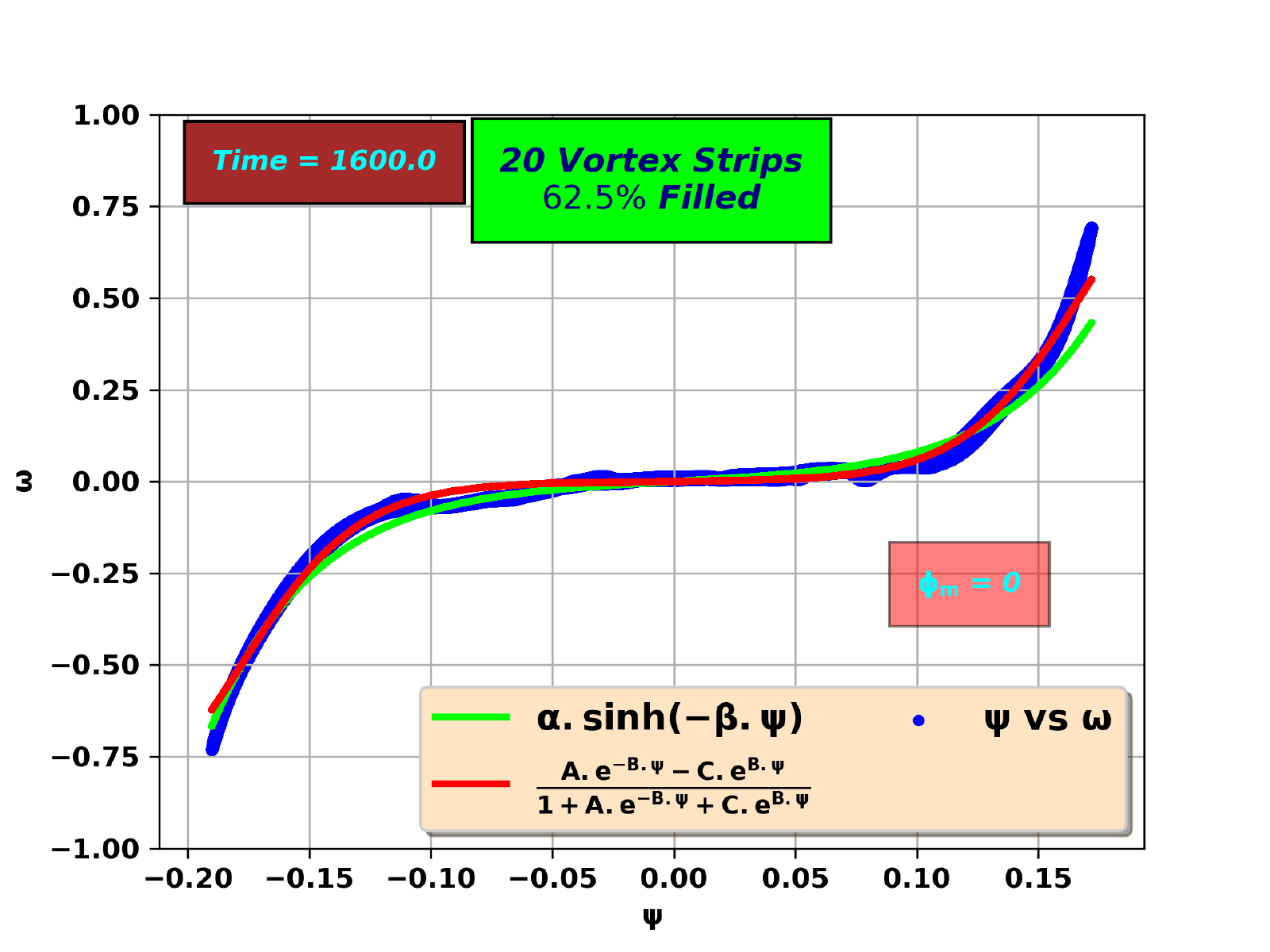}
		\caption{}
	\end{subfigure}
	\begin{subfigure}{0.45\textwidth}
		\centering
		\includegraphics[scale=0.55]{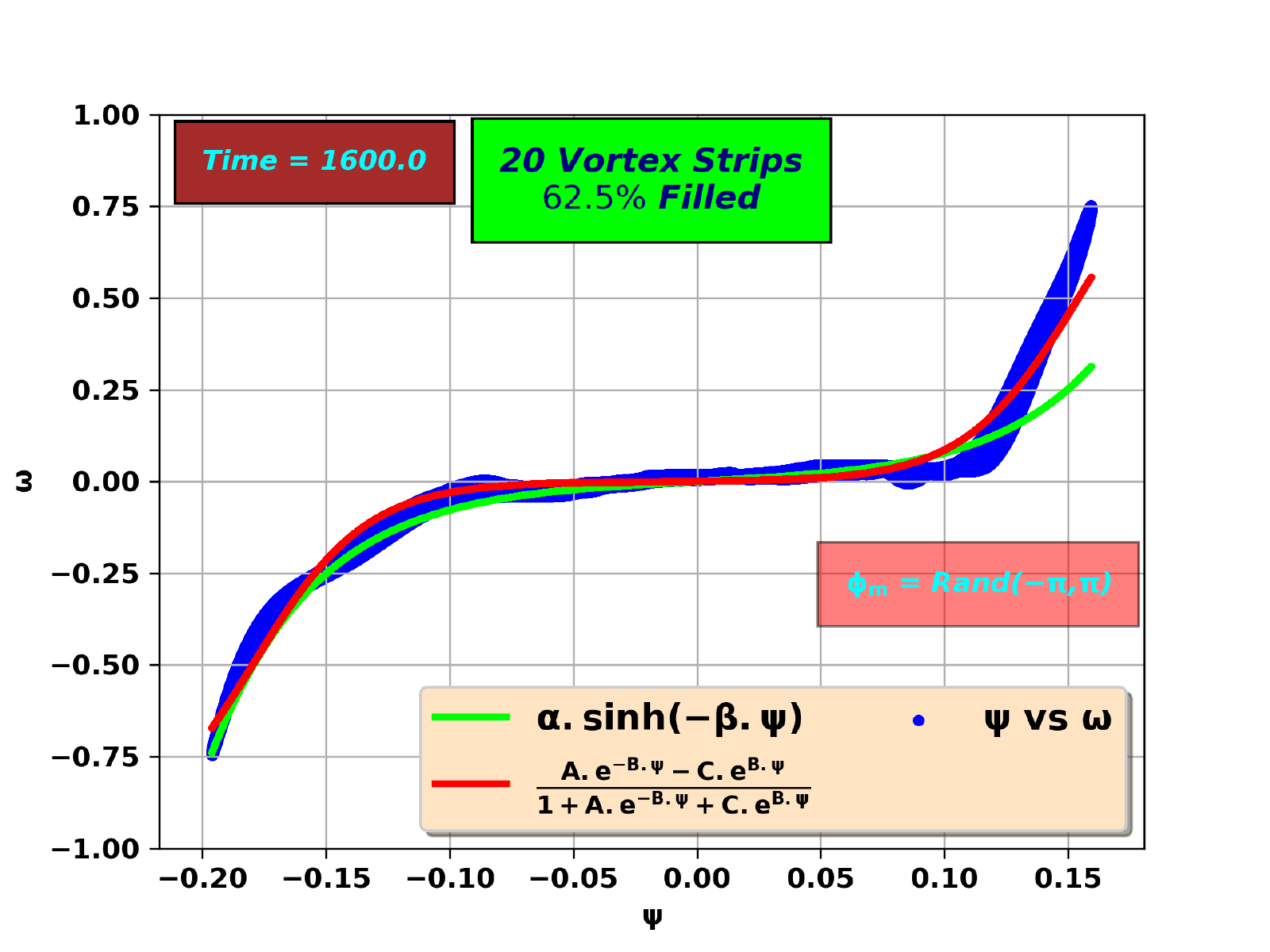}
		\caption{}
	\end{subfigure}
	\caption{$\psi$ vs $\omega$ scatter plot (at Time = 1600) for tightly packed [$62.5\%$] 20 vortex strips. Both patch vortex model (red) and the one from \textcolor{black}{a} point vortex model (green) for $\psi$ vs $\omega$ are shown. Simulation details: grid resolution $2048^2$, stepping time dt = $10^{-4}$, Reynolds number = 228576, (a) $\phi_m = 0$ (b) $\phi_m = rand(-\pi,\pi)$ (multimedia view).}
	\label{20 strips Scatter}
\end{figure*}

Even though weak capture of vortices are seen even at late times, as the quasi-steady structure of two large counter rotating vortices is reached beyond t=1000, one may expect a strong correlation to emerge between vorticity $\omega$ and stream function $\psi$ at late times.  As discussed earlier in Section \ref{physics}, a selectively decay state would result in a simple linear relationship between $\omega$ and $\psi$. Using point vortex model Pointin and Lundgren \cite{Pointin_point:1976}, Montgomery et al \cite{Montgomery_Joyce:1974}, showed based on entropy extremization that the stream function and vorticity obeys a relationship,
\begin{equation} \label{omega psi for point vortex}
\textcolor{black}{\bar {\omega}_{\textrm{PV}}} = \alpha\times Sinh(-\beta\times \bar {\psi})
\end{equation}
where as, finite size vortices KMRS theory \cite{Kuzmin:1982,Miller_PRL:1990,robert_sommeria:1991} predicted that $\bar {\omega}$ and $\bar {\psi}$ would be related by,
\begin{equation} \label{omega psi for finite vortex}
\textcolor{black}{\bar {\omega}_{\textrm{KMRS}}} = \frac{Aexp(-B\bar {\psi}) - Cexp(B\bar {\psi})}{1+\left[Aexp(-B\bar {\psi}) + Cexp(B\bar {\psi})\right]}
\end{equation}
 We observe from our simulation data that Eq. \ref{omega psi for finite vortex} shows better agreement over Eq. \ref{omega psi for point vortex}. We fit the numerically obtained $\psi$ vs $\omega$ scatter data by both the functions (Eq. \ref{omega psi for point vortex} and Eq. \ref{omega psi for finite vortex}) and identify that the theoretical prediction for finite size vortex approximation matches rather well with our late time DNS results, where as point vortex approximation does not, which is shown in Fig. \ref{20 strips Scatter} (multimedia view). Thus, our numerical observation shows quite good agreement with the theoretical prediction obtained by KMRS theory \cite{Kuzmin:1982,Miller_PRL:1990,robert_sommeria:1991}.
The set of fitting parameters for Eq. \ref{omega psi for finite vortex} are given in Table \ref{Fitting set 1} and for  Eq. \ref{omega psi for point vortex} are given in Table \ref{Fitting set 2}. Notably, as the number of strips are decreased, fit parameters (A, C and B) for Eq. \ref{omega psi for finite vortex} behave as follows: for decreasing number of strips, (A,C) values decrease while at the same time become equal to each other (i.e A = C), whereas magnitude of B increases. It is important to \textcolor{black} {observe} that the parameter B is equivalent to $\beta$ and A, C are equivalent to $\alpha$ of Eq. \ref{omega psi for point vortex}.

\begin{figure*}
	\centering
	\includegraphics[scale=0.55]{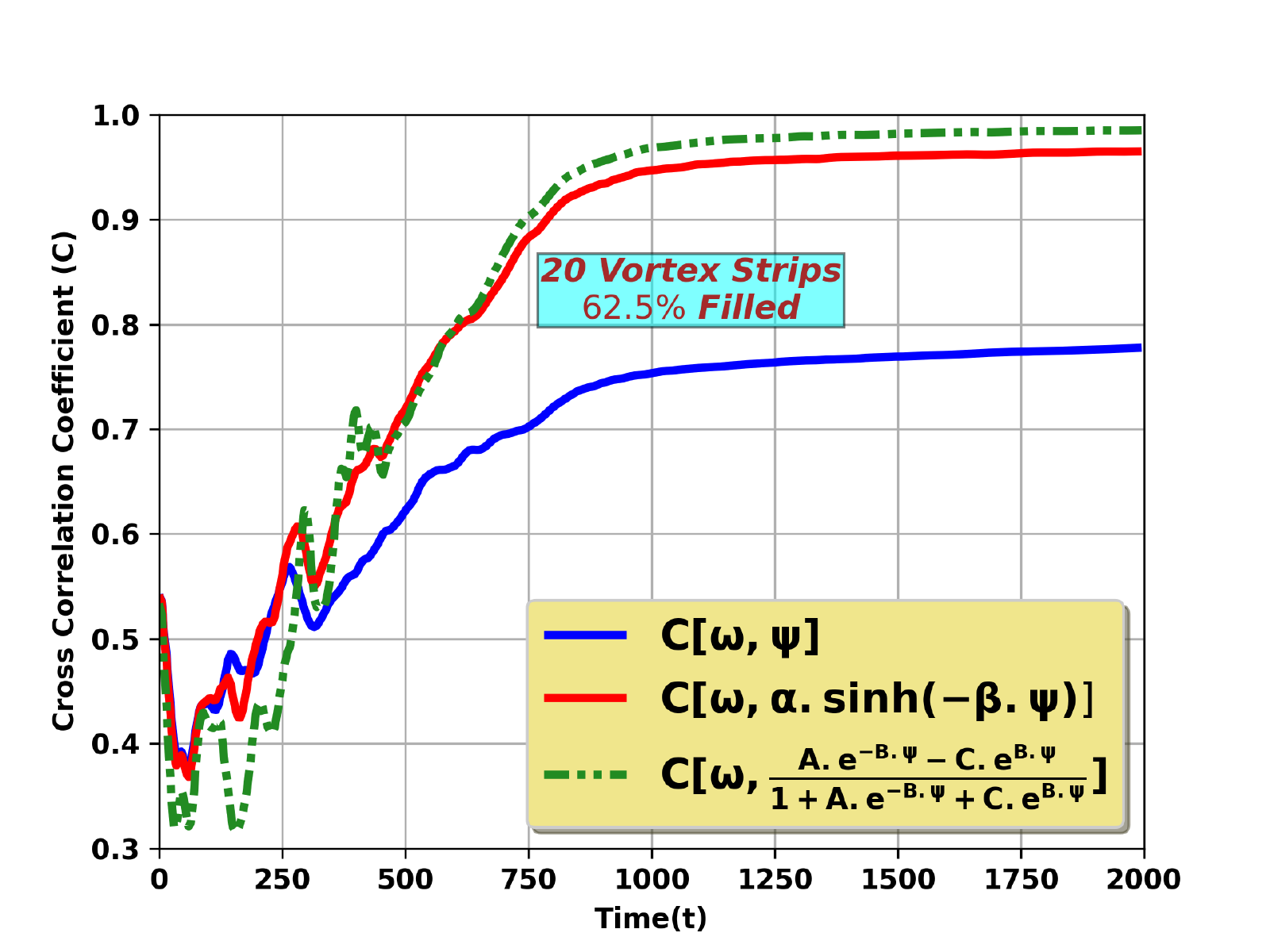}
	\caption{Dynamics of spatially averaged cross-correlations for tightly packed [$62.5\%$] 20 vortex strips between C($\omega\textcolor{black}{(x,y,t)}$, $\psi\textcolor{black}{(x,y,t)}$), $C\left(\omega\textcolor{black}{(x,y,t)}, \textcolor{black}{\bar {\omega}_{\textrm{PV}}}\textcolor{black}{(x,y,t)}\right)$ and $C\left(\omega\textcolor{black}{(x,y,t)}, \textcolor{black}{\bar {\omega}_{\textrm{KMRS}}}\textcolor{black}{(x,y,t)}\right)$. Finite size vortex approximation (Green dotted line) shows good correlation (near to 1.0) than the point vortex approximation (red solid line). Simulation details: grid resolution $2048^2$, stepping time dt = $10^{-4}$, Reynolds number = 228576.}
	\label{20 Strips CC}
\end{figure*}

Following Montgomery et al. \cite{sinh:1992}, to understand the level of  quantitative agreement of the late time DNS data with these two statistical mechanical vortex models [Eqs. \ref{omega psi for point vortex} and \ref{omega psi for finite vortex}], we calculate cross-correlation coefficient ($C$) between instantaneous vorticity field and stream function as function of time t. The cross-correlation function ($C$) between any two quantities say $\gamma (x,y,\textcolor{black}{t})$ and $\delta (x,y,\textcolor{black}{t})$ is defined as \cite{sinh:1992},

\begin{equation}
C(\gamma\textcolor{black}{(x,y,t)},\delta\textcolor{black}{(x,y,t)}) = \frac{\left<(\gamma - <\gamma>) . (\delta - <\delta>)\right> }{\left[\left<(\gamma - <\gamma>)^2\right>   . \left<(\delta - <\delta>)^2\right> \right]^\frac{1}{2}}
\end{equation}
where $\left<\right>$ represents spatial average. We calculate $C(\omega\textcolor{black}{(x,y,t)},\psi\textcolor{black}{(x,y,t)})$, $C\left(\omega\textcolor{black}{(x,y,t)}, \textcolor{black}{\bar {\omega}_{\textrm{PV}}}\textcolor{black}{(x,y,t)}\right)$ and $C\left(\omega\textcolor{black}{(x,y,t)}, \textcolor{black}{\bar {\omega}_{\textrm{KMRS}}}\textcolor{black}{(x,y,t)}\right)$ vs time and observe that $C\left(\omega\textcolor{black}{(x,y,t)}, \textcolor{black}{\bar {\omega}_{\textrm{KMRS}}}\textcolor{black}{(x,y,t)}\right)$ shows best proportionality, $C$ value very near 1 as compared to $C\left(\omega\textcolor{black}{(x,y,t)}, \textcolor{black}{\bar {\omega}_{\textrm{PV}}}\textcolor{black}{(x,y,t)}\right)$ and $C(\omega\textcolor{black}{(x,y,t)},\psi\textcolor{black}{(x,y,t)})$ [See Fig.\ref{20 Strips CC} \textcolor{black}{\& Table \ref{CC Table}}]. 

\begin{table}
	\centering
	\begin{tabular}{ |c|c|c|c|c| }
		\hline
		 \textbf{Strips} & \textbf{$\phi_m$} &  \textbf{A} & \textbf{B} & \textbf{C} \\
		 \hline
		 20 & 0 & 0.00200443 & -41.3587 & 0.000633572  \\
		 \hline
		 20 & Rand($-\pi$, $\pi$) & 0.00118647 & -42.6915 & 0.000889118 \\
		\hline
		 16 & 0 & 0.00123397 & -42.9227 & 0.000455795  \\
		\hline
		16 & Rand($-\pi$, $\pi$) & 0.000425255 & -45.8796 & 0.000647814 \\
		\hline
		8 & 0 & 8.57733$\times 10^{-5}$ & -49.3648 & 8.64778$\times 10^{-5}$ \\
		\hline
		8 & Rand($-\pi$, $\pi$) & 0.00013746 & -47.5862 & 0.000138147  \\
		\hline
		4 & 0 & 6.60245$\times 10^{-5}$ & -49.7774 & 5.55811$\times 10^{-5}$ \\
		\hline
		4 & Rand($-\pi$, $\pi$) & 6.10826$\times 10^{-5}$ & -52.7146 & 6.11048$\times 10^{-5}$ \\
		\hline
	\end{tabular}
	\caption{Fitting Parameter details for function $f(\psi) =\frac{Aexp(-B\psi) - Cexp(B\psi)}{1+\left[Aexp(-B\psi) + Cexp(-B\psi)\right]}$ with 2-dimensional DNS results at t= 1600.0. Simulation details: grid resolution $2048^2$, stepping time dt = $10^{-4}$, Reynolds number = 228576.}
	\label{Fitting set 1}
\end{table}

\begin{table}
	\centering
	\begin{tabular}{ |c|c|c|c| }
		\hline
		\textbf{Strips} & \textbf{$\phi_m$} &  \textbf{$\alpha$} & \textbf{$\beta$} \\
		\hline
		20 & 0 & 0.00826747 & -27.4554 \\
		\hline
		20 & Rand($-\pi$, $\pi$) & 0.0151563 & -23.3554 \\
		\hline
		16 & 0 & 0.010259 & -24.5965 \\
		\hline
		16 & Rand($-\pi$, $\pi$) & 0.00771058 & -27.0463 \\
		\hline
		8 & 0 & 0.00599178 & -26.1134 \\
		\hline
		8 & Rand($-\pi$, $\pi$) & 0.0066705 & -25.9249 \\
		\hline
		4 & 0 & 0.00652715 & -24.8701\\
		\hline
		4 & Rand($-\pi$, $\pi$) & 0.00919652 & -24.2253 \\
		\hline
	\end{tabular}
	\caption{Fitting Parameter details for function $f(\psi) =\alpha \sinh(-\beta \psi)$ with 2-dimensional DNS results at t= 1600.0. Simulation details: grid resolution $2048^2$, stepping time dt = $10^{-4}$, Reynolds number = 228576.}
	\label{Fitting set 2}
\end{table}

\subsection{16 Vortex Strips with total Packing Fraction $50.0\%$  - Runs 3, 4}\label{16 Strips}
\begin{figure*}
	\centering
	\begin{subfigure}{0.32\textwidth} 
		\centering
		\includegraphics[scale=0.39]{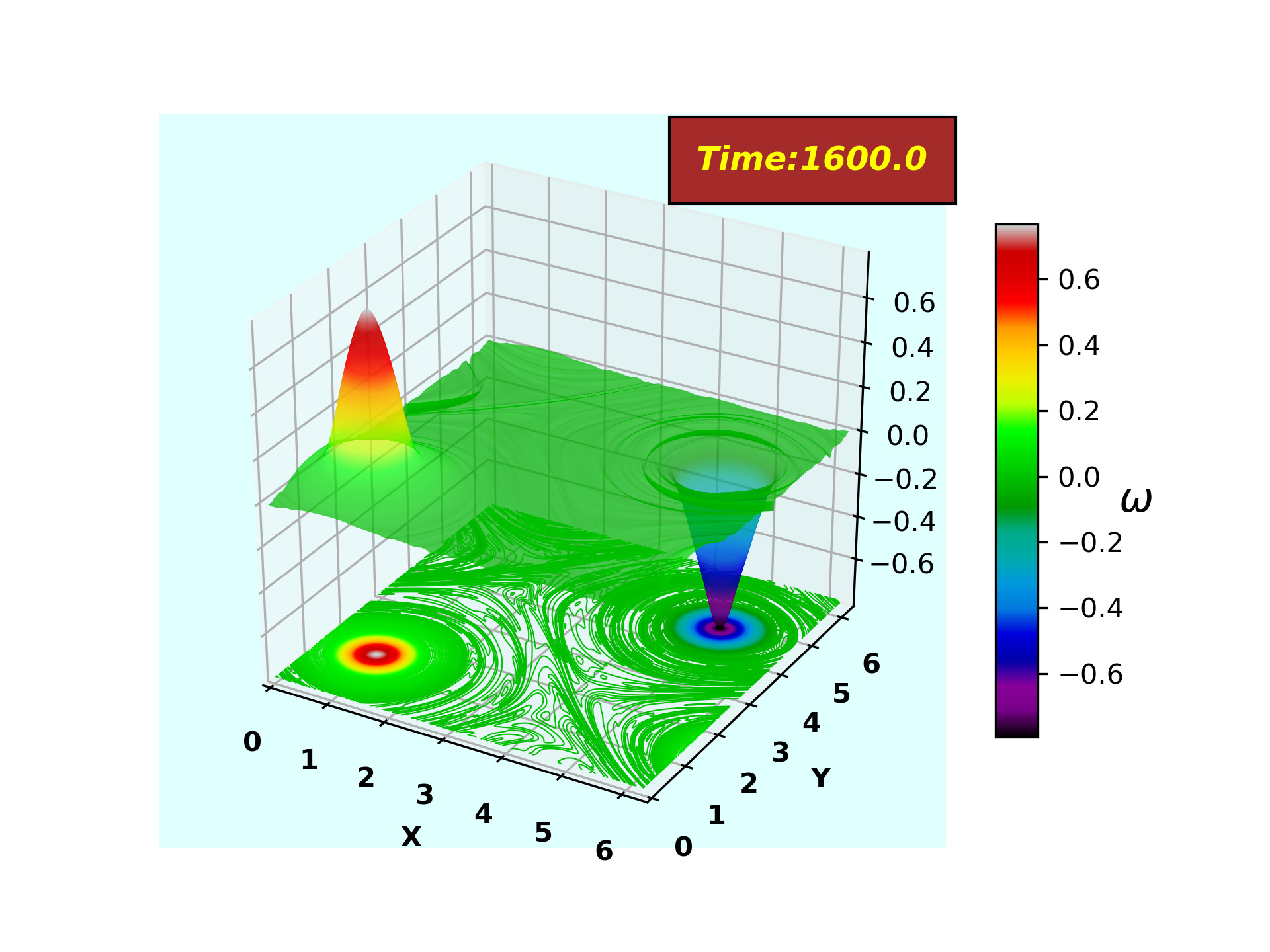}
		\caption{}
		\label{16 strips Vorticity}
	\end{subfigure}	
	\begin{subfigure}{0.32\textwidth} 
		\centering
		\includegraphics[scale=0.39]{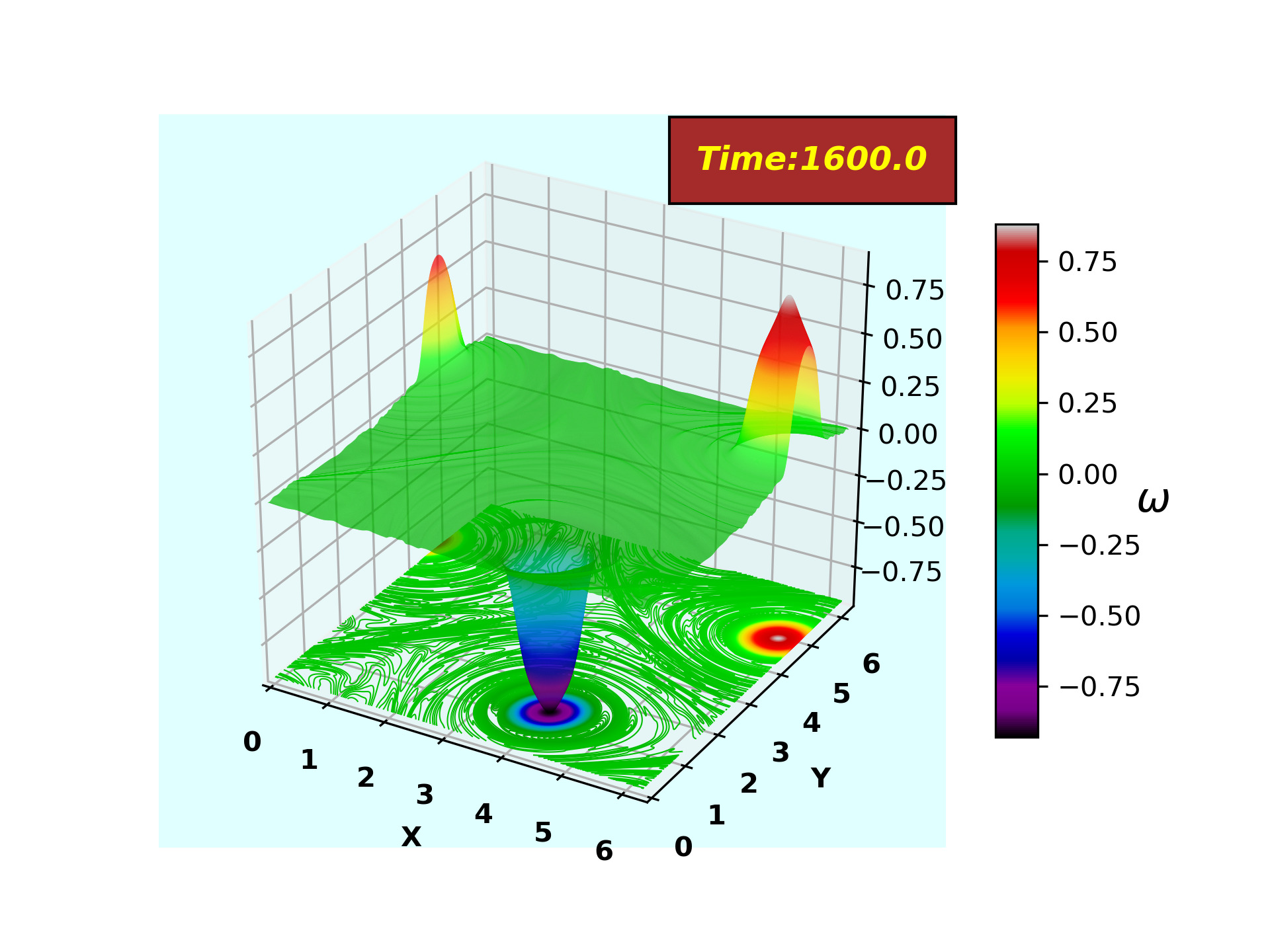}
		\caption{}
		\label{8 strips Vorticity}
	\end{subfigure} 
	\begin{subfigure}{0.32\textwidth}
		\centering
		\includegraphics[scale=0.39]{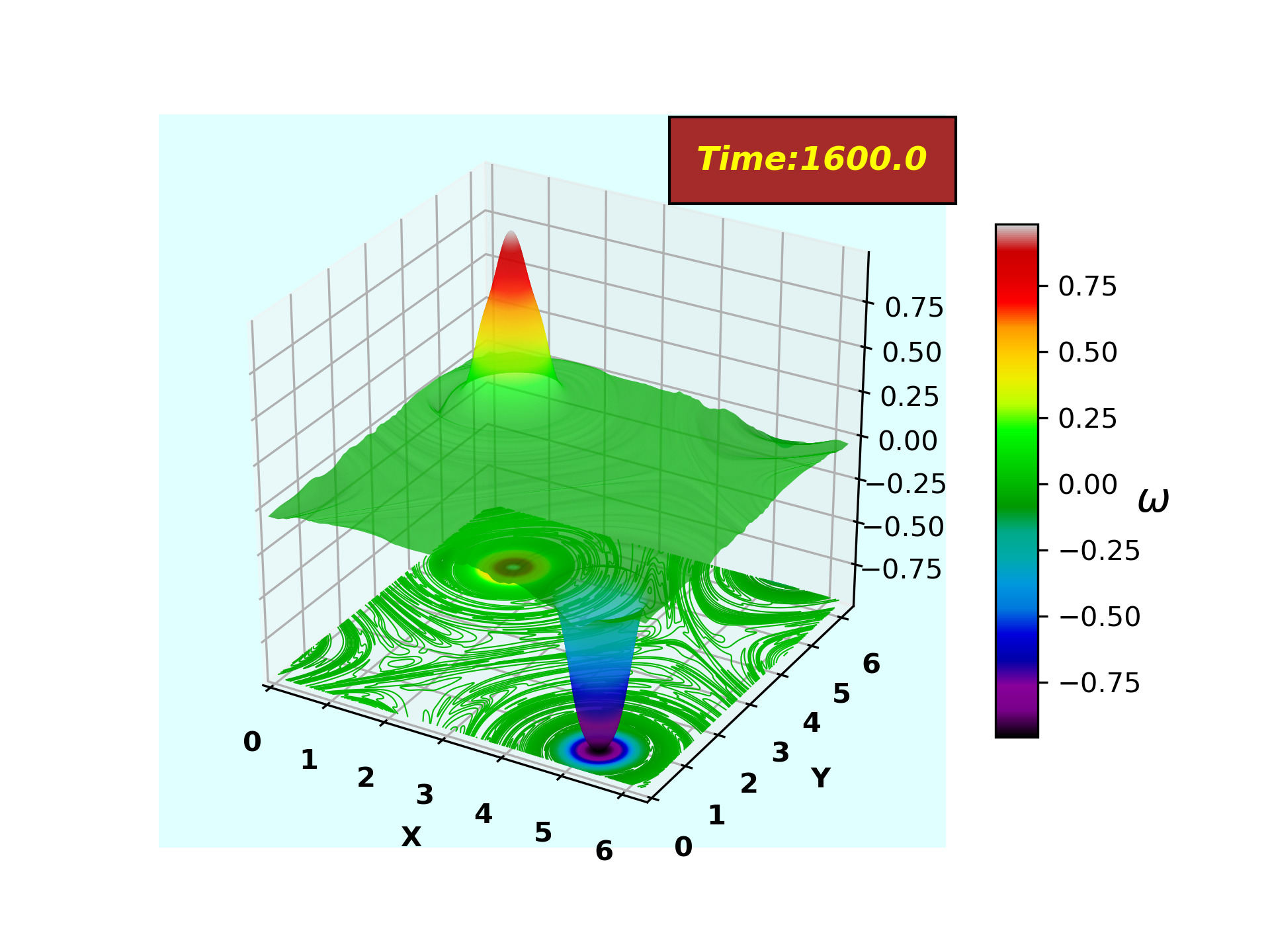}
		\caption{}
		\label{4 strips Vorticity}
	\end{subfigure} 
	\caption{\textcolor{black}{Late time state (at t=1600) of vorticity ($\omega(x,y,t)$), as time goes on similar polarity vortices merge and finally end up with one single vortex of each sign for, (a) moderately packed [$50.0\%$], (b) loosely packed [$25.0\%$], (c) very loosely packed [$12.5\%$] initial vortex configuration. Simulation details: grid resolution $2048^2$, stepping time dt = $10^{-4}$, Reynolds number = 228576.}}
\end{figure*}

We have \textcolor{black}{considered} a densely packed vorticity configuration as our initial condition in the previous subsection \ref{20 Strips} . There we ensure that the dimension\textcolor{black}{s} of each vortex strips are comparable with the unoccupied strips. As discussed earlier that from \textcolor{black}{a} statistical mechanical point of view one predicts the most probable state for finite size vortex taking the exclusion effect into account. As the initial total occupancy of vorticities of either kind is systematically reduced, this ``dilution''  plus vortex dynamics will tend to increase, on an average, the ``inter-particle'' distance and move towards an effective point vortex model.  Hence one should be able to systematically see signatures of increased agreement between our late time datum of DNS,  point vortex model predictions and that of KRMS theory with increasing dilution or reduced initial packing fraction values, whereas for the late time datum of DNS and KRMS theory may be expected agree more as compared to DNS and point vortex theory, for tight packing or increased initial packing fraction values. Hence keeping this idea in mind, we pack the initial vorticity distribution moderately i.e. only 50$\%$ of simulation domain filled by the vortex strips (rotating either clockwise or anti clockwise) and rest 50$\%$ vacant (zero vorticity region). The width of each vortex strips are considered same as earlier [See Fig. \ref{initial} (b)].\\
As seen earlier, the velocity shear destabilizes the vortex strips due to Kelvin-Helmholtz instability. Eventually the configuration evolves towards turbulence and it is strongly dominated by turbulence along with a rapid mixing of vortex layers. Due to inverse cascading of energy, system ends up with two largest vortex of either sign [See Fig. \ref{16 strips Vorticity}].

\begin{figure*}
	\centering
	\begin{subfigure}{0.32\textwidth} 
		\centering
		\includegraphics[scale=0.39]{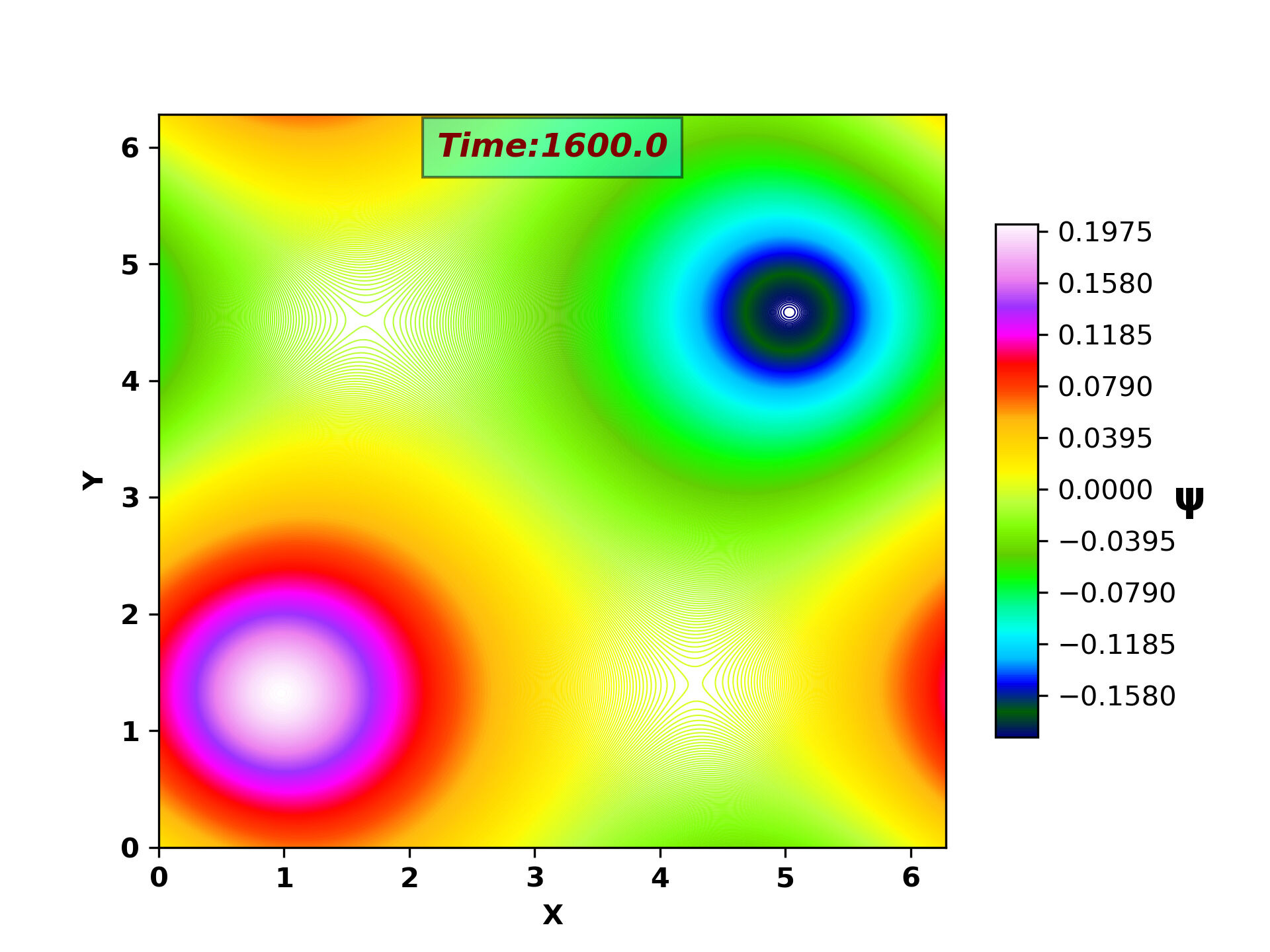}
		\caption{}
		\label{16 strips Stream function}
	\end{subfigure}	
	\begin{subfigure}{0.32\textwidth} 
		\centering
		\includegraphics[scale=0.39]{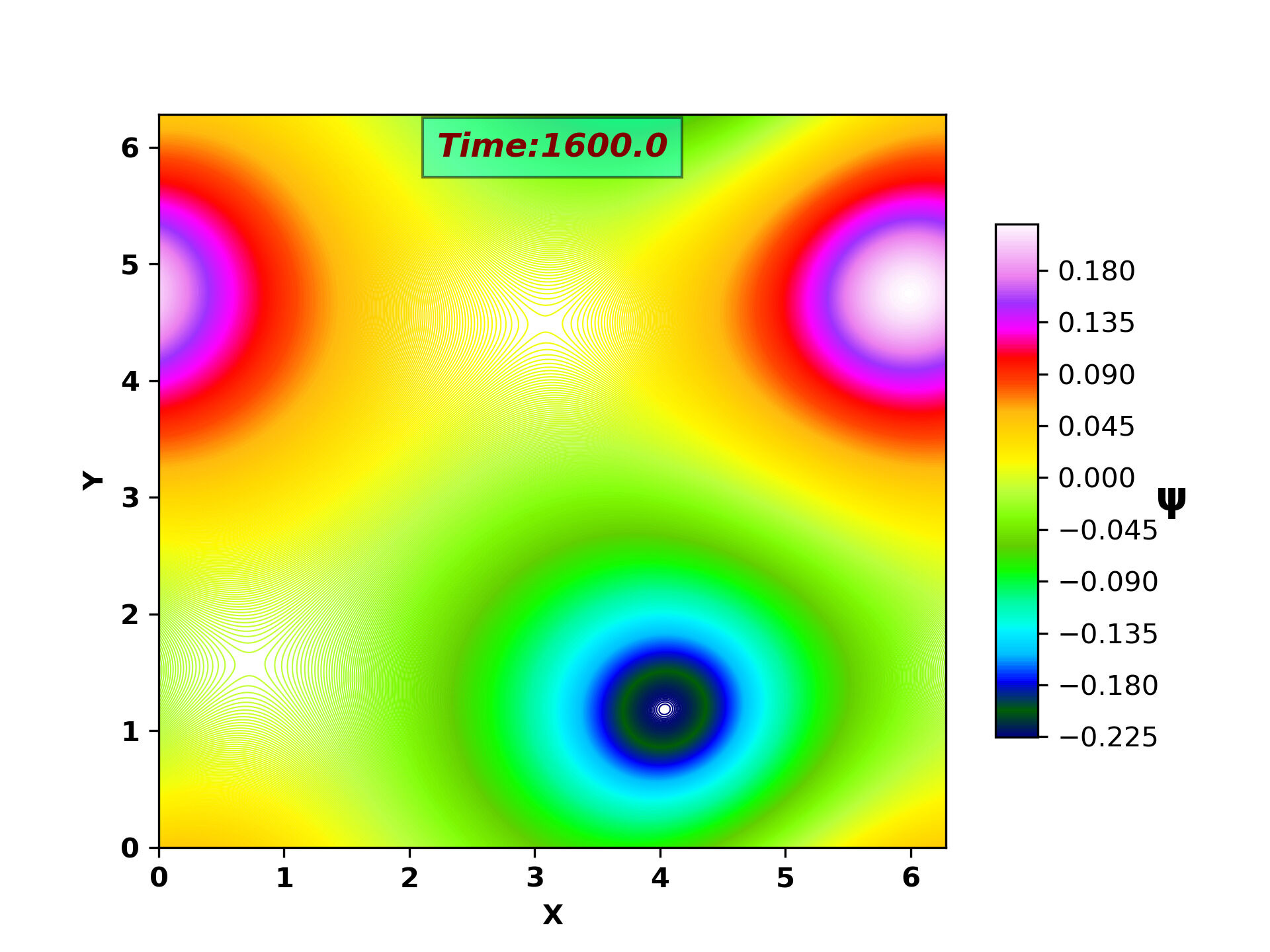}
		\caption{}
		\label{8 strips Stream function}
	\end{subfigure} 
	\begin{subfigure}{0.32\textwidth}
		\centering
		\includegraphics[scale=0.39]{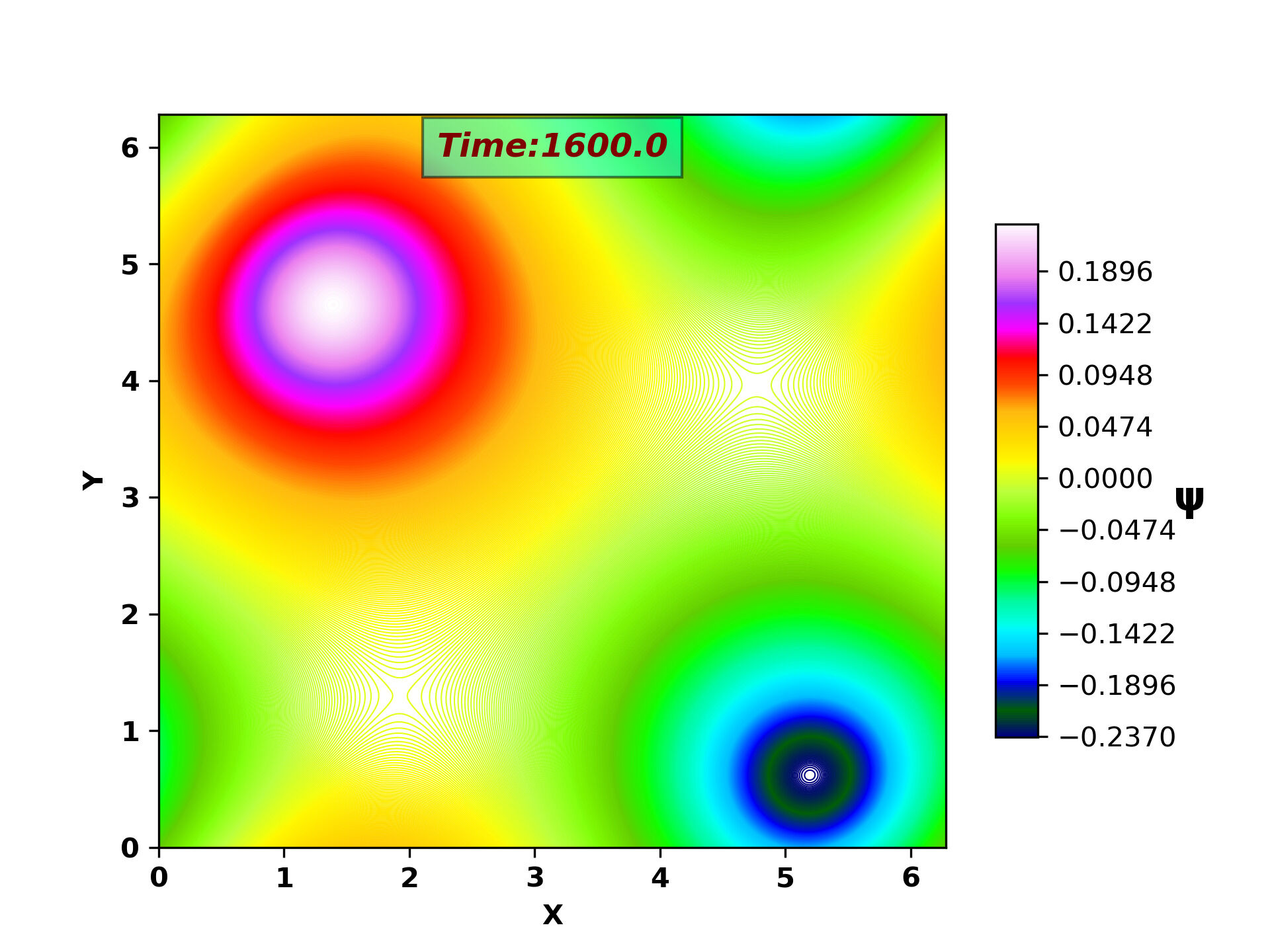}
		\caption{}
		\label{4 strips Stream function}
	\end{subfigure} 
	\caption{\textcolor{black}{Late time state (at t=1600) of stream function ($\psi$), after all the possible vortex mergers occur, the streamlines achieve Ewald potential like contours (with a basic cell containing two point vortices), also observed by Montgomery et al. \cite{montgomery_d:1991,montgomery_prl:1991} for, (a) moderately packed [$50.0\%$], (b) loosely packed [$25.0\%$], (c) very loosely packed [$12.5\%$] initial vortex configuration. Simulation details: grid resolution $2048^2$, stepping time dt = $10^{-4}$, Reynolds number = 228576.}}
\end{figure*}

\begin{figure*}
	\centering
	\begin{subfigure}{0.32\textwidth} 
		\centering
		\includegraphics[scale=0.39]{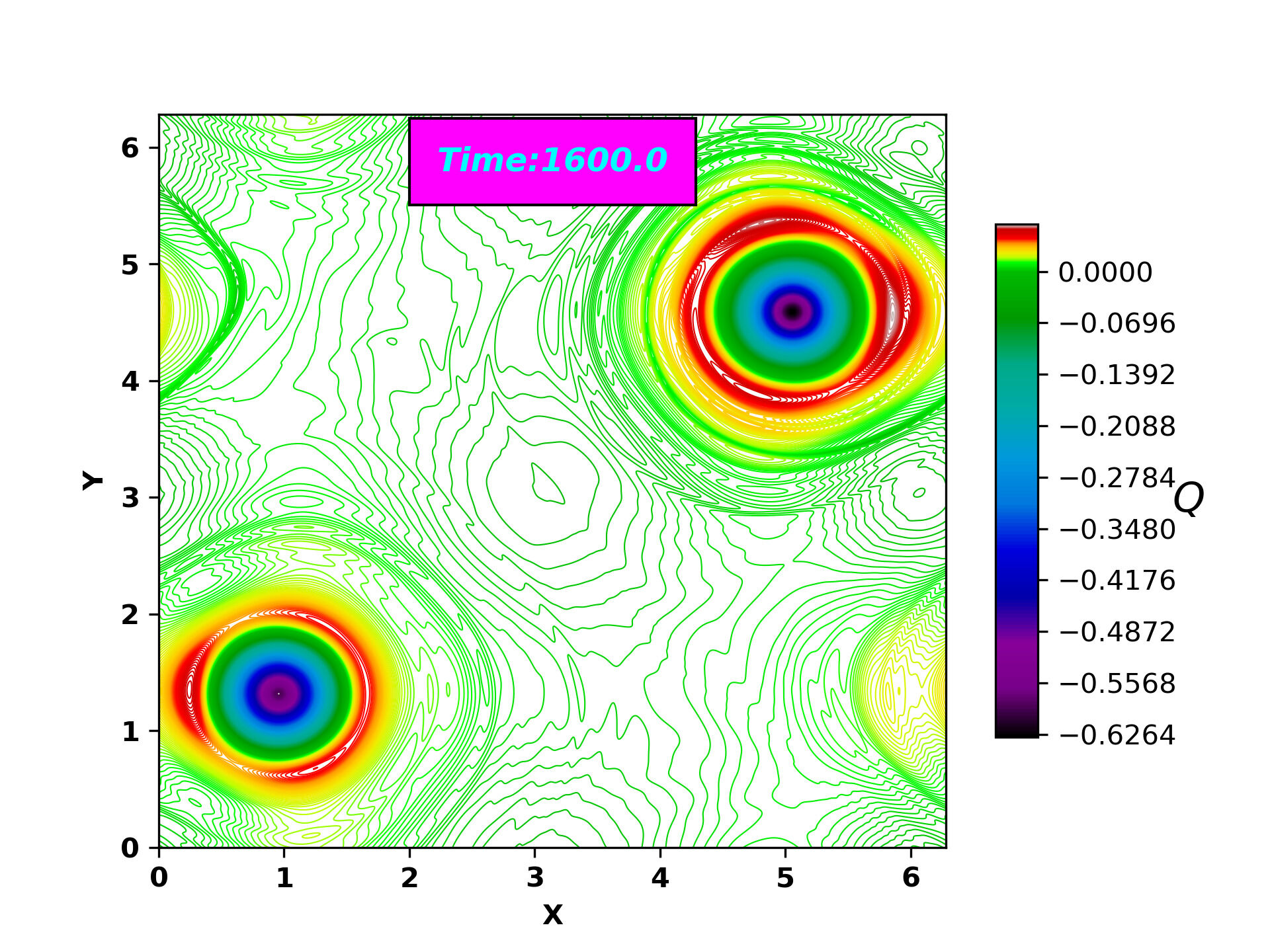}
		\caption{}
		\label{16 strips Okubo}
	\end{subfigure}	
	\begin{subfigure}{0.32\textwidth} 
		\centering
		\includegraphics[scale=0.39]{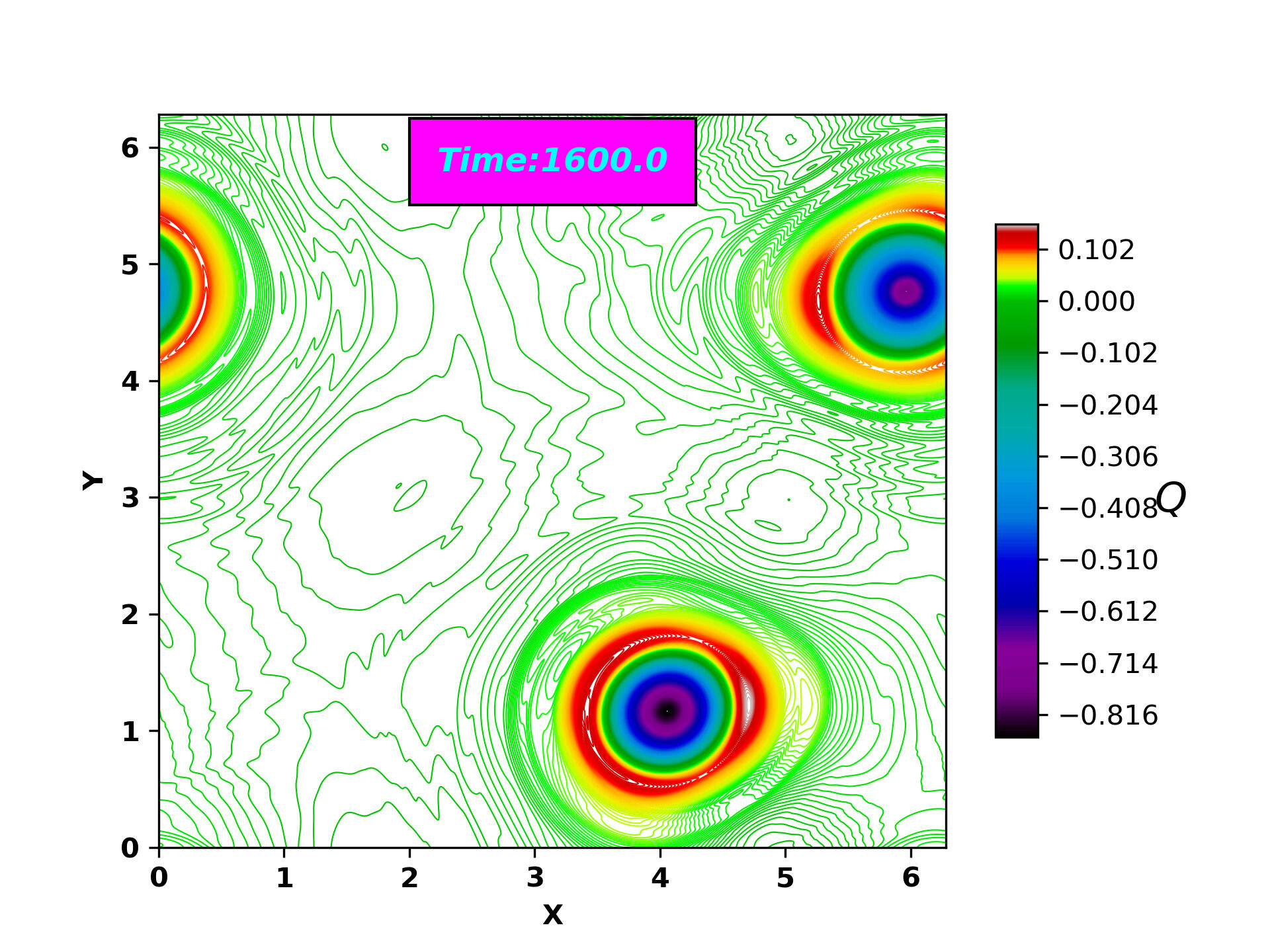}
		\caption{}
		\label{8 strips Okubo}
	\end{subfigure} 
	\begin{subfigure}{0.32\textwidth}
		\centering
		\includegraphics[scale=0.39]{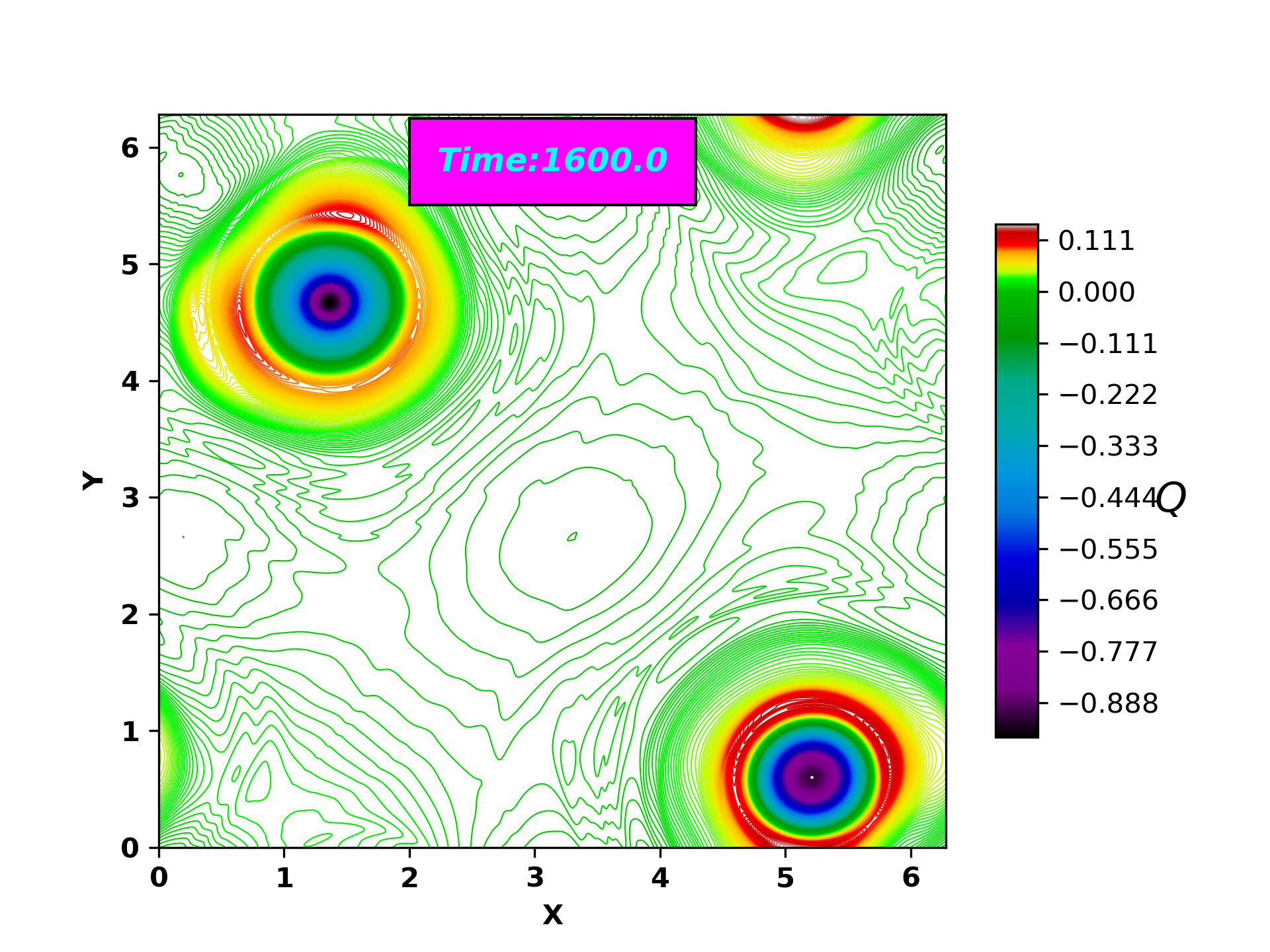}
		\caption{}
		\label{4 strips Okubo}
	\end{subfigure} 
	\caption{\textcolor{black}{Late time state (at t=1600) of Okubo-Weiss parameter ($Q(x,y,t)$), with two distinct regions (i) Vortex cores: characterized by strong negative value of $Q(x,y,t)$. (ii) Strain cells: surrounding the vortex cores, characterized by large positive value of $Q(x,y,t)$ for, (a) moderately packed [$50.0\%$], (b) loosely packed [$25.0\%$], (c) very loosely packed [$12.5\%$] initial vortex configuration. Simulation details: grid resolution $2048^2$, stepping time dt = $10^{-4}$, Reynolds number = 228576.}}
\end{figure*}

The dynamics of the stream function leads towards Ewald Potential like contour (with a basic cell containing two point vortices), observed by Montgomery et al. as well \cite{montgomery_d:1991,montgomery_prl:1991} [See Fig. \ref{16 strips Stream function}] same as earlier case.
We calculate Okubo-Weiss parameter ($Q(x,y,t)$) from our simulation and identify regions of vortex cores, characterized by strong negative value of $Q(x,y,t)$ and strain cells surrounding the vortex cores, characterized by large positive value of $Q(x,y,t)$ [See Fig.\ref{16 strips Okubo}].\\

\begin{figure*}
	\centering
	\begin{subfigure}{0.45\textwidth}
		\centering
		\includegraphics[scale=0.55]{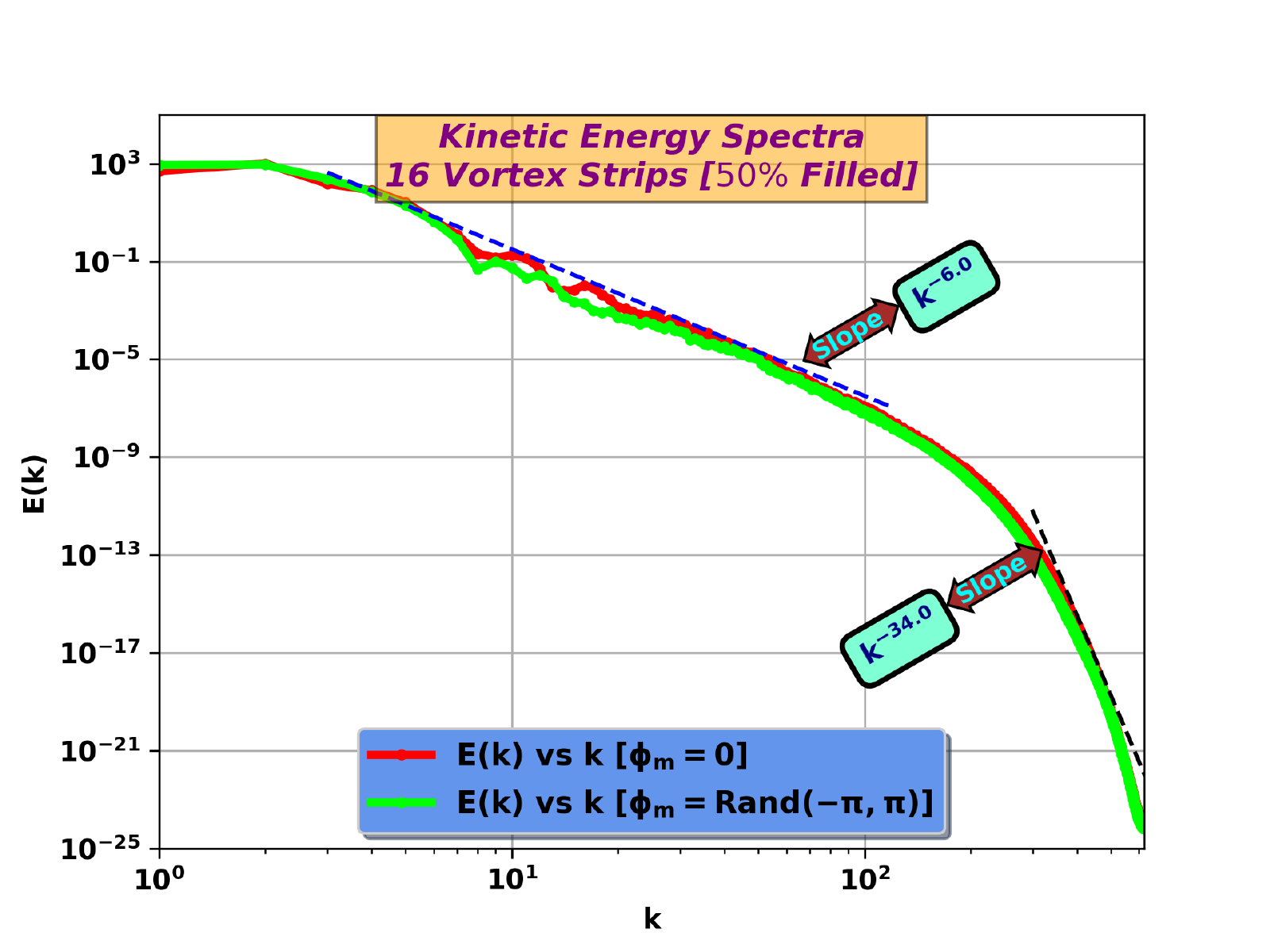}
		\caption{}
	\end{subfigure}
	\begin{subfigure}{0.45\textwidth}
		\centering
		\includegraphics[scale=0.55]{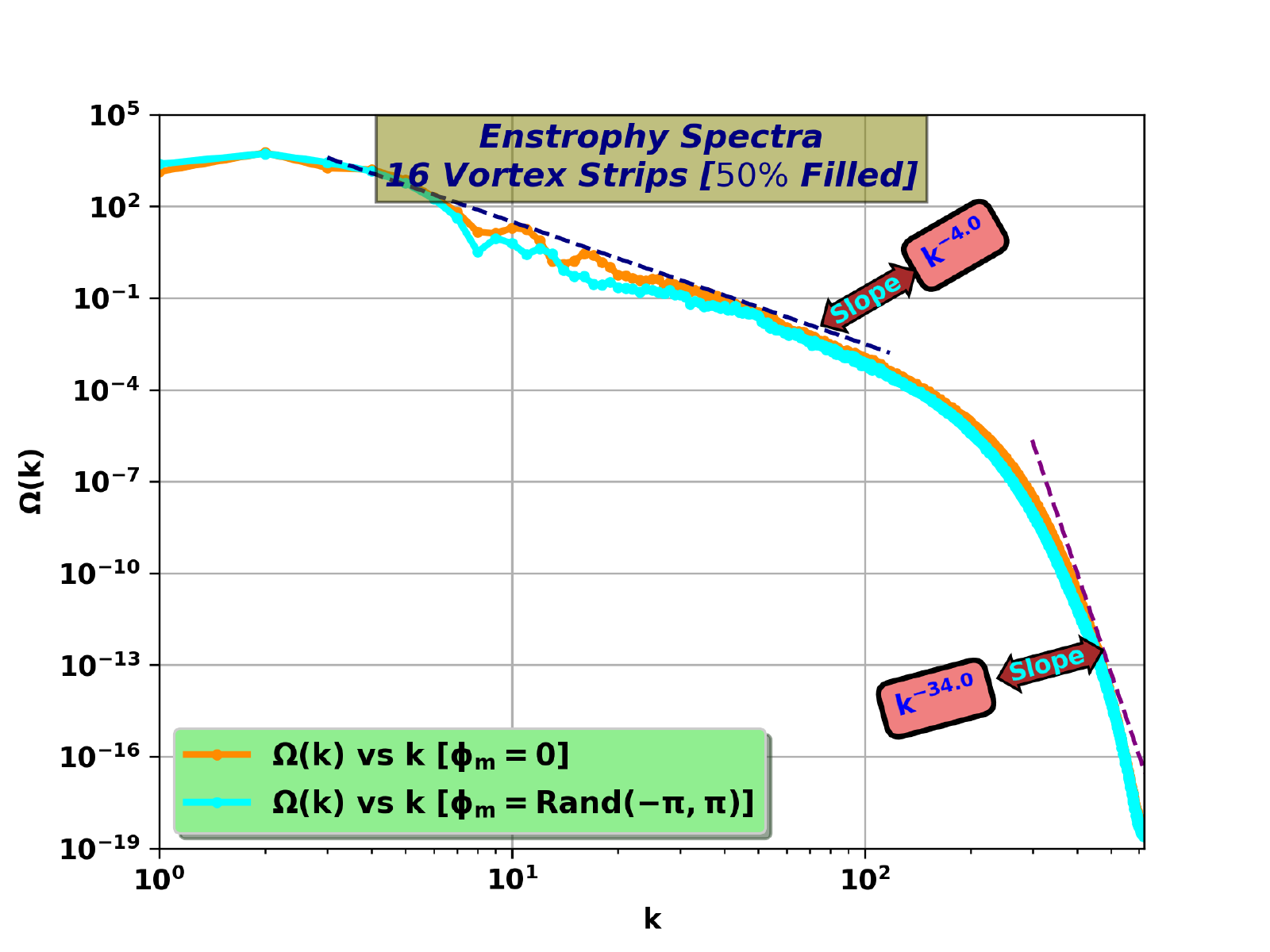}
		\caption{}
	\end{subfigure}
	\caption{\textcolor{black}{(a)} Time averaged (time average taken after saturation i.e from t=1000.0 to 1600.0) kinetic energy spectra [$\int_{0}^{\infty} E(k) dk$] showing inverse cascading (b) Time averaged (time average taken after saturation i.e from t=1000.0 to 1600.0) enstrophy spectra [$\int_{0}^{\infty} \Omega(k) dk$] showing direct cascading for moderately packed 16 Vortex strips ($50.0\%$ packed) configuration. These k-scaling\textcolor{black}{s} for kinetic energy and enstrophy agree with the earlier work \cite{gomez:1996}. Simulation details: grid resolution $2048^2$, stepping time dt = $10^{-4}$, Reynolds number = 228576.}
	\label{16 strips Spectra}
\end{figure*}

The kinetic energy spectra indicates the inverse cascading with scaling $E(k)\propto k^{-6}$ for lower $k$ and $E(k)\propto k^{-34}$ for higher $k$ [See Fig. \ref{16 strips Spectra}(a)]. However, the enstrophy spectra shows direct cascading with scaling $E(k)\propto k^{-4}$ for lower $k$ and $E(k)\propto k^{-34}$ for higher $k$ [See Fig. \ref{16 strips Spectra}(b)]. These k-scaling for kinetic energy and enstrophy agree with earlier works, for example, by Dmiturk et al. \cite{gomez:1996} for slowly decaying 2-dimensional Navier-Stokes turbulence.\\

\begin{figure*}
	\centering
	\begin{subfigure}{0.32\textwidth} 
		\centering
		\includegraphics[scale=0.39]{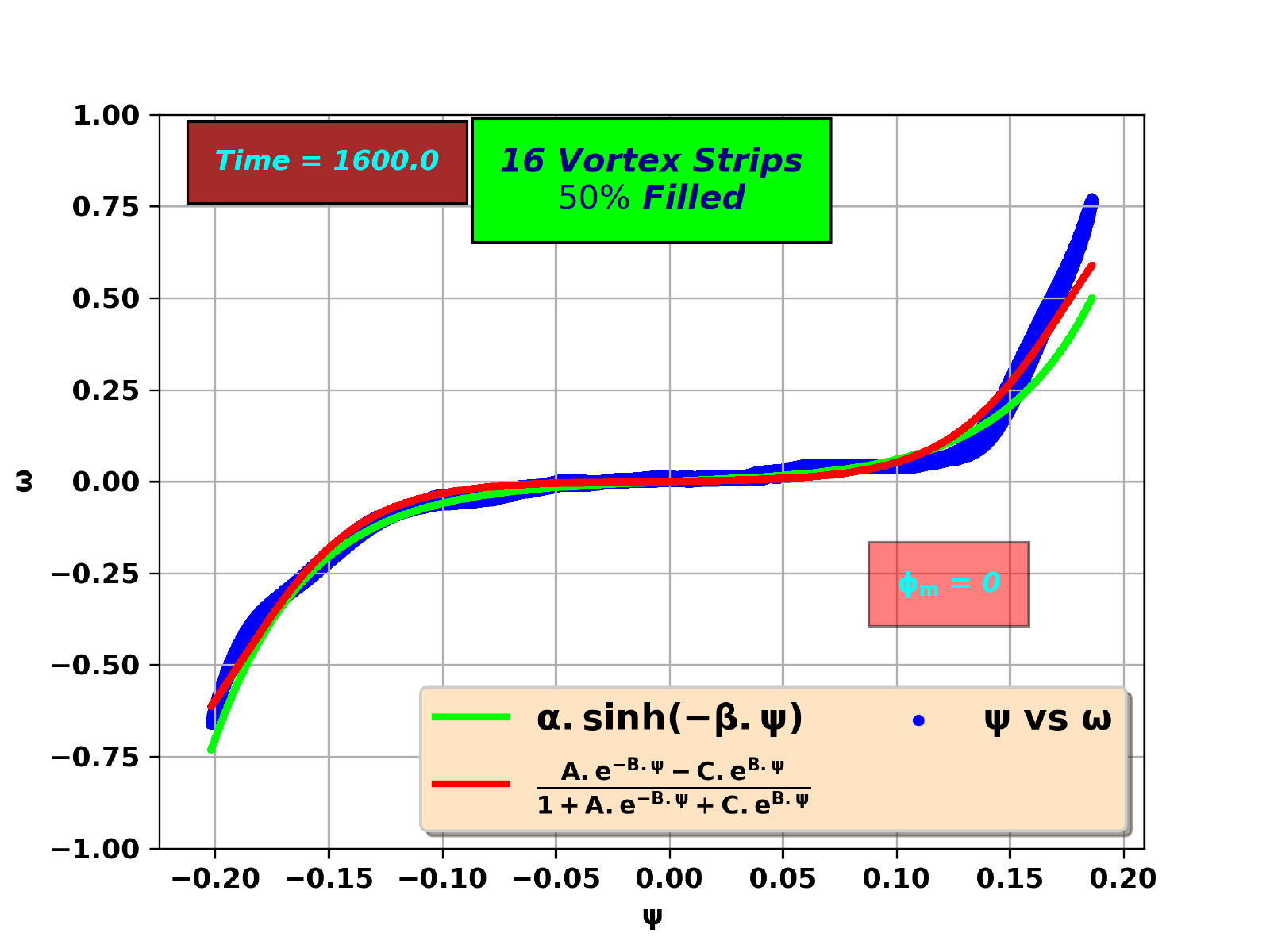}
		\caption{}
		\label{16 Strips Scatter}
	\end{subfigure}	
	\begin{subfigure}{0.32\textwidth} 
		\centering
		\includegraphics[scale=0.39]{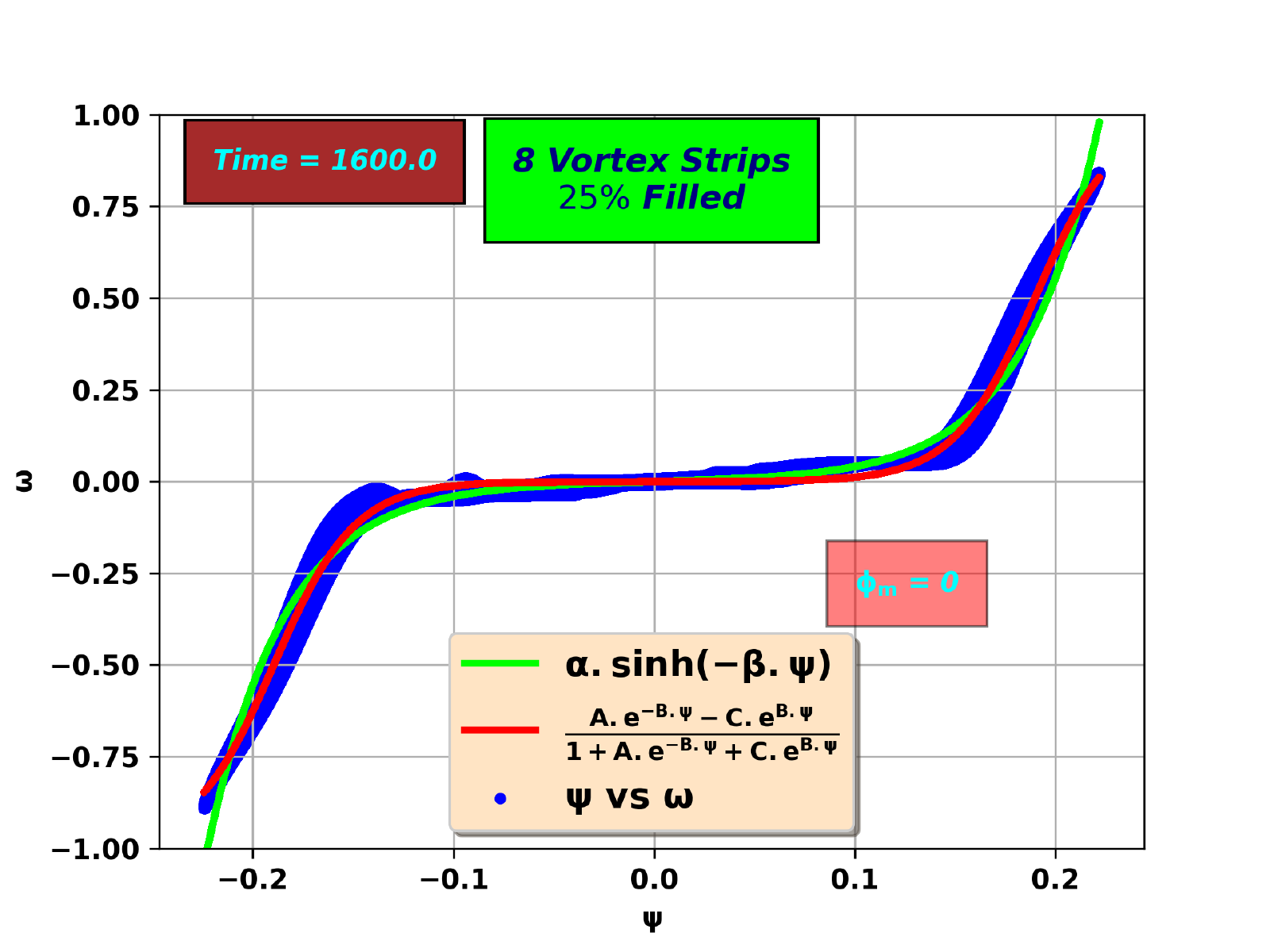}
		\caption{}
		\label{8 Strips Scatter}
	\end{subfigure} 
	\begin{subfigure}{0.32\textwidth}
		\centering
		\includegraphics[scale=0.39]{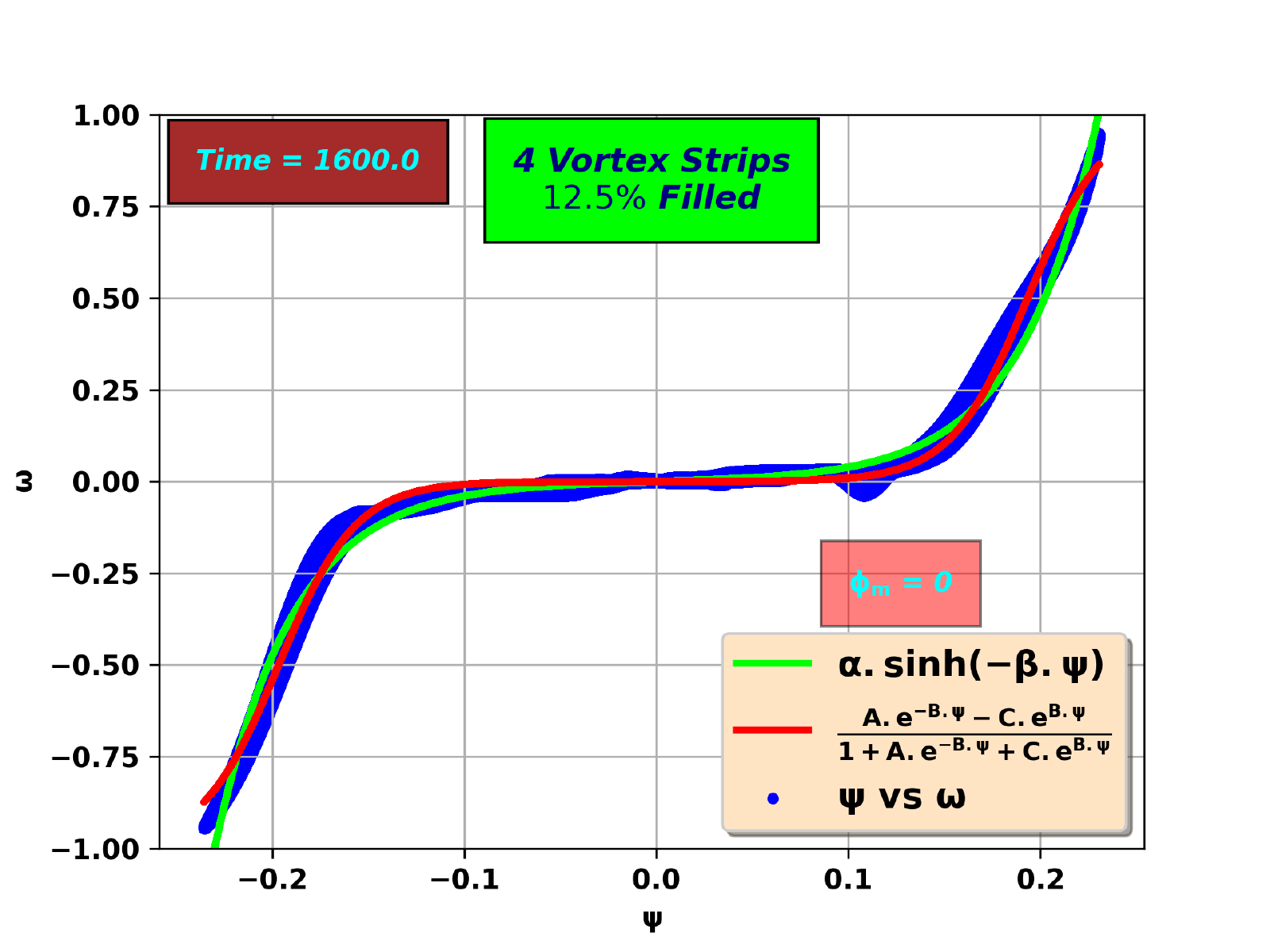}
		\caption{}
		\label{4 Strips Scatter}
	\end{subfigure} 
	\caption{\textcolor{black}{$\psi$ vs $\omega$ scatter plot (at Time = 1600) for, (a) moderately packed [$50.0\%$], (b) loosely packed [$25.0\%$], (c) very loosely packed [$12.5\%$] initial vortex configuration. Both patch vortex model (red line) and the one from \textcolor{black}{a} point vortex model (green line) for $\psi$ vs $\omega$ are shown. Simulation details: grid resolution $2048^2$, stepping time dt = $10^{-4}$, Reynolds number = 228576, $\phi_m = 0$.}}
		\label{all strips Scatter}
\end{figure*}

We plot $\psi$ vs $\omega$ for this case- one with random noise another without any random noise in perturbation and fit the function for point vortex approximation Eq. \ref{omega psi for point vortex} and function for finite size vortex approximation Eq. \ref{omega psi for finite vortex}. It is observed that the function (Eq. \ref{omega psi for finite vortex}) continues to show best proportionality with respect to Eq. \ref{omega psi for point vortex} [See Fig. \ref{16 Strips Scatter}, \ref{16 Strips Scatter phase}]. From the fitting parameter for the function (Eq. \ref{omega psi for finite vortex}) given in Table \ref{Fitting set 1}, we identify the value of $A$ and $C$ to be reducing with respect to earlier case. The decrement of  $A$ and $C$ values from former case (Case: A) strongly indicates the slow transition towards point vortex limit predicted analytically earlier by KMRS theory \cite{Kuzmin:1982,Miller_PRL:1990,robert_sommeria:1991}, as initial total circulation of either kind ($+/-$) is reduced such that total initial circulation is zero. \textcolor{black}{We have performed runs with random noise and without random noise in the phases for all the cases and have demonstrated the late time states are nearly independent of the details of the initial phase values.  In the following, we consider only a single value of phase ($\phi_m = 0$).  The details for all the cases (B, C, D) with $\phi_m = $ random noise may be found in the Appendix-\ref{Appen A} of this work.}\\

\begin{figure*}
	\centering
	\begin{subfigure}{0.32\textwidth} 
		\centering
		\includegraphics[scale=0.39]{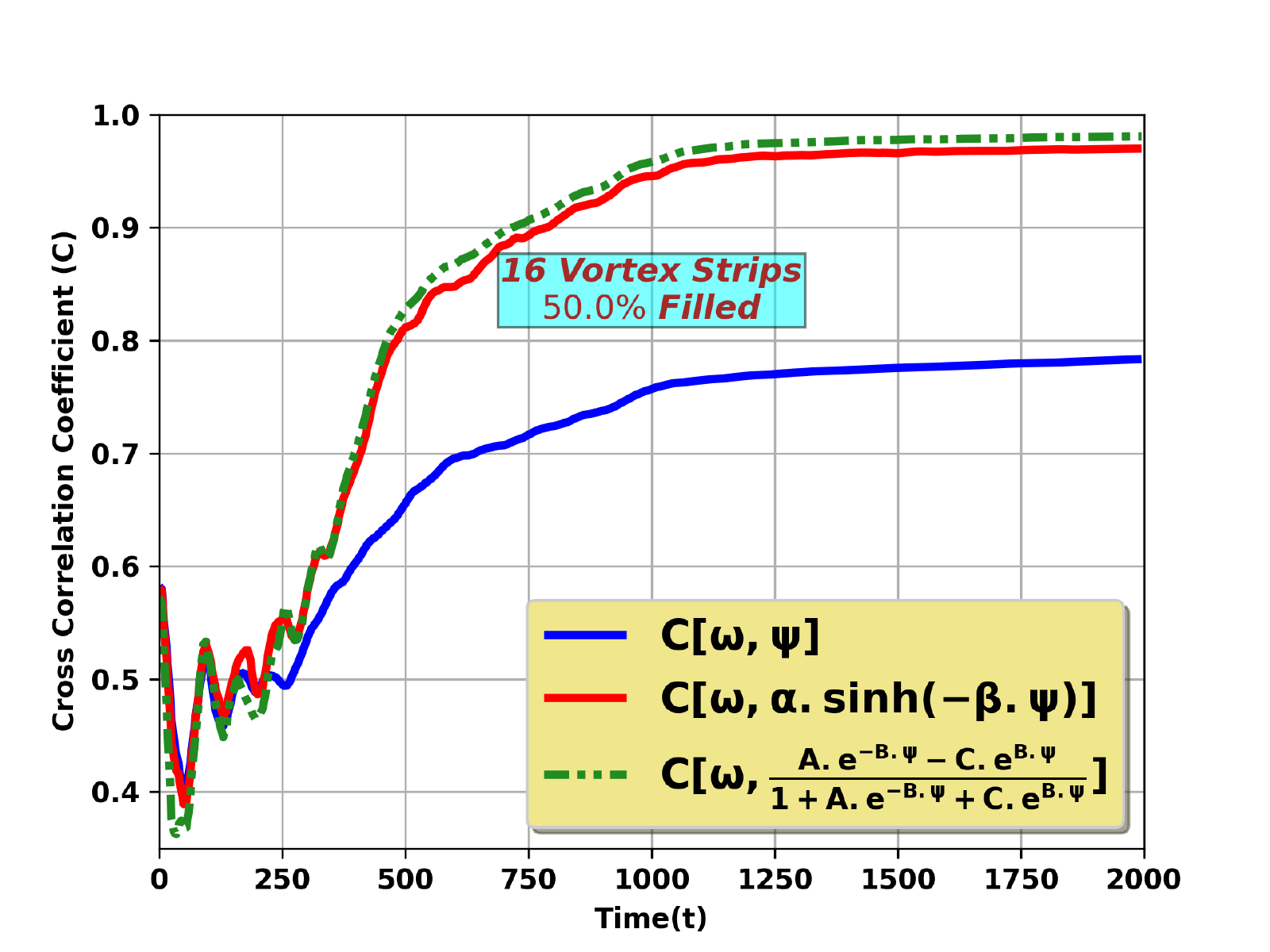}
		\caption{}
		\label{16 Strips CC}
	\end{subfigure}	
	\begin{subfigure}{0.32\textwidth} 
		\centering
		\includegraphics[scale=0.39]{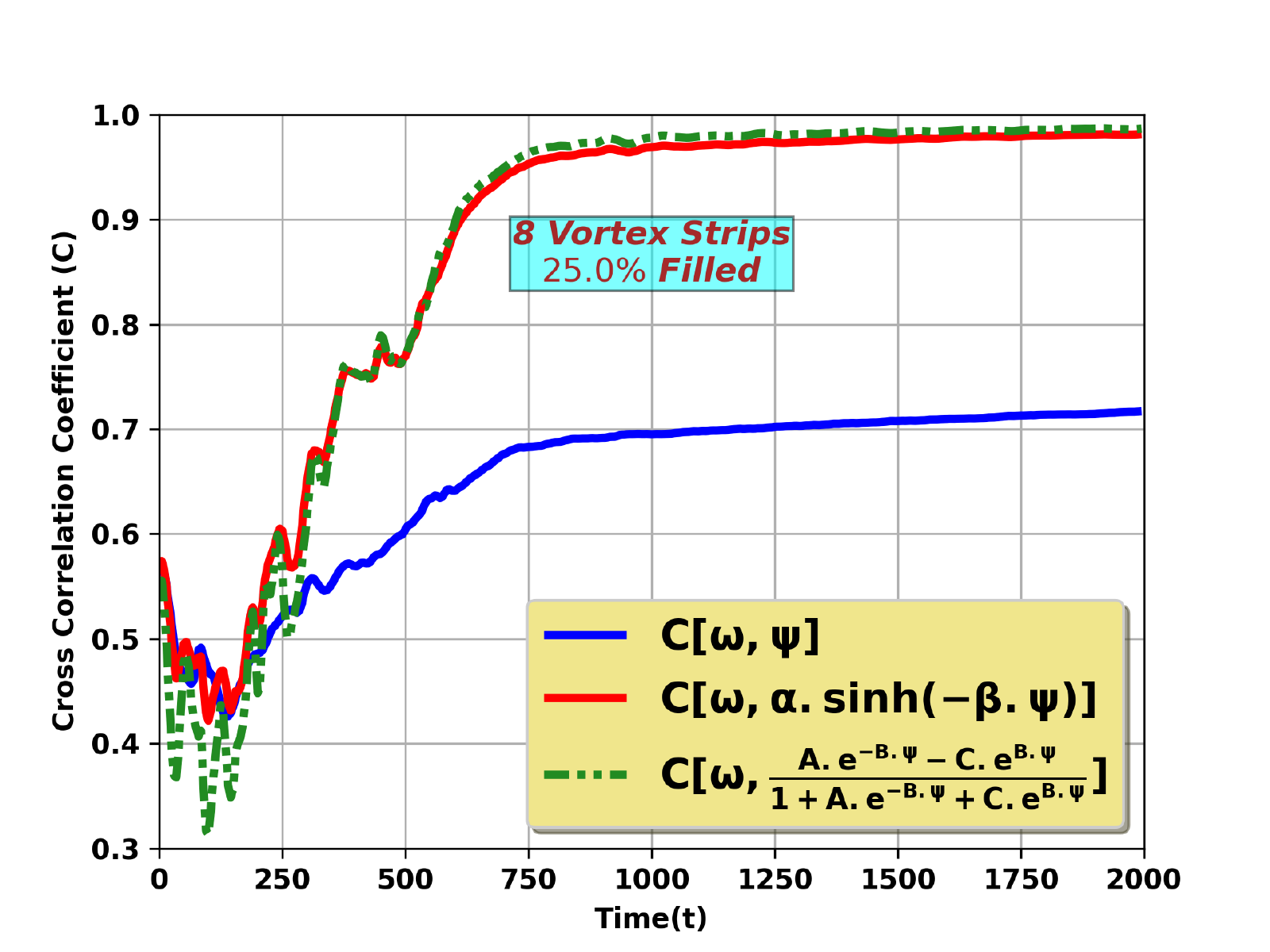}
		\caption{}
		\label{8 Strips CC}
	\end{subfigure} 
	\begin{subfigure}{0.32\textwidth}
		\centering
		\includegraphics[scale=0.39]{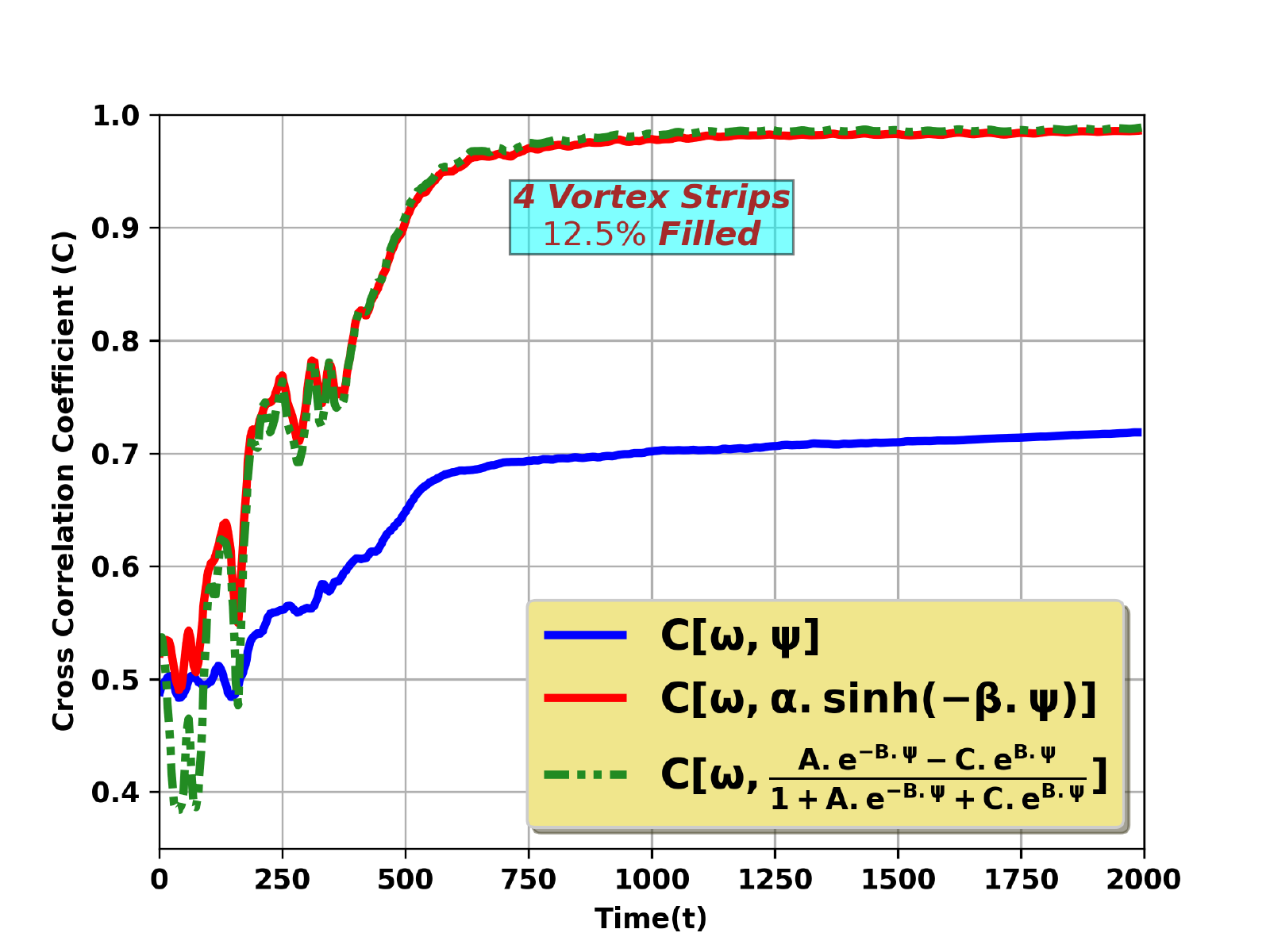}
		\caption{}
		\label{4 Strips CC}
	\end{subfigure} 
	\caption{\textcolor{black}{Dynamics of spatially averaged cross-correlations for, (a) moderately packed [$50.0\%$], (b) loosely packed [$25.0\%$], (c) very loosely packed [$12.5\%$] initial vortex configuration, between C($\omega\textcolor{black}{(x,y,t)}$, $\psi\textcolor{black}{(x,y,t)}$), $C\left(\omega\textcolor{black}{(x,y,t)}, \textcolor{black}{\bar {\omega}_{\textrm{PV}}}\textcolor{black}{(x,y,t)}\right)$ and $C\left(\omega\textcolor{black}{(x,y,t)}, \textcolor{black}{\bar {\omega}_{\textrm{KMRS}}}\textcolor{black}{(x,y,t)}\right)$. Simulation details: grid resolution $2048^2$, stepping time dt = $10^{-4}$, Reynolds number = 228576.}}
	\label{all strips CC}
\end{figure*}

 We also investigate the cross-correlations  $C(\omega\textcolor{black}{(x,y,t)},\psi\textcolor{black}{(x,y,t)})$, $C\left(\omega\textcolor{black}{(x,y,t)}, \textcolor{black}{\bar {\omega}_{\textrm{PV}}}\textcolor{black}{(x,y,t)}\right)$ and $C\left(\omega\textcolor{black}{(x,y,t)}, \textcolor{black}{\bar {\omega}_{\textrm{KMRS}}}\textcolor{black}{(x,y,t)}\right)$ as function of time t, and observe that $C\left(\omega\textcolor{black}{(x,y,t)}, \textcolor{black}{\bar {\omega}_{\textrm{KMRS}}}\textcolor{black}{(x,y,t)}\right)$ still shows best proportionality, $C$ value very near 1 as compared to $C\left(\omega\textcolor{black}{(x,y,t)}, \textcolor{black}{\bar {\omega}_{\textrm{PV}}}\textcolor{black}{(x,y,t)}\right)$ and $C(\omega\textcolor{black}{(x,y,t)},\psi\textcolor{black}{(x,y,t)})$. Though the difference between $C\left(\omega\textcolor{black}{(x,y,t)}, \textcolor{black}{\bar {\omega}_{\textrm{KMRS}}}\textcolor{black}{(x,y,t)}\right)$ and  $C\left(\omega\textcolor{black}{(x,y,t)}, \textcolor{black}{\bar {\omega}_{\textrm{PV}}}\textcolor{black}{(x,y,t)}\right)$ reduce with respect to Case: A, yet there is a quantitative difference  in the late time cross correlation values observed between the models [See Fig. \ref{16 Strips CC}, \textcolor{black}{\& Table \ref{CC Table}}].

\begin{table}
	\color{black}
	\centering
	\begin{tabular}{ |c|c|c|c|c| }
		\hline
		Cases & \textbf{Strips} & $C(\omega,\psi)$ & $C\left(\omega, \bar {\omega}_{\textrm{PV}}\right)$ &  $C\left(\omega, \bar {\omega}_{\textrm{KMRS}}\right)$ \\
		\hline
		A & 20 Strips & $77.8\%$ & $96.5\%$ & $98.5\%$ \\
		\hline 
		B & 16 Strips & $78.3\%$ & $97.0\%$ & $98.0\%$ \\
		\hline
		C & 8 Strips & $71.7\%$ & $ 98.2\%$ & $98.7\%$ \\
		\hline
		D & 4 Strips & $ 71.9\%$ & $  98.6\%$ & $98.8\%$ \\
		\hline
	\end{tabular}
	\caption{\textcolor{black}{The late-time correlation coefficients for each model and for all cases.}}
		\label{CC Table}
\end{table} 

\subsection{8 Vortex Strips with total Packing Fraction $25.0\%$  - Runs 5, 6}\label{8 Strips}
\textcolor{black} {To} achieve point vortex limit systematically, we further reduce the initial total occupancy of vorticities of either kind, which will significantly increase the ``inter-particle'' distance more than earlier. From Fig. \ref{initial} (c) it is seen that the distance between vortex strips is much more larger than earlier case. This time we only fill 25.0$\%$ of simulation domain with vortex strips and keep rest 75.0$\%$ as vorticity free domain. This configuration is almost near to point vortex limit discussed earlier.

The final state of vorticity [See Fig. \ref{8 strips Vorticity}] and stream function [See Fig. \ref{8 strips Stream function}] for this configuration similar with earlier two cases but with several interesting differences as we will see shortly.
The final state of  Okubo-Weiss parameter ($Q(x,y,t)$) from our simulation is shown in Fig. \ref{8 strips Okubo}. We identify regions of Vortex cores ($Q(x,y,t)$) and strain cells surrounding the vortex cores($Q(x,y,t)>0$) like earlier cases.

\begin{figure*}
	\centering
	\begin{subfigure}{0.45\textwidth}
		\centering
		\includegraphics[scale=0.55]{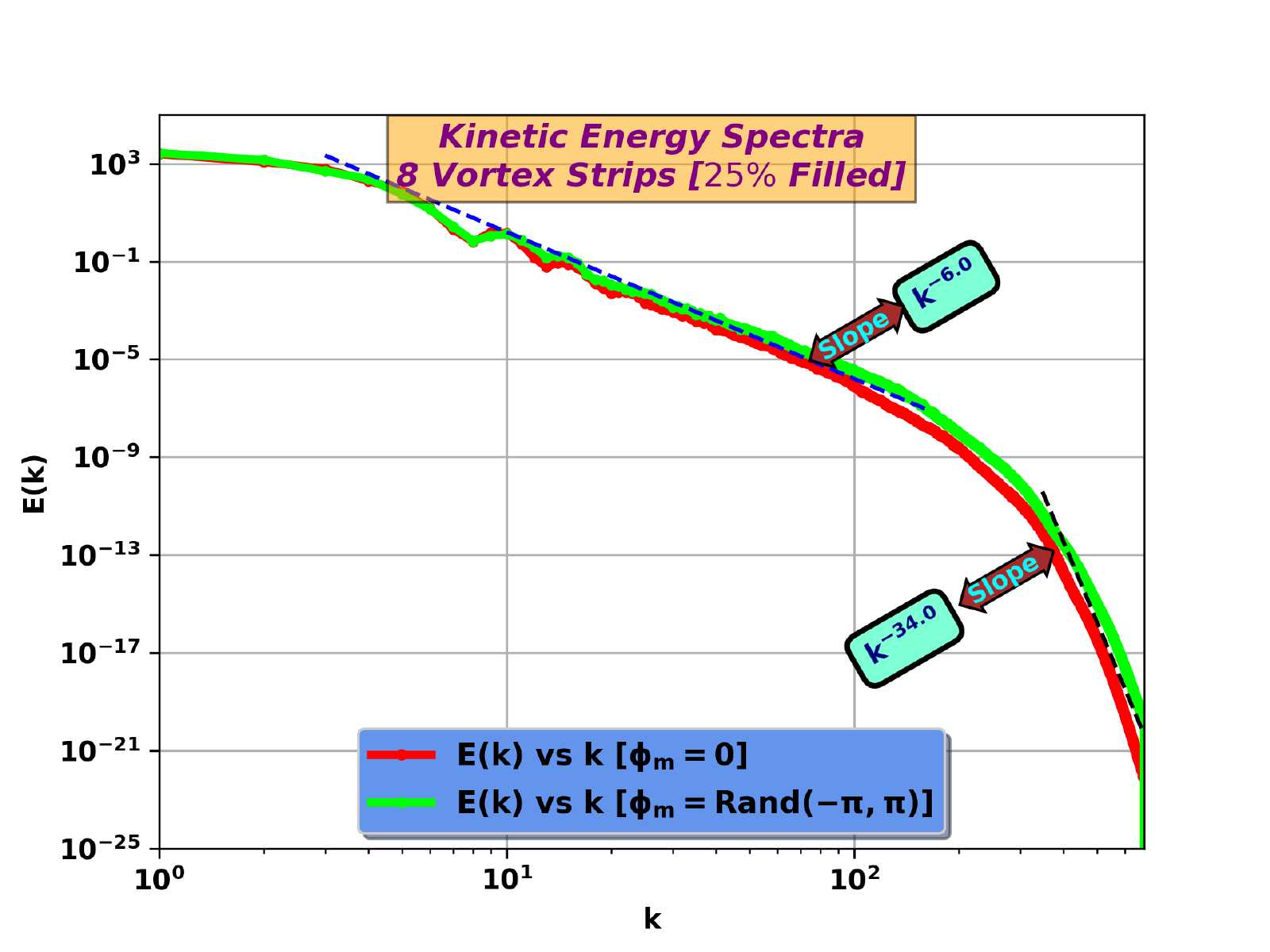}
		\caption{}
	\end{subfigure}
	\begin{subfigure}{0.45\textwidth}
		\centering
		\includegraphics[scale=0.55]{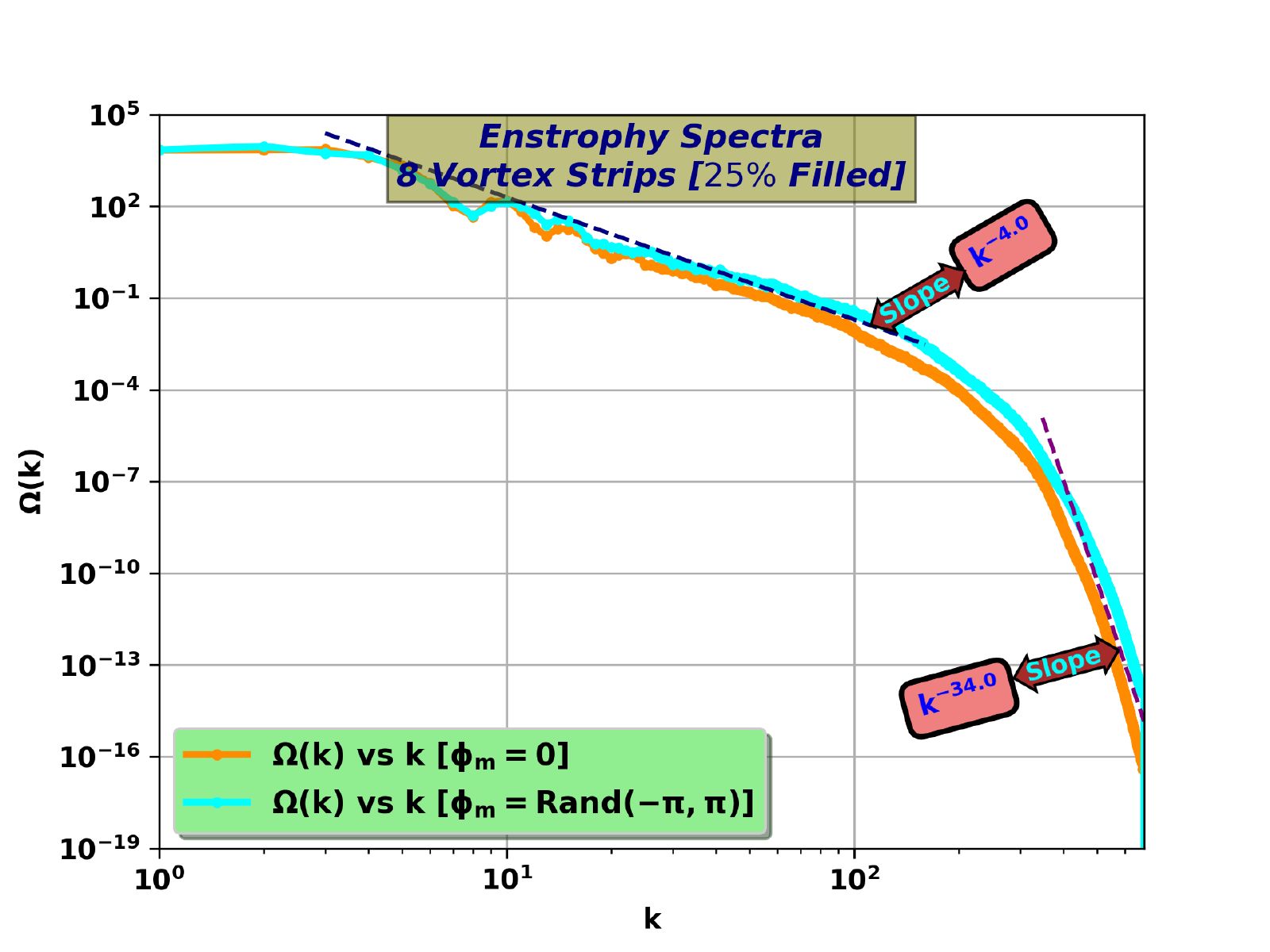}
		\caption{}
	\end{subfigure}
	\caption{\textcolor{black}{(a)} Time averaged (time average taken after saturation i.e from t=1000.0 to 1600.0) kinetic energy spectra [$\int_{0}^{\infty} E(k) dk$] showing inverse cascading (b) Time averaged (time average taken after saturation i.e from t=1000.0 to 1600.0) enstrophy spectra [$\int_{0}^{\infty} \Omega(k) dk$] showing direct cascading for loosely packed 8 Vortex strips ($25.0\%$ packed) configuration. These k-scaling\textcolor{black}{s} for kinetic energy and enstrophy agree with the earlier work \cite{gomez:1996}. Simulation details: grid resolution $2048^2$, stepping time dt = $10^{-4}$, Reynolds number = 228576.}
	\label{8 strips Spectra}
\end{figure*}

Kinetic energy spectra is also seen to indicate inverse cascading with scaling of $E(k)\propto k^{-6}$ for lower $k$ and $E(k)\propto k^{-34}$ for higher $k$ [See Fig. \ref{8 strips Spectra} (a)], whereas enstrophy spectra shows direct cascading with scaling $E(k)\propto k^{-4}$ for lower $k$ and $E(k)\propto k^{-34}$ for higher $k$ [See Fig. \ref{8 strips Spectra} (b)] like earlier two cases. These k-scaling for kinetic energy and enstrophy agree with earlier published works, for example, by Dmiturk et al. \cite{gomez:1996} for slowly decaying 2-dimensional Navier-Stokes turbulence.\\

 We plot $\psi$ vs $\omega$ and fit the scatter data  with the two earlier proposed functions (Eq. \ref{omega psi for finite vortex} and Eq. \ref{omega psi for point vortex}). It is identified that both the function fit well [See Fig. \ref{8 Strips Scatter}, \ref{8 Strips Scatter phase}]. The numerical values of the fitting parameters are given in Table \ref{Fitting set 1} and \ref{Fitting set 2}. From Table \ref{Fitting set 1}, it is seen that the value of $A$ and $C$ coefficients are almost negligible with respect to 1, approaching the the point vortex approximation, which is expected from our earlier discussion.

The cross-correlations  $C(\omega\textcolor{black}{(x,y,t)},\psi\textcolor{black}{(x,y,t)})$, $C\left(\omega\textcolor{black}{(x,y,t)}, \textcolor{black}{\bar {\omega}_{\textrm{PV}}}\textcolor{black}{(x,y,t)}\right)$ and $C\left(\omega\textcolor{black}{(x,y,t)}, \textcolor{black}{\bar {\omega}_{\textrm{KMRS}}}\textcolor{black}{(x,y,t)}\right)$ as function of time t, also indicate the same. We identify that both $C\left(\omega\textcolor{black}{(x,y,t)}, \textcolor{black}{\bar {\omega}_{\textrm{KMRS}}}\textcolor{black}{(x,y,t)}\right)$  and $C\left(\omega\textcolor{black}{(x,y,t)}, \textcolor{black}{\bar {\omega}_{\textrm{PV}}}\textcolor{black}{(x,y,t)}\right)$  \textcolor{black}{show} best proportionality, $C$ value very near 1 as compared to $C(\omega\textcolor{black}{(x,y,t)},\psi\textcolor{black}{(x,y,t)})$. \textcolor{black}{Obviously, the} difference between  $C\left(\omega\textcolor{black}{(x,y,t)}, \textcolor{black}{\bar {\omega}_{\textrm{KMRS}}}\textcolor{black}{(x,y,t)}\right)$  and $C\left(\omega\textcolor{black}{(x,y,t)}, \textcolor{black}{\bar {\omega}_{\textrm{PV}}}\textcolor{black}{(x,y,t)}\right)$ reduces further with respect to Case: A and Case: B and it is almost negligible [See Fig. \ref{8 Strips CC} \textcolor{black}{\& Table \ref{CC Table}}].


\subsection{4 Vortex Strips with total Packing Fraction $12.5\%$  - Runs 7, 8}\label{4 Strips}
Finally to investigate if the point vortex limit is reached asymptotically, we  further reduce the initial vorticity packing fraction. This time we use only 12.5 $\%$ of simulation domain filled by the vortices and rest of the domain contains zero vorticity  [Fig. \ref{initial} (d)]. The vortex evolution is found to be largely similar to the earlier cases (Case: A, B and C). The final vortex state is dominated by two vortices of either sign [See Fig. \ref{4 strips Vorticity}], also the stream function structure of late time state is found to be similar to Ewald potential contours [See Fig. \ref{4 strips Stream function}].

We also calculate the final state of  Okubo-Weiss parameter ($Q(x,y,t)$) from our simulation shown in Fig. \ref{4 strips Okubo}, which is found to be identical like earlier cases.

Like earlier cases, we plot $\psi$ vs $\omega$  and fit the scatter data with the two functions- one is derived from \textcolor{black}{a} statistical \textcolor{black}{mechanics} point of view taking finite size into account (Eq. \ref{omega psi for finite vortex}), another is the famous point vortex limit most probable state (Eq. \ref{omega psi for point vortex}). From Fig. \ref{4 Strips Scatter}, \ref{4 Strips Scatter phase} it is observed that both the function fits well and goes one upon another. The numerical values of the fitting parameters are given in Table \ref{Fitting set 1} and \ref{Fitting set 2}. From Table \ref{Fitting set 1} it is confirmed that the value of $A$ and $C$ coefficients are reduced further small amount with respect to earlier (Case: C) and it is totally negligible with respect to 1. The sequential reduction of the $A$ and $C$ values are strongly pointing \textcolor{black}{at} the conversion from finite dimensional vortex limit to point vortex limit, addressed earlier from \textcolor{black}{an} analytical point of view of KMRS theory \cite{Kuzmin:1982,Miller_PRL:1990,robert_sommeria:1991}.

For this lowest circulation case we calculate  the cross-correlations  $C(\omega\textcolor{black}{(x,y,t)},\psi\textcolor{black}{(x,y,t)})$, $C\left(\omega\textcolor{black}{(x,y,t)}, \textcolor{black}{\bar {\omega}_{\textrm{PV}}}\textcolor{black}{(x,y,t)}\right)$ and $C\left(\omega\textcolor{black}{(x,y,t)}, \textcolor{black}{\bar {\omega}_{\textrm{KMRS}}}\textcolor{black}{(x,y,t)}\right)$ as function of time t, and identify that both $C$ shows best proportionality, $C$ value very near 1 as compare to $C(\omega\textcolor{black}{(x,y,t)},\psi\textcolor{black}{(x,y,t)})$. Both the correlation overlaps, which basically signifies that the point vortex approximation has been achieved [See Fig. \ref{4 Strips CC}  \textcolor{black}{\& Table \ref{CC Table}}].

\section{Summary and Conclusion}
In this work, we have performed direct numerical simulations of 2-dimensional decaying NS turbulence at high Reynolds number using fine grid resolution. 

Our purpose is to compare the late time high Reynolds number and high grid size  DNS results with  statistical mechanical predictions of the final state of such a system obtained using entropy extremization using point vortex  and patch (or finite size) vortex models (KMRS theory) as well as with enstrophy extremization theory, with systematic control over initial vortex packing fraction values. For addressing the same, we have developed a GPU based high performance solver and have benchmarked the same with Kelvin-Helmholtz instability for two oppositely directed jets (broken-jets) with Drazin's analytical prediction\cite{drazin:1961}. After this high quality benchmarking, we have presented a series of runs for various initial total positive and negative circulation values (or packing fractions) such that total circulation is zero.

Our major findings are :

$\bullet$ For a  tightly packed vortex configuration, with high individual circulations ($C_{+}/C_{-}$) such that the initial total circulation is, $C = C_{+} + C_{-} = \int \omega dx dy = 0$, the final relaxed state is found to be quantitatively close to the most probable state predicted from \textcolor{black}{the} statistical mechanical theory based of vortex patches (i.e, KMRS theory) \cite{Kuzmin:1982,Miller_PRL:1990,robert_sommeria:1991}.

$\bullet$ To move \textcolor{black}{systematically} towards point vortex limit, system is initialized with +/- sign vortex strips keeping total circulation zero ($C = C_{+} + C_{-} = \int \omega dx dy = 0$) but with decreasing values of $C_+$ and $C_-$. We observe that as the \textcolor{black}{intial vorticity} packing fraction reduces, the late time states from our high resolution and high $R_n$ 2-dimensional DNS  agree well with KMRS \cite{Kuzmin:1982,Miller_PRL:1990,robert_sommeria:1991} predictions while substantially deviate from the predictions of point vortex theory. In the limit of small packing fraction, late time DNS results, KMRS prediction and that of point vortex theory (i.e., sinh-Poisson eqn), all three of them tend to concur. Our numerical findings validate a clear and unambigous transition of the late time 2-dimensional decaying DNS datum from the predictions of finite size vortex theory (KMRS theory) to point vortex theory as a function of initial circulation or vortex packing fraction.

 To conclude, with high resolution and at very high Reynolds number, we  investigate the systematic deviation of final relaxed state of 2-dimensional incompressible decaying NS turbulence  from \textcolor{black}{the} sinh-Poisson model predicted by statistical mechanical theory of point vortices and \textcolor{black}{show that the late time states agree} quantitatively with the predictions of KMRS theory for increasing values of initial total circulation of either kind ($C_{+}/C_{-}$) such that $C = C_{+} + C_{-} = 0$. We believe that this is a first systematic comparison of high $R_n$, high grid resolution 2-dimensional DNS results at late times and those of statistical mechanical models, which unambiguously brings out the effect of initial total circulation and classical exclusion principle in these models. \textcolor{black}{In our present work, due to very high grid resolution used, the numerical errors in evaluation of enstrophy and energy are sufficiently small and do not impact our results. However, it would indeed be interesting to use numerical methods, which are kinetic energy preserving and enstrophy preserving  by their very construction \cite{CHARNYI:2017,EDOH:2022}. We hope to attempt this interesting exercise in the near future.}\\



\section{ACKNOWLEDGMENTS}
The simulations and visualizations presented here are performed on GPU nodes and visualization nodes of Antya cluster at the Institute for Plasma Research (IPR), INDIA. One of the author S.B is thankful to Dr. Rupak Makherjee [PPPL, USA] for providing the sequential version of the solver. S.B thanks N. Vydyanathan, Bengaluru and B. K. Sharma at NVIDIA, Bengaluru, India, for extending their help with  basic GPU methods. S.B is grateful to Mr. Soumen De Karmakar at IPR for many helpful discussions regarding GPUs, and HPC support team of IPR for extending their help related to ANTYA cluster.

\bibliography{biblio}


\appendix
\section{$\psi$ vs $\omega$ scatter plot ($\phi =$ random noise)}\label{Appen A}
\begin{figure*}
	\centering
	\begin{subfigure}{0.32\textwidth} 
		\centering
		\includegraphics[scale=0.39]{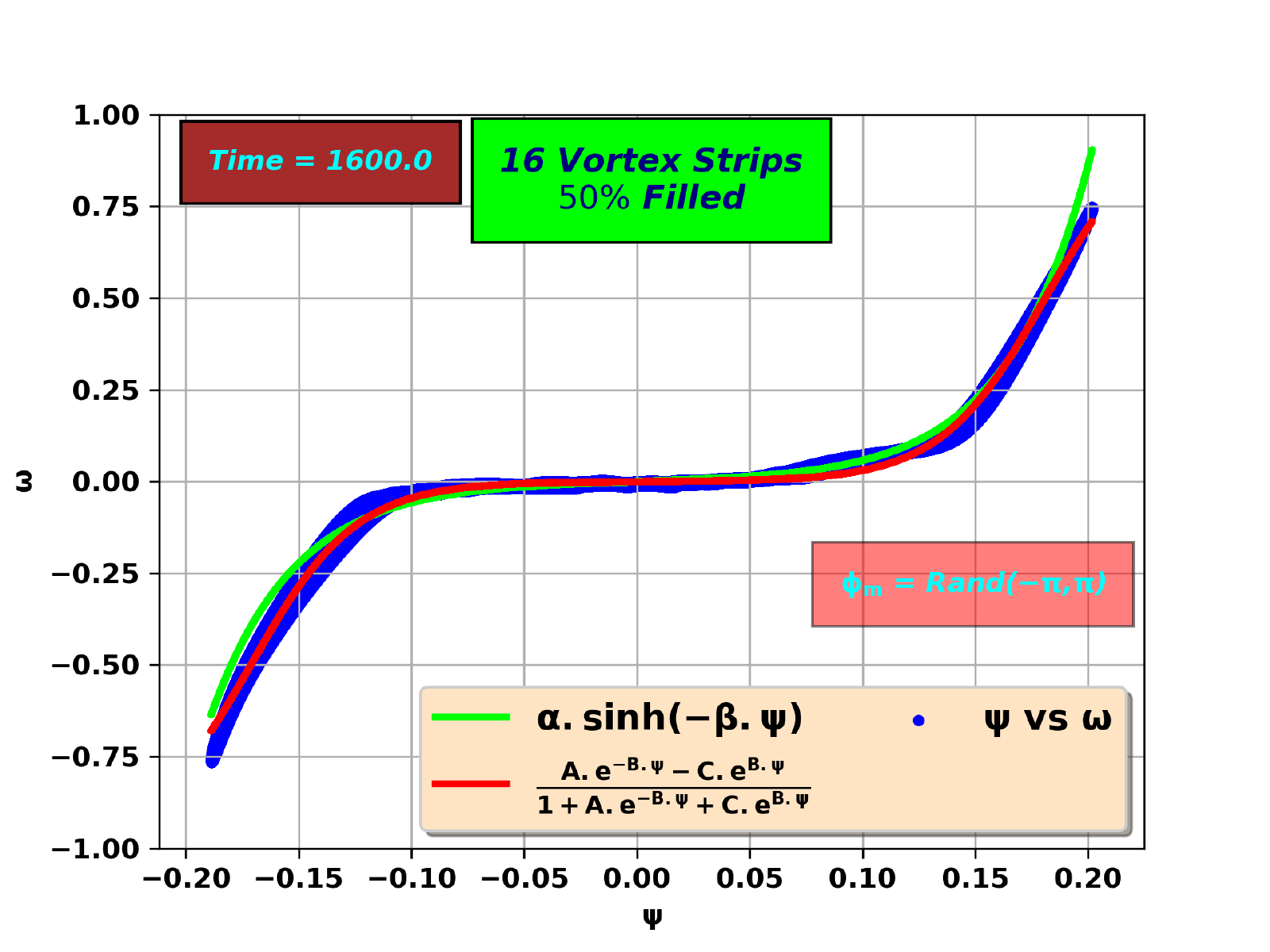}
		\caption{}
		\label{16 Strips Scatter phase}
	\end{subfigure}	
	\begin{subfigure}{0.32\textwidth} 
		\centering
		\includegraphics[scale=0.39]{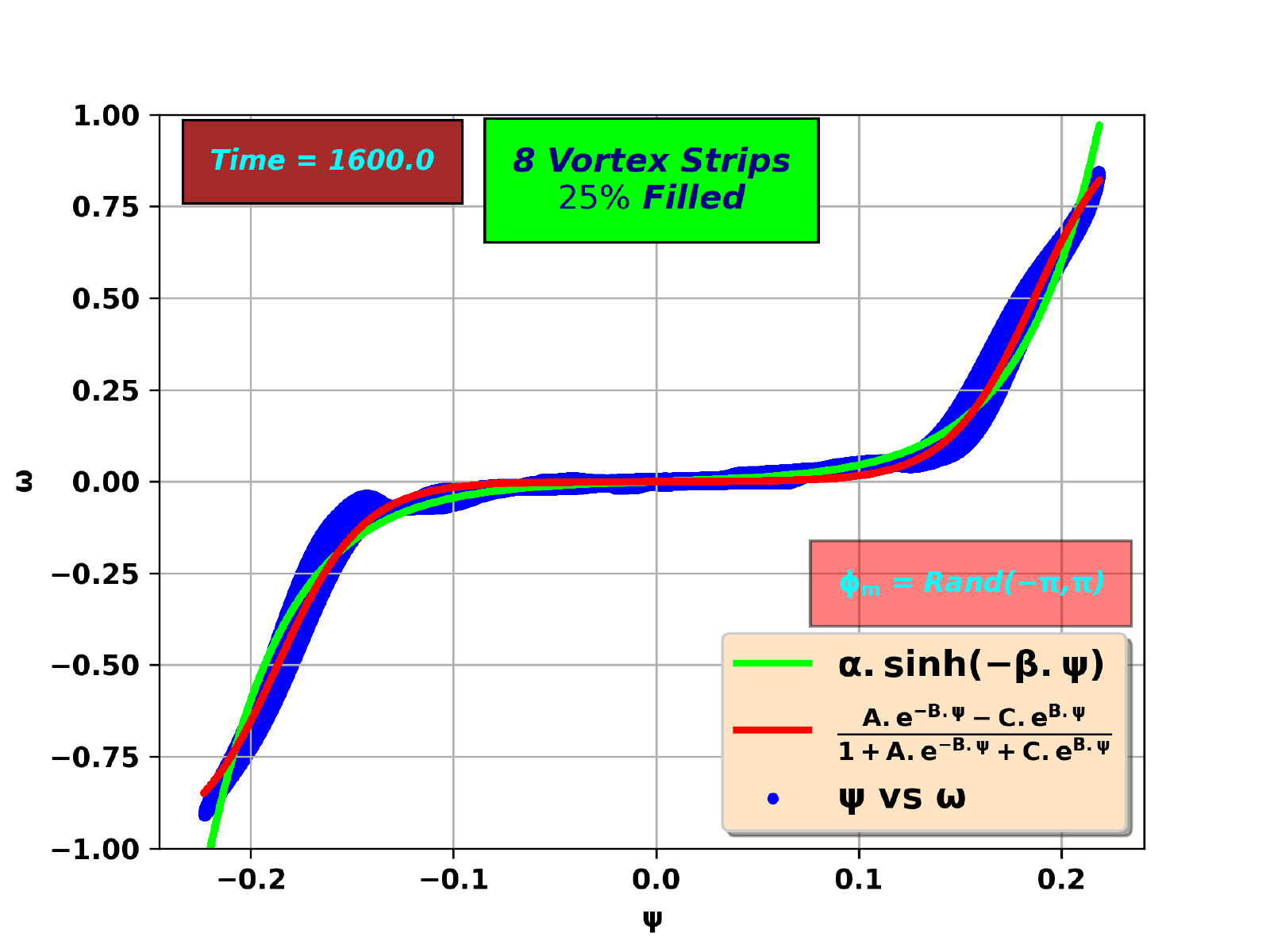}
		\caption{}
		\label{8 Strips Scatter phase}
	\end{subfigure} 
	\begin{subfigure}{0.32\textwidth}
		\centering
		\includegraphics[scale=0.39]{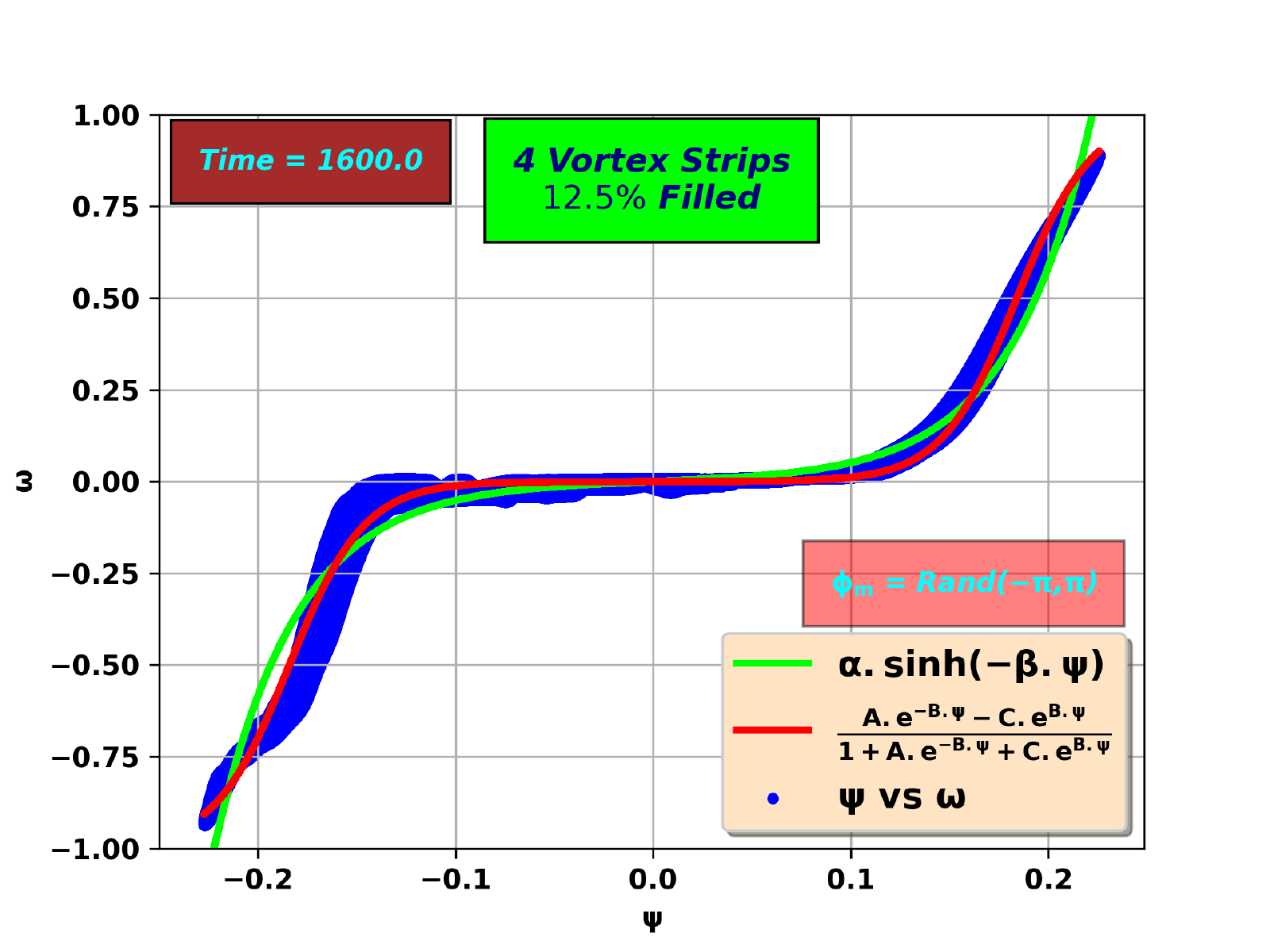}
		\caption{}
		\label{4 Strips Scatter phase}
	\end{subfigure} 
	\caption{\textcolor{black}{$\psi$ vs $\omega$ scatter plot (at Time = 1600) for, (a) moderately packed [$50.0\%$], (b) loosely packed [$25.0\%$], (c) very loosely packed [$12.5\%$] initial vortex configuration. Both patch vortex model (red line) and the one from \textcolor{black}{a} point vortex model (green line) for $\psi$ vs $\omega$  are shown. Simulation details: grid resolution $2048^2$, stepping time dt = $10^{-4}$, Reynolds number = 228576, $\phi_m = rand(-\pi, \pi)$.}}
	\label{all strips Scatter with phase}
\end{figure*}

\textcolor{black}{We plot $\psi$ vs $\omega$ for the case with random noise in perturbation and fit the function for point vortex approximation Eq. \ref{omega psi for point vortex} and function for finite size vortex approximation Eq. \ref{omega psi for finite vortex}. It is observed that the function (Eq. \ref{omega psi for finite vortex}) continues to show best proportionality with respect to Eq. \ref{omega psi for point vortex} for higher intial vorticity packing fraction [See Fig. \ref{16 Strips Scatter phase}]. As the initial vorticity packing fraction reduces, the late time states agree well with KMRS predictions [See Fig. \ref{8 Strips Scatter phase}, \ref{4 Strips Scatter phase}].}

\end{document}